%% file: main.tex
\documentclass{elsarticle}
\makeatletter
\def\ps@pprintTitle{%
  \let\@oddhead\@empty
  \let\@evenhead\@empty
  \let\@oddfoot\@empty
  \let\@evenfoot\@oddfoot}
\makeatother

\usepackage{graphicx} 
\usepackage{hyperref} 

\usepackage{amsmath} 
\usepackage{amssymb}

\usepackage{tabularx} 
\usepackage{booktabs} 
\usepackage{caption} 
\usepackage{subcaption} 

\usepackage{lipsum}  

\usepackage{xcolor} 
\usepackage{multicol}

\newcommand{\cm}[1]{\textcolor{blue}{#1}}

\newcommand{\commentout}[1]{}

\begin{document}
\begin{frontmatter}

\title{A manifold-aware Neural ODE surrogate model for stochastic induction heating with anisotropic electrical conductivity}

\author[1]{Wouter J. Schuttert\corref{cor1}}
\ead{w.j.schuttert@utwente.nl}
\cortext[cor1]{Corresponding author.}

\author[1]{Mohammed Iqbal Abdul Rasheed}

\author[1,2]{Bojana Rosi\'{c}}

\address[1]{Applied Mechanics and Data Analysis chair,
  Faculty of Engineering Technology,\\ 
  Drienerlolaan 5, 7522 NB Enschede, The Netherlands\\
  University of Twente}

\address[2]{Digital Engineering,\\  
  Lehargasse 6, 1030 Wien, Austria\\
  TU Wien }

\begin{abstract}

Induction welding plays a central role in enabling lightweight, integrated structures made from fibre-reinforced thermoplastic composites. From a modelling perspective, the induction welding process can be approximated by one-way coupled electromagnetic and heat-transfer equations. In practice, material parameters such as electrical conductivity vary significantly resulting from the deviations in the placement of fibres and hence the fibre-fibre contacts in the mesostructure are governed by consolidation quality of the material. Explicit representation of this variability on the macroscopic scale is essential to capture the closed current loops required for the induction heating process. To address this, a stochastic material model is introduced that respects the symmetric positive-definite (SPD) nature of the conductivity tensor and separates scaling and orientation uncertainties, forming the basis of a surrogate framework. The resulting stochastic conductivity model is first used to quantify the uncertainty in the induction heating process through extensive Monte Carlo simulations, providing detailed insight into the induced currents and the resulting temperature field. Subsequently, to enable efficient uncertainty propagation, SPD-aware surrogate models are trained with a subset of the simulation data, consisting of labelled material states and temperature fields. The surrogates are formulated as Constitutive Manifold Neural Networks (CMNNs) that explicitly respect the underlying SPD manifold structure and are integrated with a Neural Ordinary Differential Equation (NODE) framework to capture temporal dynamics. Several NODE integration schemes are evaluated and compared. The results reveal that stochastic conductivity substantially drives the temperature variability, with uncertainty increasing continuously over time. The proposed CMNN-NODE surrogates deliver accurate and reproducible predictions on unseen test cases and long-horizon rollouts, with the Neural ODE integrators improving run-to-run consistency.

\end{abstract}
\end{frontmatter}



\input{Symbols}
\section{Introduction}

\input{1_Introduction/introduction}\label{sec:introduction}

\section{Modelling of induction heating with stochastic electrical conductivity}

\input{2_Problem/Problem}\label{sec:problem}

\section{Surrogate modelling using constitutive manifolds and neural ODEs}

\input{3_Method/Method}\label{sec:method}

\section{Induction heating computations}

\input{4_Result/Result_Det}

\subsection{Numerical Results}

\input{4_Result/Result_Stochastic}

\subsection{Surrogate modelling for induction heating}

\input{4_Result/Result_Surrogates}

\section{Conclusion}

\input{5_Conclusion/Conclusion}

\section{Declarations}
\noindent
The datasets used and/or analysed during the current study are available from the corresponding author on reasonable request.

This research was carried out as part of the project ENLIGHTEN \cite{noauthor_enlighten_nodate} (project number N21010 - m) in the framework of the Partnership Program of the Materials Innovation institute M2i (www.m2i.nl) and the
Netherlands Organization for Scientific Research (www.nwo.nl).

W.J.S. wrote the initial manuscript, contributed to the problem formulation and methodology, set up the models, and generated the results and figures under the supervision of M.I.A.R. and B.R. B.R. contributed to the methodology, problem formulation, and manuscript writing. M.I.A.R. assisted with the realization of the case study. All authors reviewed and approved the final manuscript.

\appendix
\input{6_Appendix/mat_param}


\bibliographystyle{ieeetr}
\bibliography{references}

\end{document}

%% file: Symbols.tex

\setlength{\abovedisplayskip}{5pt}
\setlength{\belowdisplayskip}{5pt}
\setlength{\abovedisplayshortskip}{5pt}
\setlength{\belowdisplayshortskip}{5pt}

\newcommand{\refEQ}[1]{Eq.(#1)}      
\newcommand{\refFIG}[1]{Fig.(#1)}    
\newcommand{\refAPP}[1]{App.(#1)}    

\newcommand{\varElec}{\mathbf{E}}
\newcommand{\varBmag}{\mathbf{B}}
\newcommand{\varDelec}{\mathbf{D}}
\newcommand{\varHmag}{\mathbf{H}}
\newcommand{\varJcur}{\mathbf{J}}
\newcommand{\varMvec}{\mathbf{A}}
\newcommand{\varMvecDisc}{\varMvec_h}

\newcommand{\varEcharge}{\rho^{(e)}}
\newcommand{\varPermit}{\epsilon}
\newcommand{\varPermea}{\mu^{(m)}}
\newcommand{\varPermitZero}{\varPermit_0}
\newcommand{\varPermeaZero}{\varPermea_0}
\newcommand{\varEcond}{\boldsymbol{\sigma}}

\newcommand{\varFcoil}{f^{(c)}}

\newcommand{\varCelsius}{^\circ\text{C}}
\newcommand{\varHflux}{q}
\newcommand{\varHsource}{Q}
\newcommand{\varTemp}{u}
\newcommand{\varTempDisc}{\varTemp_h}
\newcommand{\varTempEle}{\varTemp_e}
\newcommand{\varTempI}{\varTemp_i}

\newcommand{\varDens}{\varrho}
\newcommand{\varHeatC}{c_p}
\newcommand{\varTcond}{\boldsymbol{\kappa}}

\newcommand{\varFEmass}{\boldsymbol{M}}
\newcommand{\varFEdamping}{\boldsymbol{C}}
\newcommand{\varFEstiffness}{\boldsymbol{K}}
\newcommand{\varFEload}{\boldsymbol{P}}

\newcommand{\varx}{\boldsymbol{x}}
\newcommand{\vartime}{t}
\newcommand{\varEspace}{\varOmega}
\newcommand{\varSigmaAlgebra}{\mathcal{F}}
\newcommand{\varProbMeas}{\mathbb{P}}
\newcommand{\varBorel}{\mathcal{B}}
\newcommand{\varProbSpace}{(\varEspace, \varSigmaAlgebra, \varProbMeas )}
\newcommand{\vardiag}{\mathfrak{d i a g}}
\newcommand{\varDiag}{\text{Diag+}}
\newcommand{\varso}{\mathfrak{s o}}
\newcommand{\varSO}{\mathrm{SO}}
\newcommand{\varReal}{\mathbb{R}}
\newcommand{\varTime}{\mathcal{T}}
\newcommand{\varDomain}{\mathcal{G}}
\newcommand{\varSymPlus}{\text{Sym}^\text{+}}
\newcommand{\varSym}{\text{Sym}}
\newcommand{\varSobo}{H^1}

\newcommand{\varEvent}{\omega}
\newcommand{\varRandomVariableNot}{\kappa}
\newcommand{\varRV}{\varRandomVariableNot(\varEvent)}
\newcommand{\varRfield}{\varRandomVariableNot(\varx,\varEvent)}
\newcommand{\varRVmean}{\Bar{\varRandomVariableNot}}
\newcommand{\varRVstd}{\Tilde{\varRandomVariableNot}}
\newcommand{\varVMFref}{\mu}
\newcommand{\varVMFcon}{\eta}

\newcommand{\varTensor}{\varEcond}  
\newcommand{\varEigVal}{\boldsymbol{\mathit{\Lambda}}}
\newcommand{\varEigVec}{\boldsymbol{V}}
\newcommand{\varLN}{\lambda}
\newcommand{\varSPIN}{\phi}
\newcommand{\varAXIS}{\boldsymbol{b}}
\newcommand{\varTensorDet}{\overline{\varTensor}}
\newcommand{\varEigValDet}{\overline{\varEigVal}}
\newcommand{\varEigVecDet}{\overline{\varEigVec}}

\newcommand{\varGRF}{\theta}
\newcommand{\varRF}{\kappa}
\newcommand{\varCov}{\text{Cov}}
\newcommand{\varCorr}{\text{Corr}}
\newcommand{\varURF}{\text{U}}
\newcommand{\varRFlog}{y}
\newcommand{\varRFvm}{\varRF^\text{vM}}
\newcommand{\varRFvmf}{\varRF^\text{vMF}}

\newcommand{\varS}{\varx}
\newcommand{\varp}{\varEvent}
\newcommand{\varP}{\boldsymbol{\varEvent}}
\newcommand{\varSp}{\varx,\varp}
\newcommand{\varSP}{\varx,\varp}

\newcommand{\varWeights}{\mathbf{\theta}}
\newcommand{\varLope}{A}
\newcommand{\varLidx}{k}
\newcommand{\varL}{\ell}                     
\newcommand{\varLcur}{\varL^{\varLidx}}      
\newcommand{\varLpre}{\varL^{\varLidx-1}}    
\newcommand{\varLn}{n}

\newcommand{\REDwjs}[1]{\textcolor{red}{#1}}   

%% file: 1_Introduction/Introduction.tex
\noindent
Meeting climate targets for a sustainable future requires major reductions in carbon emissions from the transportation sector, which motivates the replacement of metals in structural components with lightweight composite materials \cite{miller_assessment_2025}. Unidirectional carbon fibre reinforced thermoplastics (CFRTP) combine tailorable fibre orientations for specific load cases along with the ability to be reprocessed and recycled through remelting \cite{zhang_current_2020}. Fusion bonding is an increasingly adopted method to integrate CFRTP parts into larger structures in which the interface between the laminates is heated and consolidated under pressure.
A prevalent fusion bonding technique is induction welding, where heat is generated in a laminate through an alternating magnetic field originating from a coil \cite{velmurugan_study_2020}. In induction welding, the temperature history at the interface is a key indicator of bond quality, but it cannot be measured directly due to practical difficulties \cite{korycki_assembling_2022}, necessitating indirect characterisation through modelling. Hence, the industrial adoption of induction welding for CFRTP depends on predictive material models that support process design and integration into larger structures \cite{yousefpour_fusion_2004, worral_joining_2020}.

Physics-based predictive models estimate the process-indicative temperature field using coupled thermo-electromagnetic models. However, such models are typically deterministic and do not account for variations in electrical conductivity, to which the temperature field is highly sensitive \cite{grouve_simulating_2021}. In CFRTPs, electrical conductivity is inherently anisotropic and is fully characterised by a symmetric positive-definite (SPD) conductivity tensor, whose eigenvalues represent the directional conductivities and whose corresponding eigenvectors define the principal material axis aligned with the fibre orientation. Consequently, because the conductivity is greatest along the fibre direction, an applied electric field induces current that flows preferentially along the fibres. This electrical response is strongly influenced by deviations in fibre placement, the quality of fibre--fibre contacts established during consolidation, and the formation of continuous conductive pathways within the layup. However, existing models rarely represent the associated uncertainty explicitly \cite{buser_characterisation_2022}. Since the conductivity must remain symmetric and positive definite under uncertainty, an appropriate stochastic representation must preserve these properties. A probabilistic framework satisfying this requirement has been established in the computational mechanics literature \cite{guilleminot_stochastic_2013,shivanand_stochastic_2024}. A parametric formulation of this framework \cite{shivanand_stochastic_2024,schuttert_modeling_2024} has recently been applied for stationary heat problem. In the present work, we utilize this formulation to model the induction heating given the full random conductivity tensor of unidirectional CFRTPs, thereby enabling uncertainty quantification of the predicted temperature history that is time-dependent.

Predicting the temperature response corresponding to a probability distribution of electrical conductivity requires propagating uncertainty through the coupled electromagnetic--thermal finite element (FEM) model \cite{cheng_modeling_2019}. Direct uncertainty propagation using Monte Carlo sampling is computationally prohibitive, as each coupled FEM simulation is expensive and thousands of realisations are typically required. This motivates the use of surrogate models that approximate the high-fidelity solver while enabling efficient evaluation across many conductivity realisations \cite{sudret_surrogate_2017}. However, existing surrogate approaches for composite processing and induction welding of CFRTPs have largely focused on scalar quantities of interest, and no established methods are available for predicting the full spatial temperature field throughout the process history, which constitutes the high-dimensional output considered in the present work. Among the available surrogate modelling techniques, deep neural networks are particularly well suited to learning such high-dimensional nonlinear mappings. Deep neural networks have emerged as effective surrogates for expensive nonlinear physics models \cite{abdar_review_2021, tripathy_deep_2018}. However, standard architectures treat all inputs as Euclidean vectors, discarding any structure inherent to the underlying physics. In the present problem, this matters particularly because electrical conductivity is described by an SPD tensor describing fibre orientation and magnitude. The set of such tensors forms a smooth manifold rather than a flat Euclidean space, meaning that standard networks which treat conductivity inputs as ordinary vectors ignore this geometric behaviour \cite{pennec_riemannian_2006}. Manifold-aware neural networks that respect input geometry have shown improved generalisation over Euclidean counterparts in computer vision and brain imaging \cite{huang_riemannian_2017, brooks_riemannian_2019}. Constitutive manifold neural networks extend this concept to materials modelling through SPD layers that separately capture magnitude and orientation effects, mirroring the physical decomposition of the conductivity tensor \cite{schuttert_constitutive_2026}. 

In addition to representing the conductivity tensor appropriately, the surrogate must capture the transient evolution of the temperature field during induction welding. Neural ordinary differential equations (Neural ODEs) achieve this by embedding an ODE solver within the network, allowing the integration error to be controlled adaptively \cite{chen_neural_2018}. Residual networks provide a computationally efficient alternative, as their residual blocks admit a discrete-time ODE interpretation \cite{he_deep_2016}, enabling near-identity updates of the evolving temperature field.

In this work, we introduce a surrogate modelling framework for multiphysics induction heating with stochastic anisotropic electrical conductivity, in which the geometric structure of the conductivity tensor is respected throughout the entire pipeline. Electrical conductivity is represented as a random SPD-valued quantity whose magnitude and direction uncertainty are modelled independently by construction, ensuring that all sampled realisations remain physically admissible. Uncertainty is then propagated through the coupled electromagnetic--thermal model via a surrogate that combines the CMNN architecture with a Neural ODE formulation, enabling uncertainty quantification over the full spatial temperature field and process history with a tractable computational cost.

The layout of the paper is as follows. Section~2 provides the formulation for heat transfer and electromagnetics together with the stochastic model for electrical conductivity, and their respective discretisations. Section~3 explains the surrogate modelling methods. Section~4 presents results for the induction heating process under different uncertainty configurations, as well as a comparison of surrogate formulations with different time integration schemes, followed by a discussion of the results. Section~5 provides the conclusions.


%% file: 2_Problem/Problem.tex
\begin{figure}[t!]
    \centering
    \includegraphics[width=0.8\textwidth]{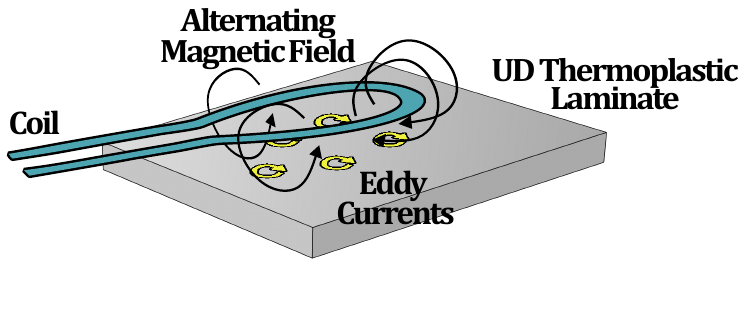}
    \caption{Schematic of the induction heating process.}
    \label{fig:IW_schematic}
\end{figure}

\noindent

Induction heating is a non-contact method for heating eletrically conductive materials and is commonly used in applications such as fusion bonding of thermoplastic composites and metals, as well as in case-hardening of metals. The technique is particularly relevant for the bonding of CFRTP, where the carbon fibres act as electrical conductors, eliminating the need for an additional susceptor to enable current flow \cite{siddique_review_2022}. The bond quality is strongly governed by the temperature history at the fusion interface, the direct measurement of which is not feasible in practice \cite{rahim_-situ_2019}. This necessitates the use of physics-based models of the coupled electromagnetic–thermal behaviour as virtual sensing tools \cite{korycki_assembling_2022}.
The model considered in this work is based on the schematic shown in \refFIG{\ref{fig:IW_schematic}}. An alternating current is supplied to a coil, generating an alternating magnetic field that induces eddy currents within the conductive carbon fibre plies. These eddy currents dissipate energy via Joule heating \cite{barazanchy_heating_2023}, resulting in a coupled electromagnetic–thermal process that governs the evolution of the temperature field relevant to bonding.

Assuming a harmonic excitation of the coil current and adopting a magneto-quasi-static (MQS) approximation, the electromagnetic field governing induction heating is described by Maxwell’s equations \cite{griffiths_introduction_1999}. The MQS assumption is justified by the fact that the characteristic dimensions (300 mm) of the welding setup are much smaller than the electromagnetic wavelength associated with the excitation frequency (approximately 877 m at 342 kHz), ensuring that wave propagation effects can be neglected. In this low-frequency regime, electromagnetic field penetration is governed by magnetic diffusion and characterised by a finite skin depth, which is typically small relative to the system dimensions. Under the MQS approximation, displacement currents are neglected in comparison with conduction currents, resulting in a diffusion-dominated electromagnetic field description and allowing electromagnetic wave propagation effects to be neglected.  Under these assumptions, the electric and magnetic fields in the spatial domain $\varDomain \subset \varReal^3$ (comprising the composite and surrounding air) are mathematically described by:
\begin{alignat}{4} 
&\nabla \cdot \varElec = \frac{\varEcharge}{\varPermit}, \label{eq:gauss_electricity}  \\ &\nabla \cdot \varBmag = 0, \label{eq:gauss_magnetism}  \\
&\nabla \times \varElec = -j 2\pi \varFcoil \varBmag, \label{eq:faraday_harmonic} \\
&\nabla \times \varBmag = \varPermea \varJcur. \label{eq:ampere_harmonic}
\end{alignat}
Here, $\varElec$ [V/m] is the electric field strength vector, $\varPermit$ [F/m] is the permittivity, $\varBmag$ [T] is the magnetic flux density vector, $\varFcoil$ [Hz] is the excitation frequency from the coil, $\varPermea$ [H/m] is the magnetic permeability, $\varJcur$ [A/m$^2$] is the current density vector and $\varEcharge$ [C/m$^3$] is the electric charge density. In bulk conductive media the last one is effectively zero, $\varEcharge = 0$, since charges redistribute rapidly enough to prevent the accumulation of net charge within the material volume \cite{griffiths_introduction_1999}. Consequently, Gauss’s law for electricity in \refEQ{\ref{eq:gauss_electricity}} becomes trivial. Note that the current density $\varJcur$ in \refEQ{\ref{eq:ampere_harmonic}} is split into the externally applied source and the induced contributions
\begin{equation} \label{eq:current_decomp}
\varJcur = \varJcur_s + \varEcond \varElec, 
\end{equation}
where $\varJcur_s$ is the prescribed coil current and $\varEcond$ [S/m] is the anisotropic conductivity tensor of the material. The latter belongs to the space of SPD matrices, $\varSymPlus(d)$, a property that follows from energy conservation principles and ensures physically consistent, balanced interactions within a material. 

Directly solving the system described by \refEQ{\ref{eq:gauss_electricity}-\ref{eq:current_decomp}} for both $\mathbf{E}$ and $\mathbf{B}$ would require handling six coupled field components, which is computationally demanding. This difficulty is commonly addressed by introducing a magnetic vector potential $\varMvec$ formulation \cite{zangwill_modern_2012}
\begin{equation} \label{eq:magnetic_potential}
    \varBmag = \nabla \times \varMvec,
\end{equation}
which reduces the number of unknowns and enforces the divergence-free constraint on the magnetic flux density automatically.
The magnetic vector potential $\varMvec$ serves as an auxiliary field from which the physical fields $\varBmag$ and $\varElec$ are recovered. Since the vector potential is not uniquely defined by \refEQ{\ref{eq:magnetic_potential}} alone, a Coulomb gauge condition 
\begin{equation}
    \nabla \cdot \varMvec = 0
\end{equation}
is imposed to remove the remaining ambiguity. Under the MQS approximation, the electric scalar potential is neglected in the conducting domain, such that the electric field is approximated by $E=-j 2 \pi \varFcoil A$. Substituting magnetic vector potential representation into Faraday's law in \refEQ{\ref{eq:faraday_harmonic}} and Ampère’s laws in \refEQ{\ref{eq:ampere_harmonic}} leads to a reduced mathematical description of the electromagnetic part of the problem:
\begin{alignat}{2}
\label{eq:deterministic_EM} 
\nabla \times \frac{1}{\varPermea} \nabla \times \varMvec(\varx) = \; &\varJcur_s -j 2\pi \varFcoil \varEcond \varMvec(\varx) &\quad &\text{a.e. in } \varDomain.
\intertext{The previous equation is subjected to boundary conditions:}
\label{eq:deterministic_EM_BC}
 &\vec{n} \times \varMvec(\varx) = \mathbf{0} &\quad &\text{on } \Gamma_{\infty} \subseteq \partial G,
\end{alignat}
where $\Gamma_{\infty}$ is the boundary at an infinite distance from the source representing wave propagation through free space. 

Under steady harmonic conditions the solution of the electromagnetic problem in Eq.~(\ref{eq:deterministic_EM}) provides the induced electric field $\varElec$, which gives rise to eddy currents in the conductive region according to
\begin{equation} \label{eq:eddy_current}
\varJcur_e(\varx) := \varEcond \varElec= -j 2 \pi \varFcoil \varEcond \varMvec(\varx).
\end{equation}
These currents constitute the source of irreversible thermal losses through Joule heating \cite{barazanchy_heating_2023}, with volumetric heat generation rate
\begin{equation} \label{eq:joule_heating}
\varHsource(\varx) = \varElec(\varx)^\top \, \boldsymbol{\varEcond} \, \varElec(\varx).
\end{equation}
This dissipation of electromagnetic energy acts as a distributed source of thermal energy in the conductive material. It therefore drives the temperature evolution in the electrically conductive CFRTP domain $\varDomain_C \subset \varDomain$, which is described by the transient heat conduction equation
\begin{alignat}{2} \label{eq:deterministic_HT}
\varDens \varHeatC \frac{\partial \varTemp (\boldsymbol{x},t)}{\partial t} &= \nabla \cdot \left( \varTcond \nabla \varTemp (\boldsymbol{x},t) \right) + \varHsource(\boldsymbol{x}) &\text{ a.e. in } \varDomain_C \times \varTime,
\intertext{over the time interval $\mathcal{T} = [0, t_e] \subset \mathbb{R}_+$ with boundary conditions:}
\label{eq:deterministic_HT_BC_D}
\varTemp(\varx, \vartime) &= \varTemp_D(\varx) \quad &\text{ on } \Gamma_D \subseteq \partial \varDomain_C, \forall \vartime \in \varTime,  \\
\label{eq:deterministic_HT_BC_N}
- \varTcond \nabla \varTemp(\varx, \vartime) \cdot \vec{n} &= \varHsource_N(\varx) \quad &\text{ on } \Gamma_N \subseteq \partial \varDomain_C, \forall \vartime \in \varTime,\\
\label{eq:deterministic_HT_BC_Conv}
- \varTcond \nabla \varTemp(\varx, \vartime) \cdot \vec{n}
&= h \,\big(\varTemp(\varx, \vartime) - \varTemp_\infty \big)
\quad &\text{ on } \Gamma_{\text{conv}} \subseteq \partial \varDomain_C, \; \forall \vartime \in \varTime.
\end{alignat}
Here, $\varDens$ [kg/m$^3$] is the material density, $\varHeatC$ [J/(kg\,K)] is the specific heat capacity, $\varTemp$ [K] is the temperature field, and $\varTcond$ [W/(m\,K)] is the anisotropic SPD thermal conductivity. The boundary $\partial \varDomain_C$ is partitioned into $\Gamma_D$, the Dirichlet boundary with prescribed temperature $\varTemp_D(\varx)$; $\Gamma_N$, the Neumann boundary with prescribed heat flux $\varHsource_N(\varx)$; and $\Gamma_{\text{conv}}$, the Robin boundary for convective heat exchange with ambient temperature $\varTemp_\infty$ and convective heat transfer coefficient $h$ [W/(m$^2$\,K)]. 

The above first-principles model is of a deterministic nature, where all parameters are assumed to be precisely known. In real induction heating applications, however, the exact values of many model parameters are either unknown or subject to variation due to the inability to perfectly control the manufacturing process of composite plies. Among all parameters, the anisotropic conductivity tensor $\varEcond$, interpreted as an effective macroscopic homogenised quantity at ply level, is particularly sensitive to the fibre architecture \cite{buser_characterisation_2022}. In particular, fibre orientation, local fibre volume fraction, waviness, and the connectivity of conductive fibre networks influence the formation of conductive pathways within the composite. These microstructural features are inherently affected by manufacturing variability and cannot be reproduced identically across specimens. As a consequence, different realisations of the microstructure may lead to different, yet equally admissible, homogenised conductivity tensors. In the present work, we do not model the microstructure as a spatially varying nor do we solve the governing equations at the microscale. Instead, a deterministic homogenisation at the ply scale is assumed, and variability arising from manufacturing imperfections is incorporated directly at the level of the effective constitutive parameters. In particular, the anisotropic conductivity tensor is treated as a random (or uncertain) homogenised quantity, representing specimen-to-specimen variations (aleatory uncertainty) of the already-averaged material response rather than explicit microstructural randomness. Consequently, the stochasticity enters only through the effective model parameters, not through a random microscale description of the material. Since $\varEcond$ directly governs the induced eddy currents and the associated Joule heating, variability in the homogenised conductivity propagates through the coupled electromagnetic–thermal system and induces variability in the resulting temperature field \cite{grouve_simulating_2021}. The present work therefore focuses on quantifying the uncertainty in the temperature field as the primary quantity of interest with respect to variations in the electrical conductivity. The magnitude of the uncertainty associated with the conductivity tensor is not chosen arbitrarily, but is informed by available data on the scatter of effective electrical properties in carbon-fibre reinforced composites reported in the literature \cite{buser_predicting_2025}. In the absence of exhaustive specimen-specific measurements, the adopted variability is interpreted as representative of typical manufacturing-induced deviations observed in comparable material systems. A sensitivity analysis is further conducted to assess the influence of the assumed level of variability on the predicted thermal response, thereby ensuring that the conclusions remain robust with respect to reasonable variations in the input uncertainty.

To introduce the variation into the coupled model in Eqs.~(\ref{eq:deterministic_EM})-(\ref{eq:deterministic_EM_BC}) and Eqs.~(\ref{eq:deterministic_HT})-(\ref{eq:deterministic_HT_BC_Conv}), a probabilistic approach is introduced, treating $\varEcond$ as a SPD tensor valued random variable \cite{shivanand_stochastic_2024}. Let $\varProbSpace$ define the probability space, where $\varEspace$ represents the sample space, $\varSigmaAlgebra$ is the associated sigma algebra, $\varProbMeas$ is the probability measure, and $\varEvent \in \varEspace$ denotes an elementary event. The conductivity tensor is then modelled as an SPD tensor valued random variable
\[
\varEcond : \varProbSpace \to \mathrm{Sym}_+^{d}(\mathbb{R}) \subset \mathbb{R}^{d\times d},
\]
with finite second moment, i.e.
\[
\mathbb{E}[\|\varEcond\|^2] < \infty,
\quad \text{so that } \varEcond \in L^2(\varProbSpace;\mathrm{Sym}_+^{d}(\mathbb{R})).
\]
Here, the set of real symmetric positive-definite matrices is denoted by $\mathrm{Sym}_+^{d}(\mathbb{R}) \subset \mathbb{R}^{d\times d}$, and for each $\omega \in \varOmega$, $\varEcond(\omega)$ denotes a realization of the conductivity tensor. Due to randomnes in $\varEcond(\omega)$, \refEQ{\ref{eq:deterministic_EM}} becomes stochastic: 
\begin{alignat}{2}
\label{eq:stochastic_EM} 
\nabla \times \frac{1}{\varPermea} \nabla \times \varMvec(\varx,\varEvent) = &\varJcur_s -j 2 \pi \varFcoil \varEcond(\varEvent) \varMvec(\varx,\varEvent) &\quad &\text{a.e. in } \varDomain, \forall \varEvent \in \varEspace,
\intertext{with boundary conditions}
\label{eq:stochastic_EM_BC}
 &\vec{n} \times \varMvec(\varx,\varEvent) = \mathbf{0} &\quad &\text{on } \Gamma_{\infty} \subseteq \partial G, \; \forall \varEvent \in \varEspace.
\end{alignat}
Similarly, the heat transfer model now includes the stochastic heat source $\varHsource(\varx, \varEvent)$ due to the coupling induced by \refEQ{\ref{eq:joule_heating}}. Entering the stochastic source term into  \refEQ{\ref{eq:deterministic_HT}} leads to:
\begin{alignat}{2} \label{eq:stochastic_HT} 
\varDens \varHeatC \frac{\partial \varTemp (\varx,\vartime,\varEvent)}{\partial \vartime} &= \nabla \cdot \left( \varTcond \nabla \varTemp (\varx, \vartime,\varEvent) \right) + \varHsource(\varx,\varEvent) &\text{ a.e. in } \varDomain_C \times \varTime, \forall \varEvent \in \varEspace,
\intertext{with boundary conditions}
\label{eq:stochastic_HT_BC_D}
\varTemp(\varx, \vartime, \varEvent) &= \varTemp_D(\varx) \quad &\text{ on } \Gamma_D \subseteq \partial \varDomain_C, \forall \vartime \in \varTime, \; \forall \varEvent \in \varEspace, \\
\label{eq:stochastic_HT_BC_N}
- \varTcond \nabla \varTemp(\varx, \vartime, \varEvent) \cdot \vec{n} &= \varHsource_N(\varx) \quad &\text{ on } \Gamma_N \subseteq \partial \varDomain_C, \forall \vartime \in \varTime, \; \forall \varEvent \in \varEspace, \\
\label{eq:stochastic_HT_BC_Conv}
- \varTcond \nabla \varTemp(\varx, \vartime, \varEvent) \cdot \vec{n} 
&= h \,\big(\varTemp(\varx, \vartime, \varEvent)) - \varTemp_\infty \big) 
\quad &\text{ on } \Gamma_{\text{conv}} \subseteq \partial \varDomain_C, \; \forall \vartime \in \varTime, \; \forall \varEvent \in \varEspace.
\end{alignat}
Note that both $\varMvec(\varx,\varEvent)$ and $\varTemp(\varx, \vartime, \varEvent)$ also depend on $\omega$, as they are solutions of the preceding coupled stochastic partial differential equations, and are therefore itself spatial and spatio-temporal random fields, respectivelly. The vector potential $A$ is an intermediate electromagnetic variable introduced for numerical convenience, whereas the temperature field represents the physically relevant output of the coupled induction heating process and is therefore the primary quantity of interest (QoI). In particular, it directly determines the thermal state of the material during processing, governing key process outcomes such as the achievement of target curing or welding temperatures, the quality of the bond formation, and the avoidance of thermal degradation due to overheating or excessive gradients. The aim of this work is to efficiently characterise the uncertainty in this QoI. It should be noted that the uncertainty in the temperature may also depend on other model parameters that could potentially vary, such as thermal conductivity. For simplicity reasons, we restrict ourselves only on the electric conductivity while all others remain at their nominal deterministic values. This choice is motivated by the fact that, in induction heating, the temperature field is significantly more sensitive to variations in electrical conductivity, which directly governs Joule heating, whereas thermal conductivity primarily affects spatial redistribution of heat. Consequently, variability in thermal conductivity is expected to have a secondary influence on the quantities of interest compared to electrical conductivity.

\input{2_Problem/Econd}

\subsection{Discretisation}\label{sec:discretisation}
\noindent
The problem introduced in \refEQ{\ref{eq:stochastic_EM}} and \refEQ{\ref{eq:stochastic_HT}} is continuous and does not admit a closed-form analytical solution in general. Therefore, discretisation is required in the spatial and temporal domains, as well as in the stochastic space. First, the spatial domain is discretised using the finite element method (FEM) \cite{bathe_finite_1996}, partitioning the domain as $\varDomain = \bigcup_{e} \mathcal{T}_e$, where $\mathcal{T}_e$ denotes an element. The solutions of \refEQ{\ref{eq:stochastic_EM}} are approximated by linear combinations of finite basis functions $N_i(\varx)$, $i = 1, \ldots, n_e$. For the electromagnetic problem, the magnetic vector potential in \refEQ{\ref{eq:stochastic_EM}} is approximated by:
\begin{equation}\label{eq:FEM_EM}
   \varMvecDisc(\varx,\varEvent) = \sum_{i=1}^{n_e} \varMvec_i(\varEvent) N_i(\varx).
\end{equation}
Hence, the continuous field $\varMvec$ is replaced by a finite-dimensional approximation $\varMvec_h$ composed of spatial basis functions with stochastic coefficient $\varMvec_i(\varEvent)$. For a single realisation of $\varEvent$, the electromagnetic system is solved in two stages: first, the current density distribution $\varJcur_s$ in the coil is obtained by computing the coil excitation through a geometry-based current distribution analysis \cite{noauthor_comsol_nodate, biro_use_1989}, after which the resulting complex-valued linear system is solved iteratively using a GMRES solver with a Vanka SSOR preconditioner \cite{saad_gmres_1986, vanka_block-implicit_1986}.

The temperature spatiotemporal random field in \refEQ{\ref{eq:stochastic_HT}} is approximated analogously to $\varMvec$ by:
\begin{equation}\label{eq:FEM_HT}
   \varTempDisc(\varx, \vartime, \varEvent) = \sum_{i=1}^{n_{eC}} \varTemp_i(\vartime, \varEvent)N_i(\varx),
\end{equation}
\cm{leading to} the discrete temperature field compactly represented by the nodal solution vector $\boldsymbol{\varTemp}_h(\vartime,\varEvent) = $\\
$\left[ \varTemp_h(\varx_1,\vartime,\varEvent), \ldots, \varTemp_h(\varx_{n_{eC}},\vartime,\varEvent) \right]^T \in \mathbb{R}^{n_{eC}}$. 

After a few derivation steps, in which both approximations are substituted back into their respective weak forms \cite{bathe_finite_1996}, and restricting attention to only the thermal system in the coupled formulation, the semi-discrete system of ordinary differential equations is obtained:
\begin{equation}\label{eq:semi_discrete}
    \varFEmass\,\dot{\boldsymbol{\varTemp}}_h(\vartime,\varEvent) + \varFEstiffness\,\boldsymbol{\varTemp}_h(\vartime,\varEvent) - \varFEload(\varEvent) = \boldsymbol{0},
\end{equation}
where the canonical matrices $\varFEmass, \varFEstiffness$ and load vector $\varFEload(\varEvent)$ are defined elementwise as
\begin{equation}
    M_{ij} = \int_{\varDomain_C} \varDens\,\varHeatC\,N_i\,N_j\,\mathrm{d}\varx,
\end{equation}
\begin{equation}\label{eq:stiffness}
    K_{ij} = \underbrace{\int_{\varDomain_C} \varTcond\,\nabla N_i\!\cdot\!\nabla N_j\,\mathrm{d}\varx}_{\text{conduction}}
    \;+\; \underbrace{\int_{\Gamma_{\!R}} h\,N_i\,N_j\,\mathrm{d}s}_{\text{Robin (convective)}},
\end{equation}
\begin{equation}\label{eq:load_vector}
    P_i(\varEvent) = \underbrace{\int_{\varDomain_C} \varHsource(\varx,\varEvent)\,N_i\,\mathrm{d}\varx}_{\text{stochastic volumetric source}}
    \;-\; \underbrace{\int_{\Gamma_{\!N}}  \varHsource_N\,N_i\,\mathrm{d}s}_{\text{Neumann}}
    \;+\; \underbrace{\int_{\Gamma_{\!R}} h\, \varTemp_\infty\,N_i\,\mathrm{d}s}_{\text{Robin (ambient)}}.
\end{equation}
Here, the Neumann and Robin boundary contributions are incorporated into $K_{ij}$ and $P_i(\varEvent)$ respectively, while the Dirichlet condition is enforced strongly on the assembled system. Since $\varHeatC$, $\varDens$, $\varTcond$, and all boundary coefficients are assumed to be independent of temperature, the matrices $\varFEmass$, $\varFEstiffness$, and load vector $\varFEload$ are not time dependent. Among them $\varFEload(\varEvent)$ is the sole source of stochasticity entering through the stochastic heat source $\varHsource(\varx,\varEvent)$. 

After spatial discretisation, time integration of the semi-discrete system \refEQ{\ref{eq:semi_discrete}} is carried out using the variable-step second-order backward differentiation formula (BDF2), which provides a balance between accuracy and stability for stiff transient problems \cite{hairer_solving_1996}. Let $\vartime_{k-1}$, $\vartime_k$, and $\vartime_{k+1}$ be three consecutive time points with variable step sizes $h_k=\vartime_k-\vartime_{k-1}$ and $h_{k+1}=\vartime_{k+1}-\vartime_k$, such that the step-size ratio reads $r=h_{k+1}/h_k$. 
Denoting the discrete temperature vector at time $\vartime_k$ by $\boldsymbol{\varTemp}_k(\varEvent)\equiv\boldsymbol{\varTemp}_h(\vartime_k,\varEvent)\in\mathbb{R}^{n{eC}}$ and using the shorthand $(\cdot)k := (\cdot)(\vartime_k)$, the BDF2 approximation of the temperature derivative at $\vartime_{k+1}$ is given by
\begin{equation} \label{eq:BDF2_dt}
\dot{\boldsymbol{\varTemp}}_{k+1} \approx
\frac{(1+2r)\,\boldsymbol{\varTemp}_{k+1} - (1+r)^2\boldsymbol{\varTemp}_{k} + 
r^2\boldsymbol{\varTemp}_{k-1}}{h_{k+1}(1+r)}.
\end{equation}
Substitution of \refEQ{\ref{eq:BDF2_dt}} into \refEQ{\ref{eq:semi_discrete}} and evaluation at $\vartime_{k+1}$ results in the following linear system:
\begin{equation} \label{eq:BDF2_lin_sys}
\Big(\tfrac{1+2r}{h_{k+1}(1+r)}\,\varFEmass + \varFEstiffness\Big)\,
\boldsymbol{\varTemp}_{k+1}(\varEvent)
= \tfrac{1+r}{h_{k+1}}\,\varFEmass\,\boldsymbol{\varTemp}_k(\varEvent)
- \tfrac{r^2}{h_{k+1}(1+r)}\,\varFEmass\,\boldsymbol{\varTemp}_{k-1}(\varEvent)
+ \varFEload(\varEvent),
\end{equation}
the solution of which has to be found.

Having established the spatial and temporal discretisations, the remaining stochastic dependence is addressed through Monte Carlo (MC) sampling \cite{le_maitre_spectral_2010}. For each realisation of $\varTensor(\varEvent_j), j=1,...,n_s$ the coupled electromagnetic-thermal system is solved to obtain the magnetic vector potential $\varMvecDisc(\varx, \varEvent_j)$ and therefore the stochastic heat source $\varHsource(\varx, \varEvent_j)$ and the corresponding load vector $\varFEload(\varEvent_j)$.
The thermal response $\varTemp_k(\varEvent_j)$ is then obtained by solving the linear system
\begin{equation}\label{eq:BDF2_lin_sys_MC}
\Big(\tfrac{1+2r}{h_{k+1}(1+r)}\,\varFEmass + \varFEstiffness\Big)\,
\boldsymbol{\varTemp}_{k+1}(\varEvent_j)
= \tfrac{1+r}{h_{k+1}}\,\varFEmass\,\boldsymbol{\varTemp}_k(\varEvent_j)
- \tfrac{r^2}{h_{k+1}(1+r)}\,\varFEmass\,\boldsymbol{\varTemp}_{k-1}(\varEvent_j)
+ \varFEload(\varEvent_j),
\end{equation}
at each time step $k = 1, \ldots, n_t$, yielding the nodal temperature trajectory $\left\{\left\{\boldsymbol{\varTemp}_h(\vartime_k, \varEvent_j)\right\}_{k=1}^{n_t}\right\}_{j=1}^{n_s}$. Since $\varFEmass$ and $\varFEstiffness$ are deterministic and time-independent, the system matrix is assembled and factorised once per realisation using the PARDISO direct sparse solver \cite{schenk_solving_2004}. The BDF2 scheme is initialised using a Backward Euler step, providing the required solution history $\boldsymbol{\varTemp}_0(\varEvent_j)$ and $\boldsymbol{\varTemp}_1(\varEvent_j)$ for the two-step recurrence. Once $n_s$ solution trajectories are collected, they are used to estimate the statistics of QoIs. 

\commentout{
Within the finite element framework, the primary unknowns are defined at the nodal level, i.e. 
$\boldsymbol{\varTemp}_k(\varEvent)$ denote the nodal degrees of freedom, and the corresponding field is obtained via the standard interpolation in Eq.~(\ref{eq:FEM_HT}). 
However, all quantities required for numerical integration and constitutive evaluation are computed at the integration (Gauss) points 
$\{\varx_q\}_{q=1}^{n_q}$. Accordingly, a superscript $(\cdot)^g$ is introduced to denote evaluation at these points, e.g.
$\boldsymbol{\varTemp}_k^g := (\varTemp_h(\varx_q,t_k,\varEvent))_{q=1}^{n_q}$ obtained by evaluating Eq.~(\ref{eq:FEM_HT}) with $N_i(\varx_q)$. As these values are used as post-processed FEM results, the vector $\boldsymbol{\varTemp}_k^g$ is henceforth understood to collect the values of the temperature field at the Gauss points at the time $t_k$, i.e.
$\boldsymbol{\varTemp}_k^g(\varEvent) \in \mathbb{R}^{n_q}$ with components $\varTemp_k(\varx_q,\vartime_k,\varEvent)$ for $q=1,\ldots,n_q$. }

To mitigate the computational cost of repeated full FEM solves across $n_s$ realisations, the goal is to build a data-driven surrogate that approximates the mapping between 
the stochastic input parameters and the corresponding temperature solution fields, thereby enabling 
efficient evaluations without repeated assembly and solution of the full finite element system. Since the underlying problem is transient, the surrogate 
is formulated in a step-wise manner, approximating the discrete time evolution of the system. Let $\boldsymbol{u}_k(\varEvent)$ denote the state over uniformly spaced time steps obtained by extrapolating the solution to successive time steps $t_k$ with constant step size $\Delta t_k = \Delta t$ for a given realization $\varEvent_j$.

Accordingly, the training dataset is defined as
\begin{equation} \label{eq:EMHT_dataset}
    \mathcal{D} := \Bigl\{\,
\bigl(\boldsymbol{u}_k(\varEvent_j),\,\varTensor(\varEvent_j)\bigr)
    \,\Bigr\}_{j=1,\,k=1}^{n_s,\,n_t},
\end{equation}
where the surrogate learns the incremental time-stepping map from the state at time $t_k$ and the 
associated response to the state at $t_{k+1}$ for each realisation. 
Although the surrogate model does not explicitly include the electromagnetic heat source as an input, its effect is implicitly incorporated through the high-fidelity coupled simulations used for training. The learned model therefore represents a data-consistent reduced evolution operator for the temperature field under the weakly coupled electromagnetic–thermal regime considered in this work. In this setting, the electromagnetic contribution acts as an implicit driving mechanism encoded in the state trajectories, rather than as an explicit forcing term.

The surrogate model is designed to approximate the discrete-time evolution of the system by learning a one-step increment map. We consider approximation of 
\begin{equation}
\Delta \boldsymbol{u}_k(\varEvent) = \boldsymbol{u}_{k+1}(\varEvent) - \boldsymbol{u}_k(\varEvent)
\end{equation}
by
\begin{equation} \label{eq:Surrogate_Map}
  \Delta \hat{\boldsymbol{u}}_k(\varEvent)
    = \varphi\!\bigl(\boldsymbol{u}_k(\varEvent),\,\varTensor(\varEvent)\bigr),
\end{equation}
where $\varphi$ is a continuous parametric function. The full trajectory is then generated recursively via
\begin{equation}
   \boldsymbol{u}_{k+1}(\varEvent)\approx \hat{\boldsymbol{u}}_{k+1}(\varEvent)
   := \boldsymbol{u}_k(\varEvent) + \Delta \hat{\boldsymbol{u}}_k(\varEvent),
\end{equation}
which is iterated for $k=0,1,\dots,n_t-1$.

The goal of this work is to construct such an approximation using a feedforward neural network $F_{\boldsymbol{\theta}}$ that is accurate not only in the one-step sense of Eq.~(\ref{eq:surrogate_target_MSE}), but also under recursive application over long time horizons, i.e.,
\begin{equation} \label{eq:NN_Map}
    \boldsymbol{u}_{k+1}(\varEvent) - \boldsymbol{u}_k(\varEvent)
    \approx F_{\boldsymbol{\theta}}\!\bigl(
        \boldsymbol{u}_k(\varEvent),\,\varTensor(\varEvent)
    \bigr).
\end{equation}
The map is constructed as a composition of affine maps and elementwise nonlinear activations:
\begin{equation} \label{eq:NNfunction}
F_{\boldsymbol{\theta}}(\boldsymbol{q}_k(\varEvent))
=
\left(
f_a^{L} \circ \varLope^{L}
\circ \cdots
\circ f_a^{1} \circ \varLope^{1}
\right)\bigl(\boldsymbol{q}_k(\varEvent)\bigr),
\end{equation}
where the input vector is defined as
\begin{equation}
\boldsymbol{q}_k(\varEvent)
:= \bigl[\,\boldsymbol{u}_k(\varEvent);\; \varTensor(\varEvent)\,\bigr].
\end{equation}
For each layer $\ell = 1,\dots,L$, the affine transformation is defined as
\begin{equation} \label{eq:layer_eq}
\varLope^{\ell}(\boldsymbol{q}^{\ell-1})
= \mathbf{W}^{\ell}\boldsymbol{q}^{\ell-1}
+ \boldsymbol{b}^{\ell},
\end{equation}
where
\[
\boldsymbol{q}^{\ell-1} \in \mathbb{R}^{n_{\ell-1}}, \quad
\boldsymbol{q}^{\ell} \in \mathbb{R}^{n_{\ell}},
\]
\[
\mathbf{W}^{\ell} \in \mathbb{R}^{n_{\ell} \times n_{\ell-1}}, \quad
\boldsymbol{b}^{\ell} \in \mathbb{R}^{n_{\ell}}.
\]
The activation function $f_a^{\ell}$ is applied elementwise:
\begin{equation}
f_a^{\ell} : \mathbb{R}^{n_{\ell}} \to \mathbb{R}^{n_{\ell}}.
\end{equation}
The initial layer variable is defined as
\begin{equation}
\boldsymbol{q}^{0} := \boldsymbol{q}_k(\varEvent)
\end{equation}
such that by recursion for $\ell = 1,\dots,L$ one has
\begin{equation}
\boldsymbol{q}^{\ell}
=
f_a^{\ell}\!\left(
\mathbf{W}^{\ell}\boldsymbol{q}^{\ell-1}
+
\boldsymbol{b}^{\ell}
\right).
\end{equation}
The set of trainable parameters is then
\[
\boldsymbol{\theta} := \{\mathbf{W}^{\ell}, \boldsymbol{b}^{\ell}\}_{\ell=1}^{L},
\]
and is learned by minimizing the mean-squared error over observed one-step transitions:
\begin{equation} \label{eq:surrogate_target_MSE}
    \mathcal{E}(\boldsymbol{\theta})
    = \frac{1}{n_s\,n_t}
      \sum_{j=1}^{n_s}\sum_{k=1}^{n_t}
      \left\|
          F_{\boldsymbol{\theta}}\!\bigl(
              \boldsymbol{u}_k(\varEvent_j),\,\varTensor(\varEvent_j)
          \bigr)
          - \bigl(\boldsymbol{u}_{k+1}(\varEvent_j) - \boldsymbol{u}_k(\varEvent_j)\bigr)
      \right\|^2.
\end{equation}
Minimisation of Eq.~(\ref{eq:surrogate_target_MSE}) yields a Monte Carlo approximation of the $L^2$-optimal predictor of the one-step increment conditioned on $(\boldsymbol{u}_k,\varTensor)$. In other words, the learned model targets the conditional mean of $\boldsymbol{u}_{k+1} - \boldsymbol{u}_k$ given the current state and stochastic input. Higher-order statistical moments of the induced solution process are not explicitly constrained by the loss function and are therefore only captured implicitly through the learned dynamics. In particular, these quantities may deteriorate over long time horizons due to error accumulation during recursive rollout and finite-sample generalisation effects.

\begin{figure}[t]               
    \captionsetup{justification=centering}
    \centering
    \includegraphics[width=\textwidth,trim=0cm 0cm 0cm 0.0cm,clip]{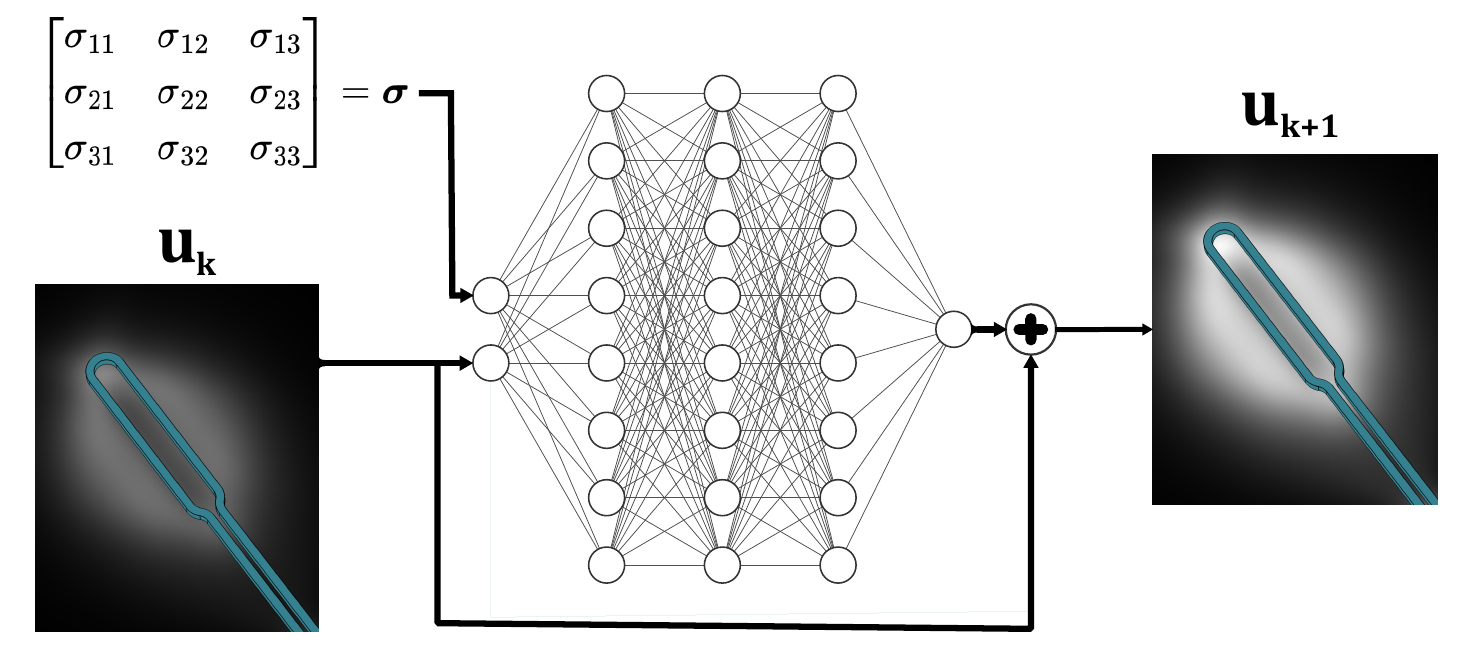}
    \caption{Neural network framework for the induction heating process with arbitrarily oriented coil in teal on
top of a black/white temperature field.}
    \label{fig:IH_NN_framework} 
\end{figure}

The previously derived overall network architecture is graphically shown in Fig.~(\ref{fig:IH_NN_framework}). The conductivity tensor and the temperature solution field at time $t_k$ act as input fields to the neural network. On the other hand, the sum of the solution at time $t_k$ and the neural network response acts as the output of the network. Therefore, the proposed model can be interpreted as a residual-type neural network.

From this perspective, the mapping $F_{\boldsymbol{\theta}}$ does not learn the full state evolution operator, but rather the residual correction to the identity map. In this setting, the presented model can be interpreted as a conditional discrete-time dynamical system in which the evolution of the state variable depends on an additional exogenous input $\varTensor(\varEvent)$. The update rule takes the form
\begin{equation}\label{resid}
\hat{\boldsymbol{u}}_{k+1}
=
\boldsymbol{u}_k
+
F_{\boldsymbol{\theta}}(\boldsymbol{u}_k;\varTensor(\varEvent)).
\end{equation}
For each fixed realization of $\varTensor(\varEvent)$, the system defines a distinct discrete-time dynamical system
\[
\hat{\boldsymbol{u}}_{k+1} = G_{\varTensor}(\boldsymbol{u}_k),
\quad
G_{\varTensor}(\cdot) := \mathrm{Id} + F_{\boldsymbol{\theta}}(\cdot;\varTensor),
\]
and the conductivity tensor acts as a parameter determining the evolution law rather than as a state variable. Numerically, the residual formulation also improves the conditioning of the learning problem with respect to the state variable $\boldsymbol{u}_k$. In particular, the Jacobian of the update map with respect to $\boldsymbol{u}_k$ is given by
\begin{equation}
\frac{\partial \hat{\boldsymbol{u}}_{k+1}}{\partial \boldsymbol{u}_k}
=
I +
\frac{\partial F_{\boldsymbol{\theta}}(\boldsymbol{u}_k;\varTensor)}{\partial \boldsymbol{u}_k}.
\end{equation}
The presence of the identity term ensures that the linearized dynamics remain close to the identity map whenever the learned increment is small, which improves the conditioning of gradient propagation through compositions of the update. In particular, this structure mitigates vanishing or ill-conditioned gradients that may arise in long iterative rollouts.

In the time discretization setting the previously described residual neural network parameterized by $\varTensor$ can be further related to the feedforward Euler discretization rule. By rewriting Eq.~(\ref{eq:semi_discrete}), one obtains the expresion for the gradient:
\begin{equation} \label{eq:euler_rhs_grad}
    \dot{\boldsymbol{\varTemp}}_{k+1} = \,\varFEmass^{-1}\bigl(\varFEload - \varFEstiffness\boldsymbol{\varTemp}_k\bigr),
\end{equation}
which further can be approximated by a feedforward neural network $\tilde{F}_{\varWeights}\!\bigl(\varTensor,\boldsymbol{\varTemp}_{k}\bigr)$ such that after explciit Euler discretization with time step $\Delta t$ one obtains
\begin{equation} \label{eq:euler_nn_2}
    \hat{\boldsymbol{\varTemp}}_{k+1} = \boldsymbol{\varTemp}_{k}
    + \Delta t \hat{F}_{\varWeights}\!\bigl(\varTensor,\boldsymbol{\varTemp}_{k}\bigr)= \boldsymbol{\varTemp}_{k}
    + \Delta \hat{\boldsymbol{\varTemp}}_{k}, \quad \Delta \hat{\boldsymbol{\varTemp}}_{k}= \Delta t \hat{F}_{\varWeights}\!\bigl(\varTensor,\boldsymbol{\varTemp}_{k}\bigr)
\end{equation}
Hence, the nodal approximation of the temperature at time step $t_{k+1}$ can be obtained by summing the previous solution and the approximated increment. The latter connects then to Eq.~(\ref{resid}) such that
\begin{equation} \label{eq:euler_nn}
  F_{\boldsymbol{\theta}}(\boldsymbol{u}_k;\varTensor(\varEvent))= \Delta t \hat{F}_{\varWeights}\!\bigl(\varTensor,\boldsymbol{\varTemp}_{k}\bigr)
\end{equation}
holds.

While residual structure can improve long-term behavior compared to direct state prediction by reducing drift, its primary advantage lies in improved optimization properties and in the imposition of a meaningful incremental modeling assumption. However, viewing the model as an iterative dynamical system also raises potential stability issues, since repeated application of the learned update may lead to error accumulation or divergence if the induced dynamics are not contractive. Thus, although the residual formulation improves trainability, it does not in general guarantee stability of the resulting discrete-time system. Next to this, the conductivity tensor $\varTensor(\varEvent)$ is used as part of the network input but takes values in the space of SPD matrices rather than in a Euclidean vector space. A naive vectorisation of its components ignores the intrinsic Riemannian geometry of the SPD manifold. This can lead to representations that are not geometrically consistent with the underlying physical object and may adversely affect learning efficiency. Therefore, manifold-aware representations are preferable, respecting the SPD structure through appropriate parameterisations or intrinsic metrics, which has been shown to improve numerical accuracy and generalisation in learning tasks involving SPD-valued data \cite{huang_riemannian_2017}.

%% file: 2_Problem/Econd.tex
\noindent

\begin{figure}[!b]         
    \captionsetup{justification=centering}
    \centering
    \begin{subfigure}[t]{0.45\textwidth}
        \includegraphics[width=\textwidth,trim=0.0cm 2.0cm 0.0cm 0.0cm,clip]{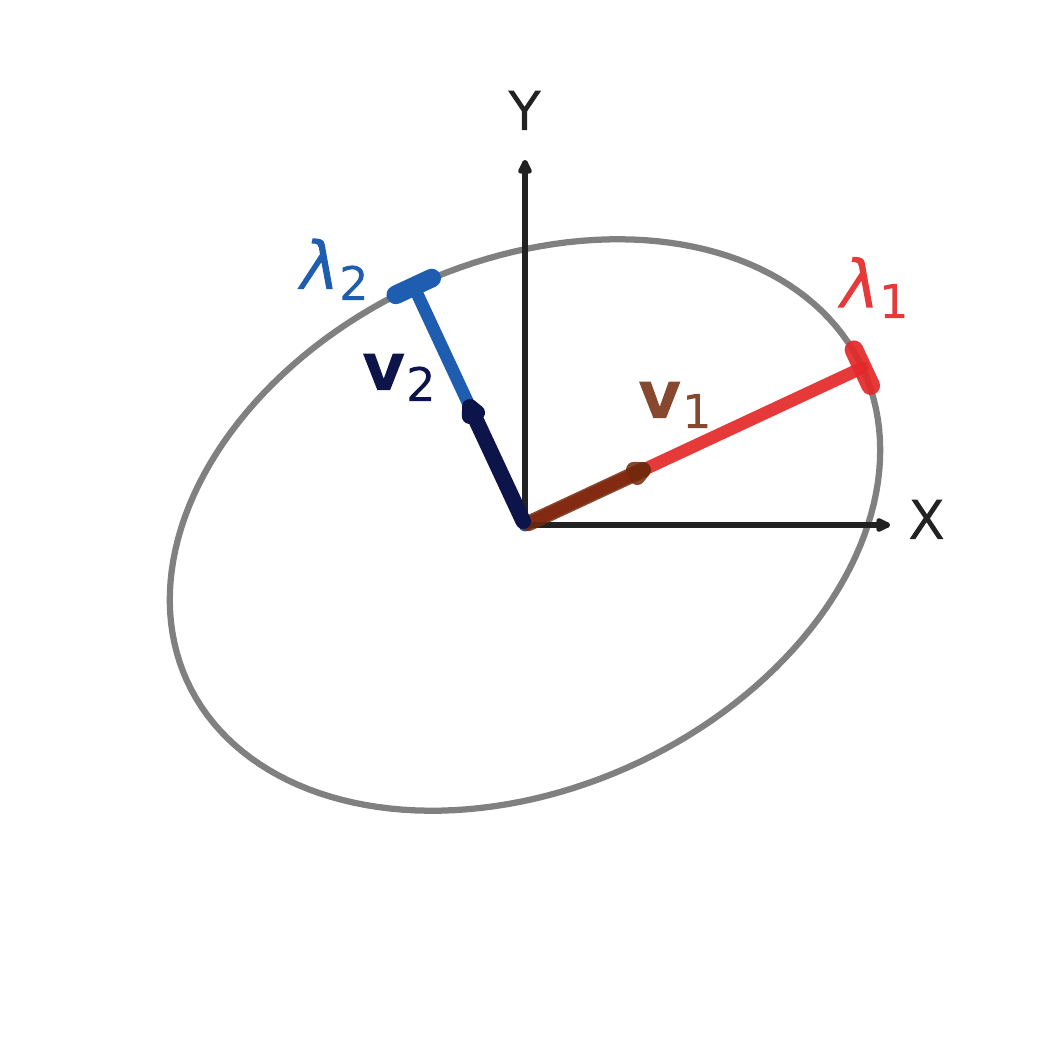}
        \caption{Geometric representation of a 2D SPD matrix as an ellipse, with principal eigenvectors $v_1$ and $v_2$ and corresponding eigenvalue $\lambda_1$ and $\lambda_2$ components.}
        \label{fig:spd_ellipse}
    \end{subfigure}       
    \begin{subfigure}[t]{0.43\textwidth}
        \includegraphics[width=\textwidth,trim=2cm 3cm 2cm 4.5cm,clip]{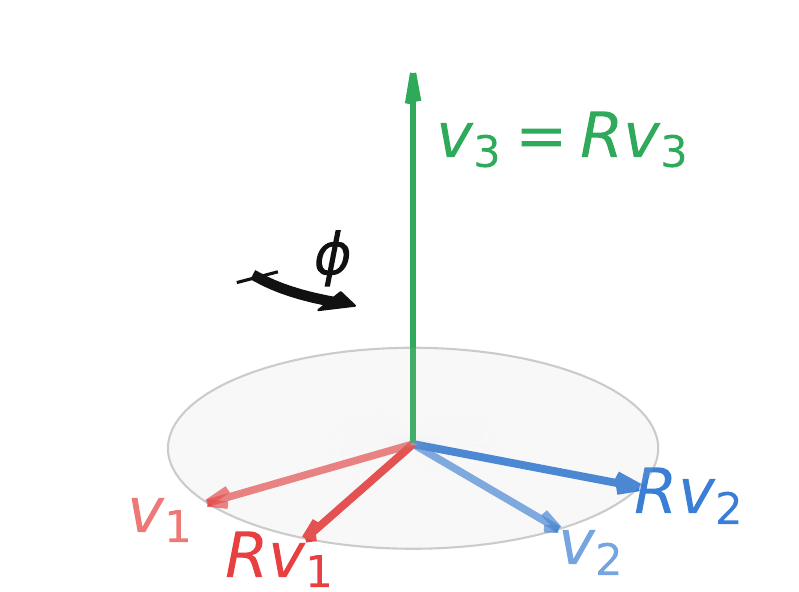}
        \caption{Realisation of the random planar rotation $\phi$ of $\varEigVec$ about $\mathbf{v}_3$. The rotation matrix $\boldsymbol{R}$ maps the reference frame $(\mathbf{v}_1, \mathbf{v}_2)$ to the rotated frame $(\boldsymbol{R}\mathbf{v}_1, \boldsymbol{R}\mathbf{v}_2)$.}
        \label{fig:phi_rot}
    \end{subfigure}
    \caption{Illustration of an SPD matrix as an ellipse (left) and the application of a planar rotation $\boldsymbol{R}$ about $\mathbf{v}_3$ (right).}
\end{figure}

In practical applications, the symmetry class of the effective material tensor cannot always be prescribed a priori with full certainty. While idealised material models assume specific symmetries (e.g. isotropic, orthotropic, or transversely isotropic behaviour), the actual effective response depends on the underlying microstructure, the homogenisation scale, and manufacturing-induced effects such as fibre misalignment, waviness, and local heterogeneities. These effects may lead to deviations from the assumed symmetry class, or to situations where multiple symmetry descriptions remain plausible at the macroscopic level \cite{shivanand_stochastic_2024}. In addition, even for a fixed symmetry class, the values of the constitutive parameters may still vary, as discussed previously. To account for both type of uncertainties mathematically, we decompose the stochastic electric conductivity tensor into 
 \begin{equation}\label{eq:eigen_decomp}
    \varEcond(\varEvent) = \varEigVec(\varEvent) \varEigVal(\varEvent) \varEigVec(\varEvent)^T, \quad \forall \omega \in \varOmega
\end{equation}
where the uncertainty in $\varEigVal$ represents the uncertainty in the parameter values here represented by eigenvalues, and $\varEigVec$ represents the uncertainty affiliated with the directional uncertainty. Geometrically, for $d = 2$ and a particular choice of $\omega$ this decomposition corresponds to an ellipse whose principal axes are aligned with the eigenvectors $\mathbf{v}_1$ and $\mathbf{v}_2$, the columns of $\varEigVec$, and whose semi-axis lengths are given by the eigenvalues $\lambda_1$ and $\lambda_2$, as illustrated in \refFIG{\ref{fig:spd_ellipse}}. 

In this formulation, material symmetry is reflected in the structure of the eigenvalues and the alignment of the eigenvectors. For instance, isotropy corresponds to equal eigenvalues, $\lambda_1=\lambda_2=\lambda_3$, and arbitrary orientation of $\varEigVec(\varEvent)$, whereas anisotropic behaviour arises from distinct eigenvalues, with symmetry classes such as orthotropy additionally imposing alignment constraints on the eigenvectors with respect to material directions. Hence, symmetry is not treated as a separate variable, but is embedded in the spectral structure of the tensor through constraints on $\varEigVal(\varEvent)$ and $\varEigVec(\varEvent)$. This viewpoint allows a unified representation of different material symmetries within a single SPD framework, where variability can be introduced consistently at the level of eigenvalues (magnitude anisotropy) and eigenvectors (directional anisotropy).
  Furthermore, following \cite{shivanand_stochastic_2024} we model each factor as
\begin{equation}
\varEigVal(\varEvent) = \exp(\boldsymbol{Y}(\varEvent)) \quad \text{and} \quad \varEigVec(\varEvent) = \boldsymbol{R}(\varEvent)\overline{\varEigVec} = \exp(\boldsymbol{W}(\varEvent))\overline{\varEigVec}
\end{equation}
in its respective linear spaces: $\boldsymbol{Y} \in \vardiag(d)$ for diagonal matrices and $\boldsymbol{W} \in \varso(d)$ for skew-symmetric matrices. Note that such a model for the eigenvalues restricts the stochastic $  \varEcond(\varEvent) $ to the SPD class, irrespectively of the choice of the distribution for $\boldsymbol{Y}$. The directional uncertainty is then encoded in $\boldsymbol{W}(\omega)$ representing orientational variation with respect to the fixed principal direction $\overline{\varEigVec}$. The stochastic model is intended for moderate variability around a reference configuration, such that eigenvalue ordering is preserved almost surely and no eigenvalue crossing occurs. The full tensor $\varTensor(\varEvent)$ is then constructed as
\begin{equation}\label{eq:stochastic_tensor}
    \varTensor(\varEvent) = \exp(\boldsymbol{W}(\varEvent)) \; \overline{\varEigVec} \; \exp(\boldsymbol{Y}(\varEvent)) \; \overline{\varEigVec}^T \; \exp(\boldsymbol{W}(\varEvent)^T), \quad \forall \varEvent \in \varEspace.
\end{equation}
The eigenvalue components in $\boldsymbol{Y}(\varEvent)$ are modelled by a Gaussian distribution,
\begin{equation}\label{eq:stochastic_eigval}
    \boldsymbol{Y}(\varEvent) = \mathrm{diag}(y_i(\varEvent)) \quad \text{with} \quad y_i(\varEvent) \sim \mathcal{N}(\mu_i, \sigma_i^2), \quad i = 1, \ldots, d,
\end{equation}
which corresponds to a lognormal distribution for the eigenvalues $\lambda_i$ themselves. The mean $\mu_i$ and standard deviation $\sigma_i$ are chosen appropriately to match the statistics of the eigenvalues. The orientation related tensor is modelled through a skew-symmetric matrix
\begin{equation}\label{eq:stochastic_eigvec_1}
   \boldsymbol{W}(\varEvent) = \mathrm{skw}(\boldsymbol{w}) = \begin{bmatrix}
   0 & -w_3(\varEvent) & w_2(\varEvent) \\
   w_3(\varEvent) & 0 & -w_1(\varEvent) \\
   -w_2(\varEvent) & w_1(\varEvent) & 0
   \end{bmatrix}, \quad \boldsymbol{W}(\varEvent) = -\boldsymbol{W}^T(\varEvent),
\end{equation}

\noindent
where $\boldsymbol{w}(\varEvent)$ is the random variable giving the axis--angle representation of a rotation, decomposed as $\boldsymbol{w}(\varEvent) = \phi(\varEvent)\,\varAXIS$ into a scalar rotation angle $\phi(\varEvent)$ and a unit rotation axis $\varAXIS$. The axis-angle representation is used to parametrise small stochastic rotations around a reference configuration, such that the exponential map provides a locally well-defined mapping from $\mathfrak{so}(3)$ to $SO(3)$. Each composite ply is thin --- 0.14\,mm thick compared to its $300 \times 300$\,mm in-plane dimensions --- so that fibre orientation, and hence orientational uncertainty, is effectively restricted to the $XY$-plane. Consequently, the rotation axis $\varAXIS$ is fixed along $\boldsymbol{e}_3$, and the admissible rotations reduce to a one-parameter subgroup of $SO(3)$, parametrised by a single scalar angle $\phi(\varEvent) \in S^1$. This is illustrated in \refFIG{\ref{fig:phi_rot}}, where $\boldsymbol{R}$ maps the in-plane reference frame $(\mathbf{v}_1, \mathbf{v}_2)$ to the rotated frame $(\boldsymbol{R}\mathbf{v}_1, \boldsymbol{R}\mathbf{v}_2)$. To sample the random orientation, a von Mises (vM) distribution $\mathcal{VM}(\varVMFref, \varVMFcon)$ is employed, with probability density function
\begin{equation}\label{eq:VM}
    f(\phi; \varVMFref, \varVMFcon) = C \exp\!\left(\varVMFcon \cos(\phi - \varVMFref)\right),
\end{equation}
where $\phi(\varEvent) \in S^1$ is the scalar rotation angle, $\varVMFref$ is the mean angle, $\varVMFcon$ is the concentration parameter, and $C$ is the normalising constant. The vM distribution is the maximum-entropy distribution for directional data on the circle \cite{mardia_directional_2000}, making it the natural analogue of a Gaussian for orientational uncertainty. Taken together, this parameterisation provides the flexibility to represent different levels of material symmetry through constraints on the spectral structure of the conductivity tensor. Isotropy is recovered when all eigenvalues are equal, whereas anisotropic behaviour arises when the eigenvalues are distinct. In addition, orthotropic-type behaviour is obtained when the eigenvectors are aligned with fixed material directions, while deviations from this alignment introduce orientational variability. By treating selected eigenvectors as random, the orientation of principal material directions can be modelled probabilistically, thereby accounting for possible deviations from idealised symmetry assumptions, as discussed in detail in \cite{shivanand_stochastic_2024}.

%% file: 3_Method/Method.tex
\noindent




\noindent Given the continuous version of the thermal equation
\begin{equation} \label{eq:semi_discrete_rhs}
    \frac{d\boldsymbol{\varTemp}_h(\vartime)}{d\vartime}
    = \varFEmass^{-1}\bigl(\varFEload - \varFEstiffness\boldsymbol{\varTemp}_h(\vartime)\bigr),
\end{equation}
we can replace the right hand side with the continuous neural network approximation $f_{\xi}$ to obtain neural ordinary differential equation (ODE) \cite{queiruga_continuous--depth_2020}
\begin{equation} \label{eq:neural_ode}
    \frac{d\boldsymbol{\varTemp}_h(\vartime)}{d\vartime}
    = f_{\xi}\!\bigl(\varTensor,\boldsymbol{\varTemp}_h(\vartime)\bigr).
\end{equation}
Continuous-depth parametrisations of this kind have been shown to yield smoother latent vector fields than discrete-step surrogates trained on the same data \cite{queiruga_continuous--depth_2020}. In this framework, $f_\xi$ represents a learned approximation of the underlying vector field that generates the system dynamics. The available data consist only of discrete-time snapshots, which motivates the introduction of a numerical time discretization to connect the continuous model with observations. 
The advancement of solution from $\vartime_k$ to $\vartime_{k+1}$ is obtained by integrating \refEQ{\ref{eq:neural_ode}} over the time interval
\begin{equation} \label{eq:ode_integrate}
    \hat{\boldsymbol{\varTemp}}_{k+1} = \boldsymbol{\varTemp}_k
    + \int_{\vartime_k}^{\vartime_{k+1}}
      f_{\xi}\!\bigl(\varTensor,\boldsymbol{\varTemp}_h(\vartime)\bigr)\,d\vartime.
\end{equation}
Since the integral is generally intractable in closed form, it is approximated by a numerical time-stepping scheme. The choice of time integration scheme therefore does not merely serve a numerical purpose, but introduces an inductive bias on the learned dynamics by determining how the continuous-time Neural ODE is observed through discrete-time data. This choice may significantly affect the stability of the resulting learned evolution, since explicit schemes can amplify modeling or discretization errors, particularly for stiff system as in our case. Consequently, the choice of integration scheme must balance accuracy, computational cost, and stability of the learned surrogate dynamics.

Regardless of the choice of time integrator, the learned vector field $f_{\xi}$ does not inherit the classical stability properties associated with schemes such as BDF2 applied to the underlying FEM system. 
The underlying semi-discrete FEM system is linear and can be written in the form
\[
\dot{\boldsymbol{\varTemp}}_h = \boldsymbol{A}\boldsymbol{\varTemp}_h + \boldsymbol{b},
\quad \boldsymbol{A} = -\varFEmass^{-1}\varFEstiffness.
\]
For this class of systems, classical time integration schemes such as BDF2 provide well-characterised stability properties, which are directly linked to the spectrum of $\boldsymbol{A}$ and guarantee appropriate damping of stiff modes \cite{hairer_solving_1996}. In contrast, replacing the physical right-hand side by a neural approximation $f_{\xi}$ yields a learned dynamical system that does not, in general, preserve the linear structure or spectral properties of $\boldsymbol{A}$. Consequently, classical stability results for BDF2 on linear systems do not directly transfer to the learned model, even if it is trained on data generated by a linear PDE.
As a result, stability of long-term rollouts depends on how well $f_{\xi}$ approximates the underlying operator in a spectral sense, rather than solely on the properties of the numerical integrator.
As a consequence, approximation errors may accumulate under recursive time stepping, and the stability of long-term rollouts depends jointly on the properties of the learned vector field and the numerical integration scheme. In practice, rollout stability is therefore typically assessed empirically and may require additional regularisation or architectural constraints on $f_{\xi}$.

If time integraton is achieved by the explicit Euler method then we have the scheme corresponding to the residual neural network as presented in the previous section. In such a case, the stability is conditional and may require very small step sizes in order to achieve prescribed accuracy, as discussed later. Similarly, the use of classical fourth-order Runge--Kutta scheme (RK4), advancing the solution by a weighted average of gradients at four different points per step:
\begin{equation} \label{eq:rk4_network_discrete_time}
\begin{aligned}
   s_1 &= f_{\xi}\!\bigl(\varTensor,\boldsymbol{\varTemp}_k\bigr), \\[2pt]
   s_2 &= f_{\xi}\!\bigl(\varTensor,\boldsymbol{\varTemp}_k + \tfrac{\Delta\vartime}{2}\,s_1\bigr), \\[2pt]
    s_3 &= f_{\xi}\!\bigl(\varTensor,\boldsymbol{\varTemp}_k + \tfrac{\Delta\vartime}{2}\,s_2\bigr), \\[2pt]
   s_4 &= f_{\xi}\!\bigl(\varTensor,\boldsymbol{\varTemp}_k + \Delta\vartime\,s_3\bigr), \\[6pt]
   \boldsymbol{\varTemp}_{k+1} &= \boldsymbol{\varTemp}_k
       + \tfrac{\Delta\vartime}{6}\bigl(s_1 + 2s_2 + 2s_3 + s_4\bigr).
\end{aligned}
\end{equation}
will not improve significantly the stability of stiff system. Here, the quantities $\boldsymbol{s}_1,\ldots,\boldsymbol{s}_4$ correspond to evaluations of the neural vector field $f_{\xi}$ at intermediate states obtained within the time interval $[\vartime_k,\vartime_{k+1}]$. These intermediate evaluations provide a higher-order approximation of the flow generated by the continuous-time Neural ODE. For sufficiently smooth vector fields, the Runge--Kutta scheme achieves fourth-order accuracy in time with respect to the exact solution of the non-stiff ODE. 

On the other hand, Adams–Moulton methods \cite{hairer_solving_1996} are implicit multistep schemes that are generally much more stable than explicit methods for stiff systems. Because they use future solution values, they allow much larger time steps without numerical instability. This makes them well-suited for our problem. However, they require solving a system of equations at each step, which increases computational cost compared to explicit methods. In this paper we use predictor–corrector strategies such as Adams--Bashforth--Moulton (ABM) schemes \cite{hairer_solving_1993} based on an explicit Adams--Bashforth step that provides an initial prediction for $\hat{\boldsymbol{\varTemp}}_{k+1}$, which is subsequently refined using an implicit Adams--Moulton corrector through fixed-point or iterative solution of the resulting nonlinear equation. 
The Adams--Moulton corrector of third order reads
\begin{equation} \label{eq:adams_moulton2_nn}
    \boldsymbol{\varTemp}_{k+1} = \boldsymbol{\varTemp}_k
    + \frac{\Delta\vartime}{12}\Bigl(
        5\,f_{\xi}\!\bigl(\varTensor,\boldsymbol{\varTemp}_{k+1}\bigr)
      + 8\,f_{\xi}\!\bigl(\varTensor,\boldsymbol{\varTemp}_k\bigr)
      -   f_{\xi}\!\bigl(\varTensor,\boldsymbol{\varTemp}_{k-1}\bigr)
    \Bigr),
\end{equation}
and defines a nonlinear implicit system for $\boldsymbol{\varTemp}_{k+1}$, since nonlinearity arises both from the neural network evaluation and from the implicit dependence on the unknown state. The resulting system is solved using fixed-point iteration. Convergence depends on the contractivity of the combined neural operator and time-integration mapping, which is influenced by the learned dynamics and the step size relative to the stability region of the scheme.

The choice of the third-order Adams–Moulton scheme is motivated by its favorable balance between accuracy and stability for moderately stiff dynamics arising in neural-operator-based models. Compared to fully implicit linear-stencil methods such as BDF2, the Adams–Moulton formulation naturally incorporates multiple past time levels, which improves temporal accuracy without requiring higher-stage nonlinear solves. Although the method leads to an implicit nonlinear system at each step due to the neural-network parameterized right-hand side, this structure is advantageous in our setting, as it allows consistent reuse of neural evaluations within the fixed-point iteration.

From a stability perspective, the scheme provides sufficient damping for the targeted dynamics when operated within an appropriate step-size regime, while maintaining higher-order accuracy than more dissipative implicit alternatives. In practice, this makes it well-suited for neural surrogates where the dominant error stems from model approximation rather than time-integration stiffness alone, and where accurate temporal resolution is essential. Compared to the BDF2 scheme used in the reference FEM solver, predictor--corrector methods such as ABM introduce reduced numerical damping and a stronger dependence of stability on the predictor accuracy and corrector convergence. As a consequence, stiff modes that are naturally suppressed by BDF2 may reappear, and the overall stability of the time integration becomes more sensitive to the chosen time step and iterative refinement strategy.

To assess stability of the learned dynamics, we evaluate the induced discrete flow map 
\[
\Phi_{\Delta t}^{\xi}(\boldsymbol{u}_k)
=
\boldsymbol{u}_k
+
\int_{t_k}^{t_k+\Delta t}
f_{\xi}\!\bigl(\boldsymbol{u}(t), \varTensor\bigr)\,dt
\]
associated with the neural vector field and the corresponding time discretization. Local stability is then analysed via the spectral radius of the Jacobian
\[
\rho\!\left(\frac{\partial \Phi_{\Delta t}^{\xi}}{\partial \boldsymbol{\varTemp}}\right),
\]
where values exceeding unity indicate local amplification of perturbations. For stability analysis of the nonlinear learned dynamics, the system is linearized along the trajectory by evaluating the Jacobian of the neural vector field at the current state,
\[
J_k = \frac{\partial f_{\xi}}{\partial \boldsymbol{\varTemp}}(\boldsymbol{\varTemp}_k).
\]
This yields a locally linear approximation of the evolution, allowing the stability of the discrete flow to be assessed through the spectral properties of the associated linearized update operator.

Beyond this local characterisation, two empirical probes assess whether the amplification predicted by the Jacobian spectrum manifests under repeated application of the learned update. The first is a one-step amplification factor,
\[
G(\varepsilon)
=
\frac{\mathrm{RMSE}(\delta \boldsymbol{\varTemp}_{\mathrm{out}})}{\mathrm{RMSE}(\delta \boldsymbol{\varTemp}_{\mathrm{in}})},
\]
where $\delta\boldsymbol{\varTemp}_{\mathrm{in}}$ is the input perturbation, drawn as isotropic Gaussian noise of magnitude $\varepsilon$ (a fraction of the input scale), and $\delta\boldsymbol{\varTemp}_{\mathrm{out}}$ is the resulting change in the one-step prediction. Sweeping $\varepsilon$ extends the analysis beyond the infinitesimal regime probed by the Jacobian: $G>1$ signals that the learned map expands perturbations, and its dependence on $\varepsilon$ reveals whether this expansion persists into the finite-amplitude, nonlinear regime.

The second probe evaluates how these effects accumulate over a trajectory. An initial-condition perturbation is propagated through the recursive application of the learned update, and the growth of the deviation $\lVert \boldsymbol{\delta}(t) \rVert$ is tracked over the rollout horizon. Where $G(\varepsilon)$ isolates a single step, this measures how perturbations compound along the composed flow map, and thus whether local expansion becomes unbounded trajectory error. Together with the spectral analysis, these probes establish whether the learned dynamics inherit the perturbation-contracting behaviour of the reference operator $(-\varFEmass^{-1}\varFEstiffness)$.

\subsection{Training and parameterization}

The neural network parameters $\xi$ are trained in a supervised learning setting using observed time trajectories of the semi-discrete FEM solution. Given training data
$\{\sigma(\omega_k), \boldsymbol{\varTemp}_k(\varEvent_j)\}_{k=1}^{n_t}$ for realizations $j=1,\ldots,n_s$, the network is used to approximate the continuous-time vector field governing the evolution. The training objective is defined as the minimization of the discrepancy between predicted and observed states:
\begin{equation} \label{eq:gen_loss_func}
\mathcal{E}(\xi)
=
\frac{1}{n_s n_t}
\sum_{j=1}^{n_s}\sum_{k=1}^{n_t}
\left\|
\hat{\boldsymbol{\varTemp}}_{k+1}(\varEvent_j)
-
\boldsymbol{\varTemp}_{k+1}(\varEvent_j)
\right\|^2,
\end{equation}
and classical gradient-based approaches are used for the parameter learning.

The approximation accuracy as well as stability depend on the choice of neural network parametrization $f_\xi(\cdot)$. When the vector field is parameterized as a feedforward neural network
\begin{equation}\label{ffn_clas}
f_\xi(\boldsymbol{u}) = \mathrm{FFN}_\xi(\sigma, \boldsymbol{u}),
\end{equation}
the resulting implicit system is governed entirely by the nonlinear neural evaluation. The corresponding nonlinear solve at each time step is given by
\begin{equation}
\boldsymbol{u}_{k+1}
=
\boldsymbol{u}_k
+
\frac{\Delta t}{12} \Big(
5 \mathrm{FFN}_\xi(\sigma,\boldsymbol{u}_{k+1})
+ 8 \mathrm{FFN}_\xi(\sigma,\boldsymbol{u}_k)
 -\mathrm{FFN}_\xi(\sigma,\boldsymbol{u}_{k-1})
\Big).
\end{equation}
Hence, the accuracy and computational costs are driven by the algorithm used for solving nonlinear equations \cite{kelley_iterative_1995}.
In this case, stability is determined by the spectral properties of the Jacobian of the learned vector field,
\(
J_f = \frac{\partial f_\xi}{\partial \boldsymbol{u}},\)
and the stability region of the time integration scheme.

To improve the learning, one may approximate the vector field by a residual neural network \cite{chen_neural_2018}. In such a case we do not approximate the vector-field directly, but we add additonal identity-dependent linear drift such that
\begin{equation} \label{eq:semi_discrete_rhs1} \frac{d\boldsymbol{\varTemp}_h(\vartime)}{d\vartime}
    = \boldsymbol{\varTemp}_h(\vartime) +g_\xi (\boldsymbol{\varTemp}_h.\varTensor),
\end{equation}
holds. This structure can improve optimization stability compared to a plain FFN by reducing the burden on the network to represent the full transformation from scratch. By using a structural bias toward exponential growth or decay, the spectrum of the Jacobian is shifted and thereby the stability properties of the discretized system are altered. Consequently, the residual-vector-field formulation imposes stronger constraints on stability and requires the learned correction term to compensate for the inherent linear drift. In this case we have:
\begin{equation}
f_\xi(\boldsymbol{u}) = \boldsymbol{u} + g_\xi(\sigma,\boldsymbol{u}),
\end{equation}
such that the Adams--Moulton scheme becomes
\begin{align}
\boldsymbol{u}_{k+1}
&=
\boldsymbol{u}_k
+
\frac{\Delta t}{12}\Big(
5 (\boldsymbol{u}_{k+1} + g_\xi(\sigma,\boldsymbol{u}_{k+1}))
+ 8 (\boldsymbol{u}_k + g_\xi(\sigma,\boldsymbol{u}_k))
- (\boldsymbol{u}_{k-1} + g_\xi(\sigma,\boldsymbol{u}_{k-1}))
\Big).
\end{align}
Hence, the residual parameterization introduces an additional linear coupling in the unknown state $\boldsymbol{u}_{k+1}$, effectively shifting the spectrum of the discrete operator. This modifies the conditioning of the nonlinear system and can affect convergence of fixed-point iterations used in practice. In particular, the Jacobian of the residual formulation becomes
\begin{equation}
J = I + \frac{\partial g_\xi}{\partial \boldsymbol{u}},
\end{equation}
which induces an eigenvalue shift relative to the FFN parameterization. As a consequence, stability is influenced not only by the learned nonlinear correction $g_\xi$, but also by the inherent linear identity contribution \cite{he_deep_2016}, which may amplify or damp perturbations depending on the spectral properties of the system and the chosen time step.

If the network is parametrized by a sequence of residual blocks
\begin{equation}
\boldsymbol{h}_{j+1} = \boldsymbol{h}_j + g_{\xi,j}(\boldsymbol{h}_j),
\qquad j = 0,\dots,L-1,
\end{equation}
with initial condition $\boldsymbol{h}_0 = g_\xi^0(\sigma,\boldsymbol{u}_h)$, the output of the network is
\begin{equation}
f_\xi(\boldsymbol{u};\sigma) = \boldsymbol{h}_L.
\end{equation}
By recursive substitution, this yields the expansion
\begin{equation}
f_\xi(\boldsymbol{u};\sigma)
=
\boldsymbol{u}
+
\sum_{j=0}^{L-1}
g_{\xi,j}(\boldsymbol{h}_j),
\end{equation}
where each intermediate state satisfies
\begin{equation}
\boldsymbol{h}_j
=
\boldsymbol{u}
+
\sum_{j=0}^{j-1} g_{\xi,\ell}(\boldsymbol{h}_\ell).
\end{equation}
In the limit of small residual steps, the architecture can be interpreted as an Euler discretization of an underlying continuous transformation in depth:
\begin{equation}
\frac{d\boldsymbol{h}(s)}{ds} = g_\xi(\boldsymbol{h}(s)), 
\qquad s \in [0,1],
\end{equation}
with $s$ denoting the artificial depth variable.
Substituting the deep residual network into the Neural ODE yields
\begin{equation}
\dot{\boldsymbol{u}}(t)
=
f_\xi(\boldsymbol{u}(t);\sigma)
=
\boldsymbol{u}(t)
+
\sum_{j=0}^{L-1} g_{\xi,j}(\boldsymbol{h}_j(\boldsymbol{u}(t))).
\end{equation}
Thus, the vector field itself is generated by an internal discrete dynamical system in feature space.
Compared with a single residual block or a standard feedforward network, a deep residual architecture provides greater expressive power by representing the vector field as a composition of incremental nonlinear corrections. The identity skip connections improve gradient propagation during training and bias each block toward learning only deviations from the identity mapping. Moreover, the recursive residual structure admits a natural continuous-depth interpretation, making it particularly well suited for Neural ODEs, where it can be viewed as a discretization of an underlying continuous transformation in network depth. Their stability is given by 
the Jacobian of the vector field over the chain rule as
\begin{equation}
J_f(\boldsymbol{u})
=
\frac{\partial f_\xi}{\partial \boldsymbol{u}}
=
\prod_{k=0}^{L-1}
\left(
I + J_{g_{\xi,k}}(\boldsymbol{h}_k)
\right), J_{g_{\xi,k}}(\boldsymbol{h}_k)
=
\frac{\partial g_{\xi,k}}{\partial \boldsymbol{h}}(\boldsymbol{h}_k).
\end{equation}
This shows that the Jacobian is a product of near-identity transformations.
Due to the multiplicative structure of the Jacobian in deep residual networks, stability depends on the cumulative effect of each residual block. If each block satisfies a mild contractivity condition
\begin{equation}
\rho\!\left(I + J_{g_{\xi,k}}\right) \approx 1,
\end{equation}
then the overall mapping remains well-conditioned. However, repeated composition may lead to either attenuation or amplification of perturbations depending on the spectral properties of the individual blocks.

\section{Manifold-aware Neural ODE}

The neural network architecture described above is not yet optimal, as it assumes that the physical parameterization in the form of $\varTensor$ is provided as a vector in a Euclidean space. However, the tensor field is modeled as a SPD-valued random variable, as in \refEQ{\ref{eq:stochastic_tensor}}, taking values on the Riemannian manifold $\textrm{Sym}_{+}^d(\mathbb{R})$ of SPD matrices.
A naive vectorisation of $\varTensor$ in Euclidean coordinates as in Fig.~(\ref{fig:IH_NN_framework}) disregards the intrinsic Riemannian structure. This leads to distortions in both geometric distances and the statistical structure of the induced distribution.
To preserve this structure, we employ the Constitutive Manifold Neural Network (CMNN) architecture \cite{schuttert_constitutive_2026}, which incorporates SPD-valued inputs through manifold-consistent mappings that respect the geometry of $\textrm{Sym}_{+}^d(\mathbb{R})$ . In contrast to Euclidean vectorisation, this approach operates directly on the manifold, thereby preserving the relevant invariance properties. 

We introduce the constitutive manifold neural ODE approximation such that the formulation in Eq.~(\ref{ffn_clas}) reads:
\begin{equation} \label{eq:neural_ode1}
    \frac{d\boldsymbol{\varTemp}_h(\vartime)}{d\vartime}
    = f_{\xi}\!\bigl(f_\zeta(\varTensor),\boldsymbol{\varTemp}_h(\vartime)\bigr),
\end{equation}
where \(f_\zeta(\varTensor)\) enforces hard constraints on the physical parameterization of the conductivity tensor. 
The Strength--Angular (StrAng) layer $f_\zeta(\cdot)$ enforces the SPD constraint by decomposing $\varTensor$ into physically meaningful components---strength and orientation---and mapping them into a Euclidean representation in a geometry-preserving manner. It is defined as the composition of three transformations \cite{schuttert_constitutive_2026}:
\begin{alignat}{1} \label{eq:StrAng}
    f_\zeta(\varTensor) &:= \bigl(\varrho^{\text{flat}} \circ \varrho^{\text{log}} \circ \varrho^{\text{eig}}\bigr)(\varTensor), \\
\intertext{where}
    \varrho^{\text{eig}}  &: \varSymPlus(d) \xrightarrow{\mathrm{eig}}
        \varDiag^{+}(d) \times \varSO(d), \\
    \varrho^{\text{log}}  &: \varDiag^{+}(d) \times \varSO(d)
        \xrightarrow{(\log,\,\log)}
        \mathfrak{d}(d_1) \times \mathfrak{so}(d_2), \\
    \varrho^{\text{flat}} &: \mathfrak{d}(d_1) \times \mathfrak{so}(d_2)
        \xrightarrow{\mathrm{flatten}}
        \mathbb{R}^{d_1 + d_2}.
\end{alignat}
Here, $\varDiag^{+}(d)$ denotes the space of positive diagonal matrices containing the eigenvalues of $\varTensor$, while $\varSO(d)$ denotes the rotation group containing the corresponding eigenvector matrices. The logarithmic maps transfer these quantities to their associated tangent spaces: the diagonal tangent space $\mathfrak{d}(d)$ for the conductivity strengths and the Lie algebra $\mathfrak{so}(d)$ for the orientations. The final flattening operation converts these tangent-space representations into a Euclidean vector suitable as input to the neural network.
The mappings are applied sequentially, beginning with $\varrho^{\text{eig}}$, which performs the eigendecomposition as defined in \refEQ{\ref{eq:eigen_decomp}} and thereby decomposes $\varTensor$ into its eigenvalues and eigenvectors. This separates the conductivity into magnitudes along the principal directions and the corresponding fiber orientations \cite{feragen_geometries_2017}. The resulting factors are then mapped to their respective tangent spaces through $\varrho^{\text{log}}$, where the scalar logarithm is applied to the positive eigenvalues and the matrix logarithm to the rotation component, effectively providing a local linearisation of the underlying manifold structure. Finally, $\varrho^{\text{flat}}$ combines these tangent-space representations into a single Euclidean vector, making the result compatible with standard neural network layers. Since the eigendecomposition is not unique due to permutations of eigenvalue–eigenvector pairs, an explicit ordering rule (\ref{app:unique_decomp}) is imposed to fix a canonical representative. This ensures that $f_\zeta(\cdot)$ is well-defined for rotations up to $\tfrac{\pi}{4}$, which lies well within the angular regime considered in this work.

This construction guarantees that the resulting tensor is symmetric positive definite by design, while providing a structured Euclidean embedding that separates spectral magnitude and rotational degrees of freedom.
Here, $d_1$ and $d_2$ denote the dimensions of minimum number of scaling and orientation components that are set by symmetry class and are modelled as uncertain, see \refEQ{\ref{eq:stochastic_tensor}}. Hence, components held fixed do not contribute to this embedding, so that a purely orientational uncertainty gives $d_1 = 0$.

The network is trained as described previously using an $L_2$ loss function. The material tensor layers are not optimized during training; instead, the physical constraints are enforced explicitly through a hard parametrization.

%% file: 4_Result/Result_Det.tex
\begin{figure}[t]               
    \captionsetup{justification=centering}
    \centering
    \begin{subfigure}[b]{0.34\textwidth}
        \includegraphics[width=\textwidth]{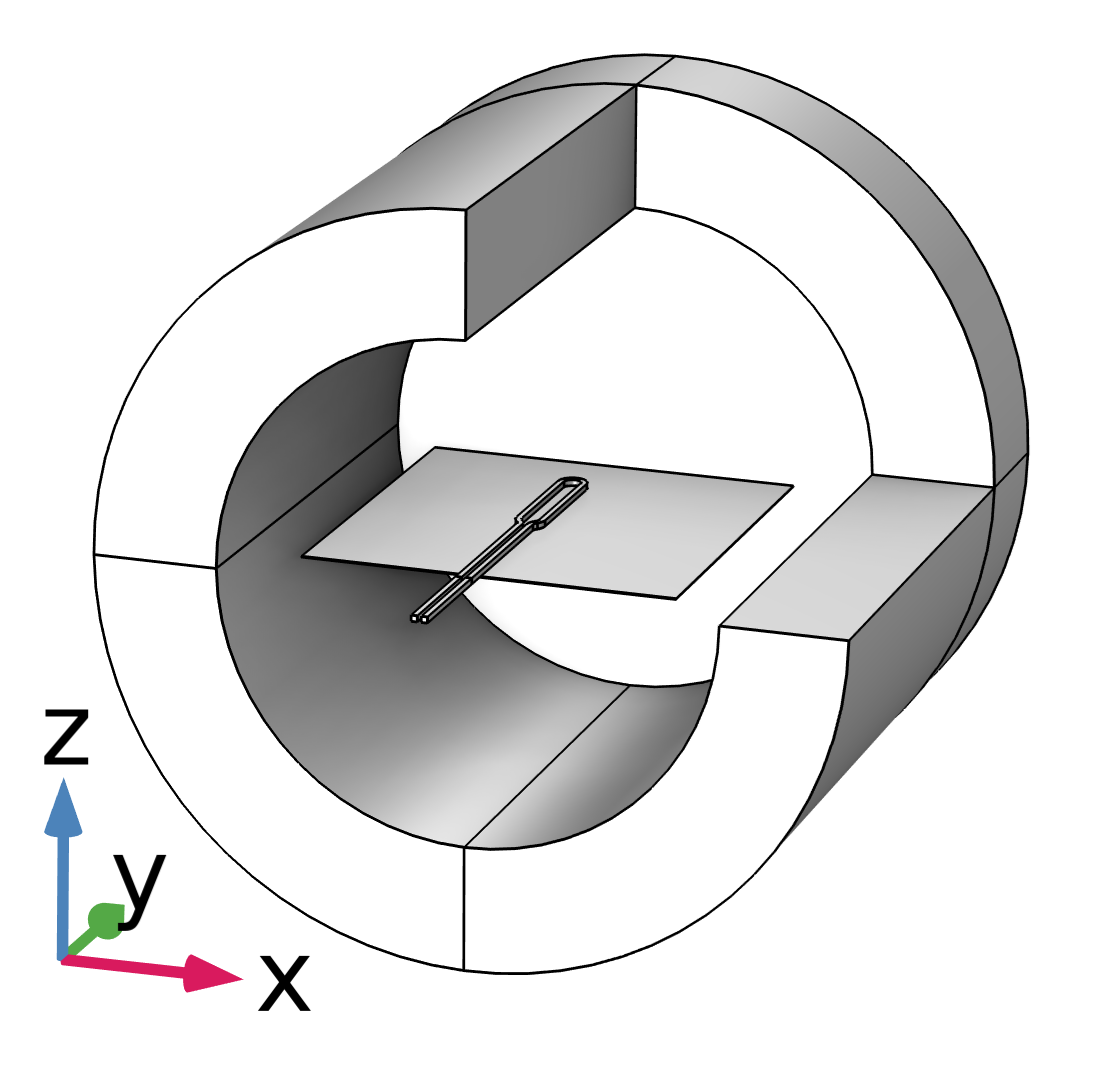}
        \caption{Domain $\varDomain$ including the air, coil and CFRTP}
        \label{fig:iw4e_em_domain}
    \end{subfigure}
    \begin{subfigure}[b]{0.3\textwidth}
        \includegraphics[width=\textwidth]{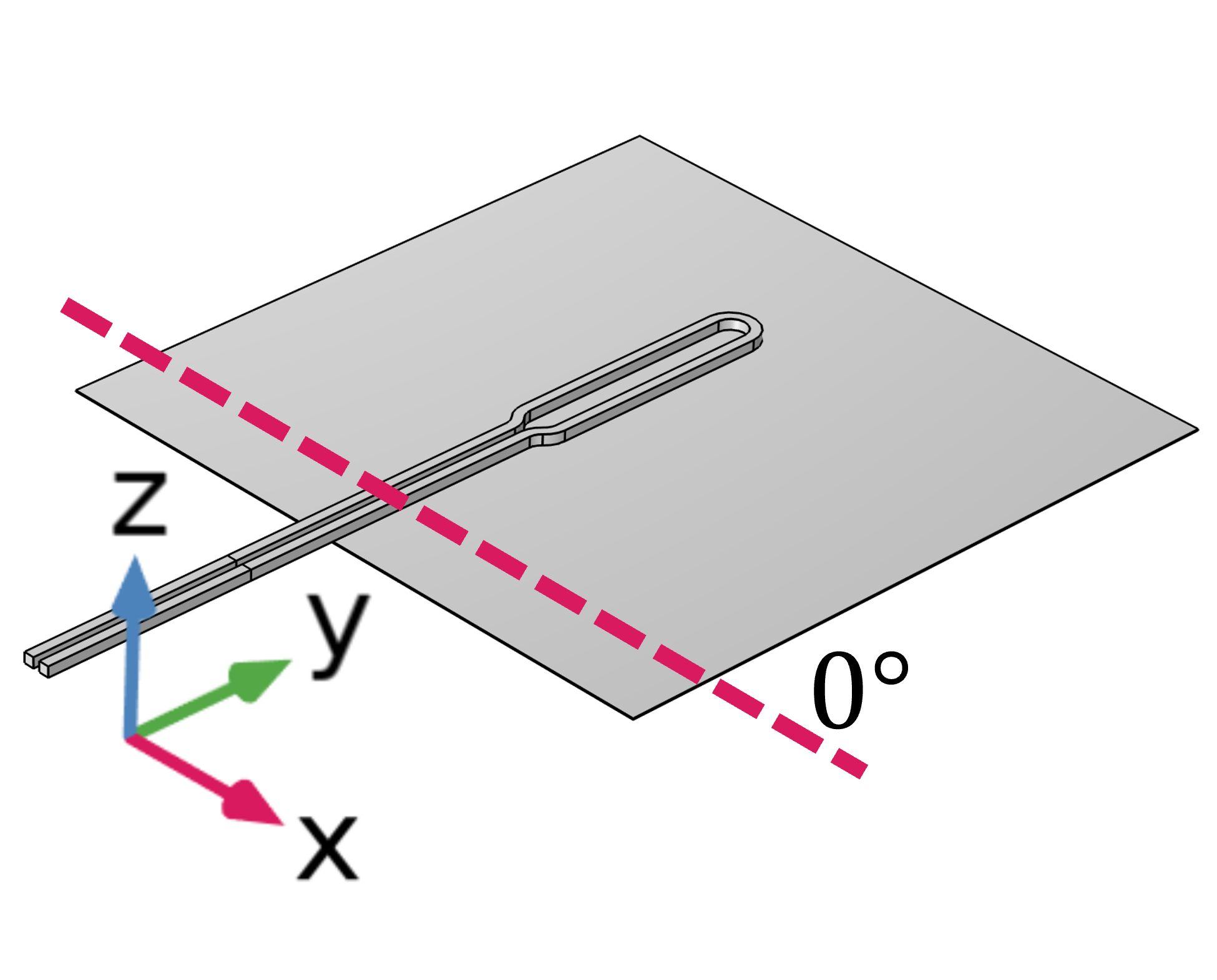}
        \caption{Domain $\varDomain_C$ which is the CFRTP plate and an indication of direction for the fibres.}
        \label{fig:iw4e_heat_domain}
    \end{subfigure}       
    \begin{subfigure}[b]{0.34\textwidth}
        \includegraphics[width=\textwidth]{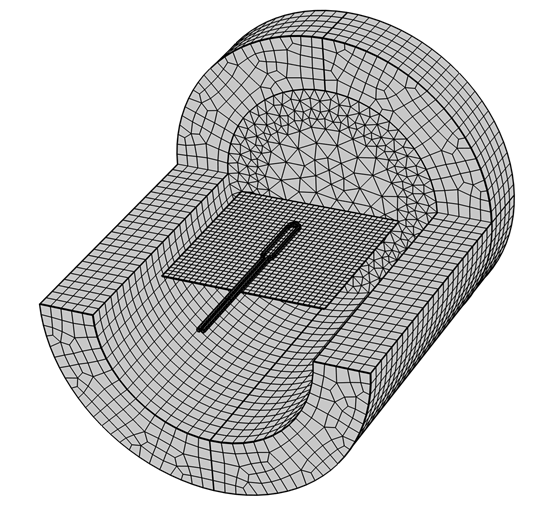}
        \caption{Discretisation}
        \label{fig:iw4e_discretisation}
    \end{subfigure}
    \caption{Simulations domains}
\end{figure}

\noindent
The induction heating model introduced in Section \ref{sec:problem} is used to study how stochastic electrical conductivity influences the temperature distribution of the CFRTP during heating. As a first step, the deterministic case is considered, with \refEQ{\ref{eq:deterministic_EM}} weakly coupled to \refEQ{\ref{eq:deterministic_HT}} \cite{keyes_multiphysics_2013}. The electromagnetic domain $\varDomain$ is shown in \refFIG{\ref{fig:iw4e_em_domain}}, where the geometry is constructed in COMSOL Multiphysics and a coil is placed horizontally above a square 300 $\times$ 300 $\times$ 0.14 CFRTP laminate. The surrounding air domain is enclosed by an infinite-element boundary layer, enforcing the boundary condition in \refEQ{\ref{eq:deterministic_EM_BC}}. The heat-transfer subdomain $\varDomain_C$ is illustrated in \refFIG{\ref{fig:iw4e_heat_domain}}. The domain boundaries are subjected to Robin boundary conditions with an ambient temperature of $T_\infty = 30\,^\circ\mathrm{C}$ and a heat-transfer coefficient of $h = 10\,\mathrm{W\,m^{-2}\,K^{-1}}$.

Deterministic process parameters are summarised in Table \ref{tab:parameters}. The laminate is assumed to be made of Toray Cetex\textsuperscript{\textregistered} TC1200 carbon-fibre-reinforced polyether ether ketone (PEEK), a thermoplastic based composite commonly used in aerospace applications \cite{buser_characterisation_2022}, with a symmetric layup $[45^\circ, -45^\circ]_s$. Each ply has a thickness of 0.14 mm. For simplicity heat capacity $\varHeatC$, density $\varDens$ and thermal conductivities $\varTcond$ are assumed to be constant with temperature. The  $\varTcond$ listed in Table \ref{tab:parameters} are in the principal material coordinate system of the composite ply, while the electrical conductivity tensor $\overline{\varTensor}$ of the ply shown in \refEQ{\ref{eq:initial_econd}} is expressed in global coordinates visible in \refFIG{\ref{fig:iw4e_heat_domain}}, with its decomposition into eigenvalues and eigenvectors given as
\begin{equation}\label{eq:initial_econd}
\begin{split}
    \hspace{-1.5cm}
    \bar{\varTensor} &= \begin{bmatrix}
        19153.98 & 19146.98 & 0 \\
        19146.002 & 19153.02 & 0 \\
        0  & 0 & 0.507 \\
    \end{bmatrix}_{ply}[S/m] \\[6pt]
    \text{where} \quad
    \bar{\varEigVal}_{\varTensor} &= \begin{bmatrix}
        38300 \\ 6.04 \\ 0.507
    \end{bmatrix}_{ply}[S/m] \;,\quad
    \bar{\varEigVec}_{\varTensor} = \begin{bmatrix}
        0.7071 & \pm 0.7071 & 0 \\
        \pm 0.7071 & 0.7071 & 0 \\
        0  & 0 & 1 \\
    \end{bmatrix}_{ply}
\end{split}.
\end{equation}

The FEM model is implemented in COMSOL Multiphysics \cite{noauthor_comsol_nodate}, using the solvers and time integration scheme detailed in Section~\ref{sec:discretisation}. The time solutions are interpolated in postprocessing onto a uniform grid over $[0,10]\,\text{s}$ with $\Delta t = 0.5\,\text{s}$, yielding temperature trajectories at 21 time steps.
\begin{figure}[t]               
\captionsetup{justification=centering}
    \centering
    \begin{subfigure}[t]{0.49\textwidth}
        \includegraphics[width=\textwidth,trim=0cm 0cm 0.7cm 0.7cm,clip]{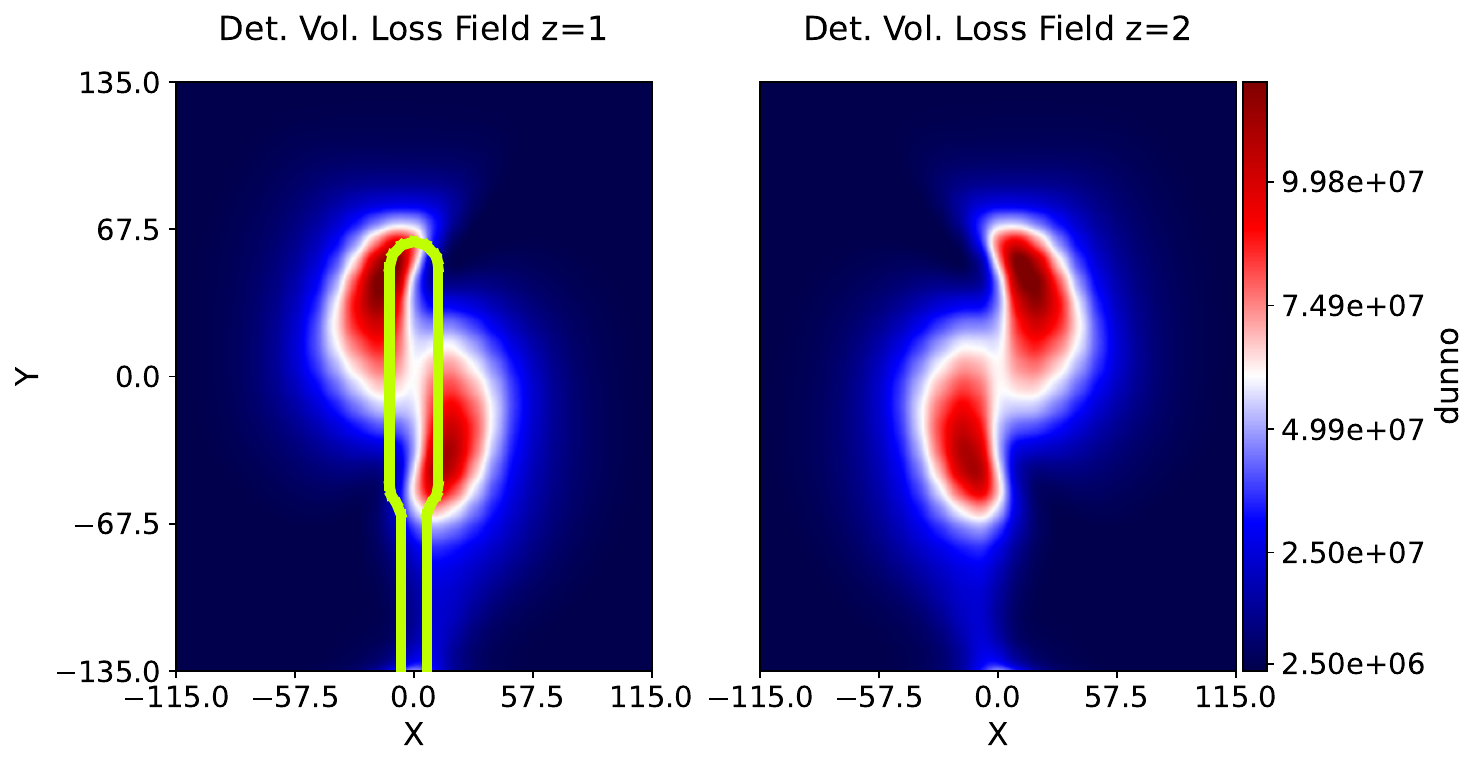}
        \caption{Volumetric loss plot in the 1st and 2nd ply. Overlay of the coil is included in green on the left plot.}
        \label{fig:det_source}
    \end{subfigure}
    \hfill
    \begin{subfigure}[t]{0.49\textwidth}
        \includegraphics[width=\textwidth,trim=0cm 0cm 0cm 0.7cm,clip]{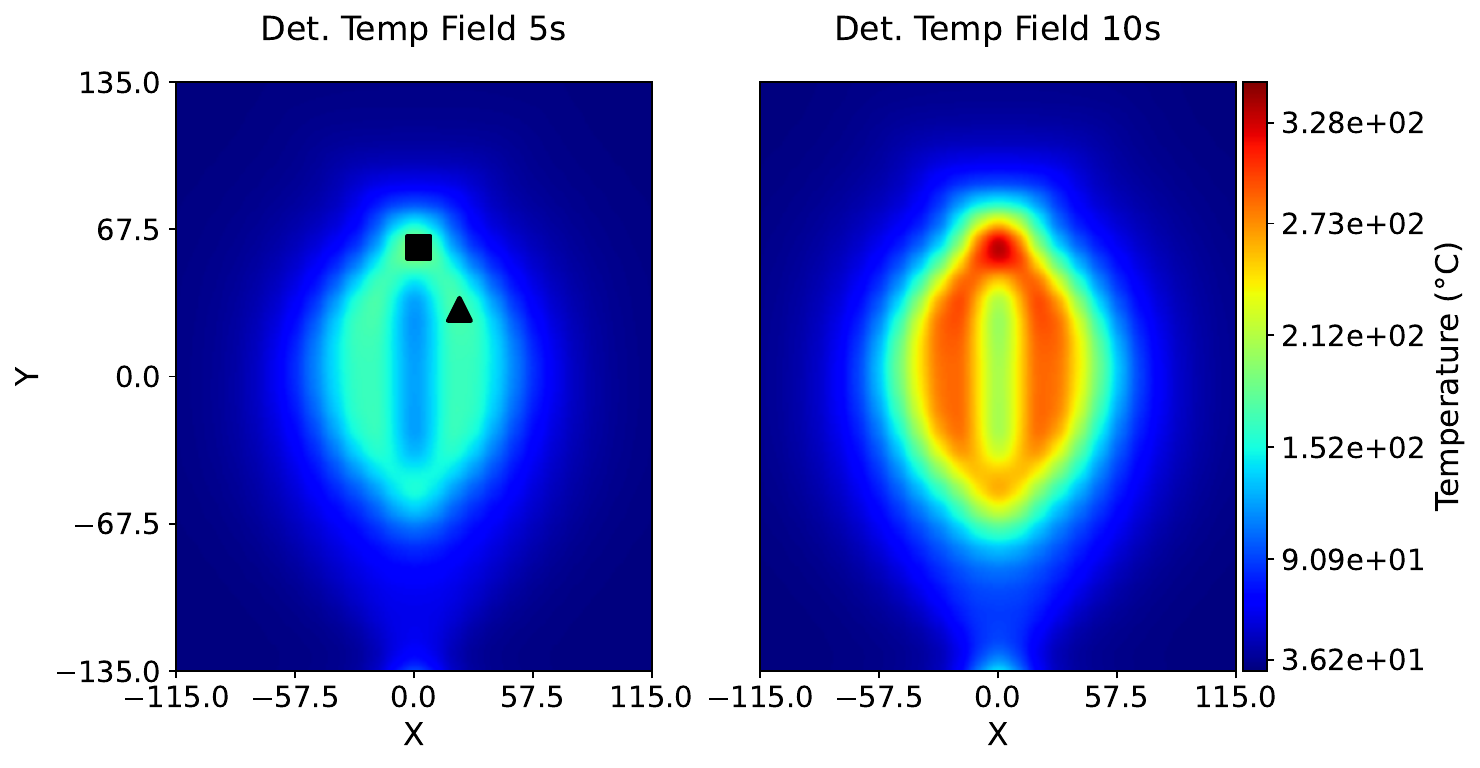}
        \caption{Temperature plot at $t=5s$, $t=10s$ in the second ply.}
        \label{fig:det_temp}
    \end{subfigure}
    \caption{Deterministic simulation results.}
    \label{fig:det_results}
\end{figure}

\begin{table}[b]
    \centering
  \caption{Deterministic process parameters, where $(\kappa_{1}, \kappa_{2}, \kappa_{3})$ denotes the thermal conductivity along the fibre direction, in-plane transverse to the fibre, and the out-of-plane direction respectively.}
  \setlength{\tabcolsep}{6pt}
  \renewcommand{\arraystretch}{1.15}
  \begin{tabular}{@{}l c c@{}}
    \toprule
    Parameter & Value & Unit \\
    \midrule
    $\varJcur_s$     & 495 RMS & [A] \\
    $\varFcoil$      & 342     & [kHz] \\
    $\varPermea_r$   & 1       & \\
    $\varPermit_r$   & 3.7     & \\
    L$\times$W$\times$H & 300 $\times$ 300 $\times$ 0.14 & [mm] \\
    Layup            & $[45^\circ,-45^\circ]_s$ & \\
    \bottomrule
  \end{tabular}
  \hspace{1.5em}
  \begin{tabular}{@{}l c c@{}}
    \toprule
    Parameter & Value & Unit \\
    \midrule
    $\varHeatC$      & 946  & [J/(kgK)] \\
    $\varDens$       & 1601 & [kg/m$^3$] \\
    $\kappa_{1}$  & 6.4  & [W/(mK)] \\
    $\kappa_{2}$  & 0.81 & [W/(mK)] \\
    $\kappa_{3}$  & 0.74 & [W/(mK)] \\
    $t_e$            & 10   & [s] \\
    \bottomrule
  \end{tabular}
  \label{tab:parameters}
\end{table}

The volumetric loss calculated with the EM analysis is presented in \refFIG{\ref{fig:det_source}}, where the domain is cropped to the region of interest for clarity and the coil geometry is overlaid in green for reference. The resulting field is strongly influenced by the fibre orientation. For the $45^\circ$ ply, the upper-left and lower-right regions exhibit the highest heat losses, whereas this pattern is mirrored vertically in the $-45^\circ$ ply. The spatial correspondence between the regions of elevated loss and the coil geometry is clearly visible from the overlay, confirming that the induced eddy currents are concentrated beneath the coil windings. In addition, the straight lower section of the coil appears to contribute non-negligible volumetric losses up to the lower boundary of the domain.

The heat-transfer results, shown in \refFIG{\ref{fig:det_temp}}, illustrate the evolution of the heating pattern at 5s and 10s, which clearly follows the geometry of the coil. The ply-wise antisymmetry observed in the EM analysis does not persist in the thermal results, as the plies are sufficiently thin to allow rapid heat diffusion across layers, leading to an almost symmetric temperature distribution. Furthermore, the left sub-figure indicates two locations marked by a square and a triangle which are later used for the evaluation of probability density functions. The square denotes the point reaching the highest overall peak temperature, while the triangle identifies the point with the highest peak temperature along the coil's parallel section.

%% file: 4_Result/Result_Stochastic.tex
\begin{figure}[t]               
    \captionsetup{justification=centering}
    \centering
    \includegraphics[width=\textwidth,trim=0cm 0cm 0cm 0.7cm,clip]{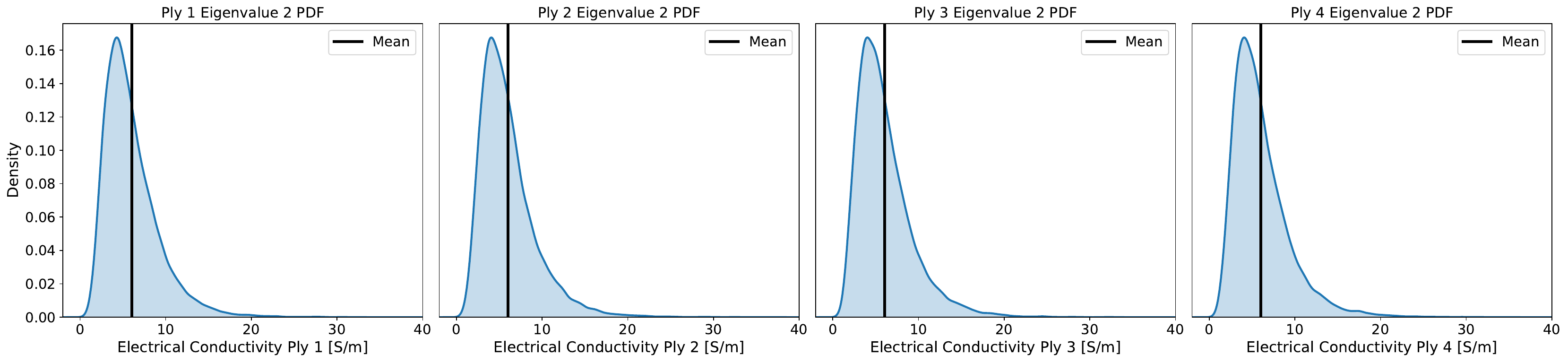}
    \caption{Independent distributions of the second eigenvalue of electrical conductivity for all four plies with scaling-orientation uncertainty. }
    \label{fig:sclori_eig_2_comp} 
\end{figure}

\begin{figure}[!t]               
    \captionsetup{justification=centering}
    \centering
    \includegraphics[width=\textwidth,trim=0cm 0cm 0cm 0.7cm,clip]{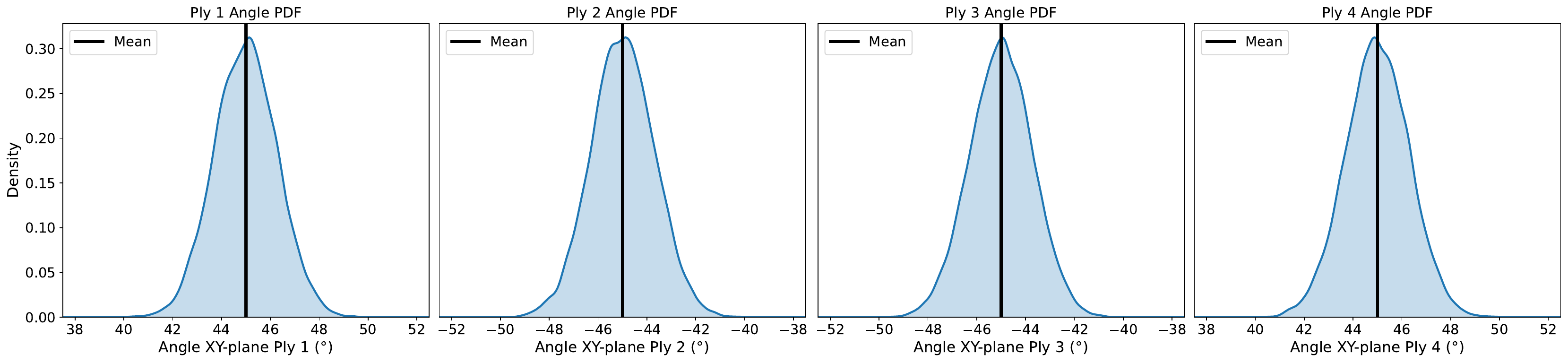}
    \caption{Independent distributions of the rotation angle in the XY plane for all four plies with scaling-orientation uncertainty. }
    \label{fig:sclori_angle_comp} 
\end{figure}
\noindent
For the stochastic analysis, the governing systems of equations defined in \refEQ{\ref{eq:stochastic_EM}} and \refEQ{\ref{eq:stochastic_HT}} are solved under stochastic electrical conductivity inputs. For $\varTensor(\varEvent)$ we consider three scenarios: (i) only the scaling is uncertain; (ii) only the orientation is uncertain; (iii) both scaling and orientation are uncertain. This separates the relative contributions of scaling and orientation uncertainty, and any interaction between them. To isolate the effect of $\varTensor(\varEvent)$, the thermal conductivity is held deterministic at the values listed in Table~\ref{tab:parameters} across all $\varEvent$.

\begin{table}[b]
  \caption{Mean and standard deviation of the lognormal distributions for the eigenvalues of electrical conductivity tensor in principal material orientation (S/m), obtained from \cite{buser_predicting_2025}.}
  \centering
  \setlength{\tabcolsep}{6pt}
  \renewcommand{\arraystretch}{1.15}
  \begin{tabular}{lcc}
    \toprule
    Eigenvalue & Mean $[S/m]$ & Standard Deviation $[S/m]$ \\
    \midrule
    $\lambda_1$ (fibre)     & 38300 & 1460 \\
    $\lambda_2$ (transverse)     & 6.04 & 3.27\\
    $\lambda_3$ (out-of-plane)    & 0.507 & 0.285 \\
    \bottomrule
  \end{tabular}  
  \label{tab:Eigenvalues}
\end{table}

In the case when scaling $\varEigVal$ is considered to be uncertain, and the angle of orientation is fixed to the ply orientation $\overline{\varEigVec}_{ply}$ the eigenvalues of $\varTensor(\varEvent)$ are modelled as independent lognormal random variables according to \refEQ{\ref{eq:stochastic_eigval}}, with the full tensor reconstructed via \refEQ{\ref{eq:stochastic_tensor}}. Since the laminate consists of four plies each with three independent eigenvalues, a single realisation requires $4 \times 3 = 12$ independent draws. The stochastic parameters of the eigenvalues, obtained from \cite{buser_predicting_2025}, are summarised in Table~\ref{tab:Eigenvalues}. The distribution of the second eigenvalue across the four plies is shown in \refFIG{\ref{fig:sclori_eig_2_comp}}. It is presented because its relative spread is large enough that the lognormal character is clearly exhibited, with the strictly positive support and positive skew both visually apparent.

In the second simulation case we fix scaling parameters $\varEigVal$ to values $\overline{\varEigVal}_{ply}$, and only vary the orientation uncertainty. In particular, the principal fibre direction of each ply is modelled as a random variable, with rotation about the Z-axis only. The uncertainty reduces to a single random angle per ply, sampled independently from a von Mises distribution on the unit circle, see \refEQ{\ref{eq:VM}}, parameterised by the nominal mean fibre direction and a concentration parameter $\kappa = 2000$, chosen such that the resulting angular spread (standard deviation $\approx 1.3^\circ$) matches the global absolute fibre misalignment reported in \cite{gomarasca_characterising_2021}. The resulting distribution of fibre orientations across the four plies is shown in \refFIG{\ref{fig:sclori_angle_comp}}, where samples remain tightly clustered around their respective mean orientations. 

In the third case, both uncertainty sources are activated simultaneously, with the eigenvalues drawn from the lognormal model (\refEQ{\ref{eq:stochastic_eigval}}, Table~\ref{tab:Eigenvalues}) and the per-ply fibre angles from the von Mises distribution (\refEQ{\ref{eq:VM}}, $\kappa = 2000$) of the two preceding cases. Scaling and orientation draws are assumed independent within and across plies, bringing the total to $4 \times 3 + 4 = 16$ independent draws per realisation.

To obtain solutions for all three stochastic cases, Monte Carlo sampling is performed using $n_s = 20{,}000$ realisations, as per the procedure described in the discretisation, see Section \ref{sec:discretisation}.

\begin{figure}[!]               
    \captionsetup{justification=centering,margin=0cm}
    \centering
    \vspace{-2cmcm}
    \begin{subfigure}[t]{0.32\textwidth}
        \includegraphics[width=\textwidth, trim=0.0cm 0.0cm 0.0cm 0.7cm,clip]{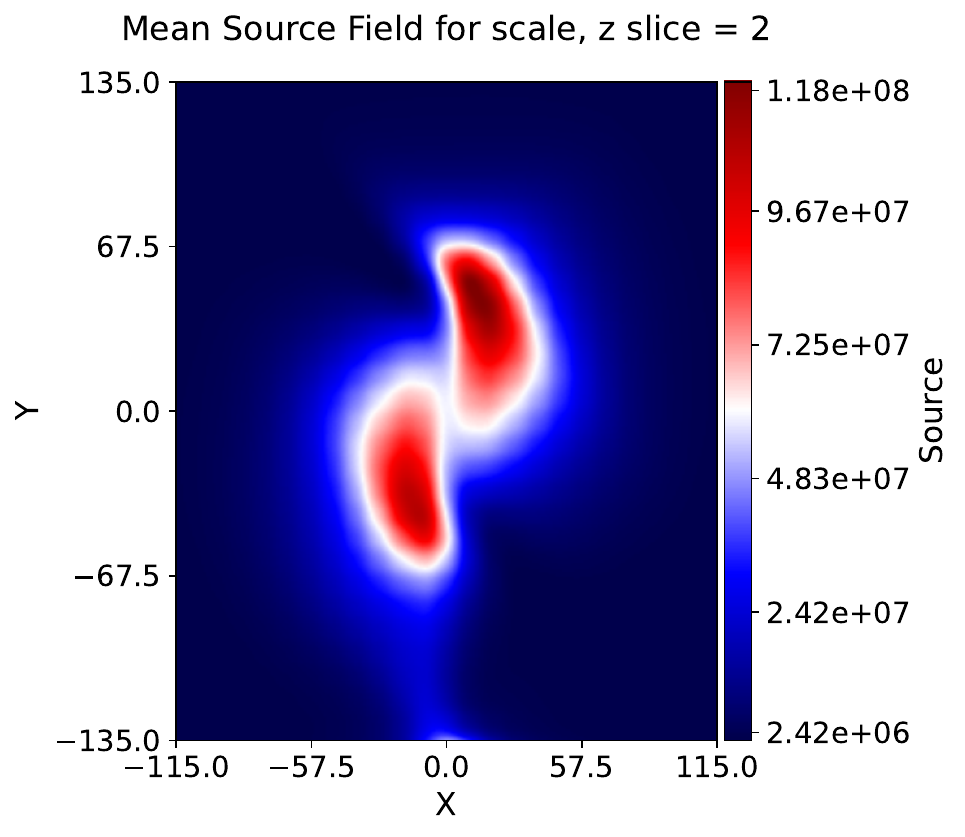}
        \caption{Mean of volumetric loss field for scaling uncertainty.}
        \label{fig:ref_qh_mean_scl}
    \end{subfigure}    
    \hfill
    \begin{subfigure}[t]{0.32\textwidth}        
        \includegraphics[width=\textwidth, trim=0.0cm 0.0cm 0.0cm 0.7cm,clip]{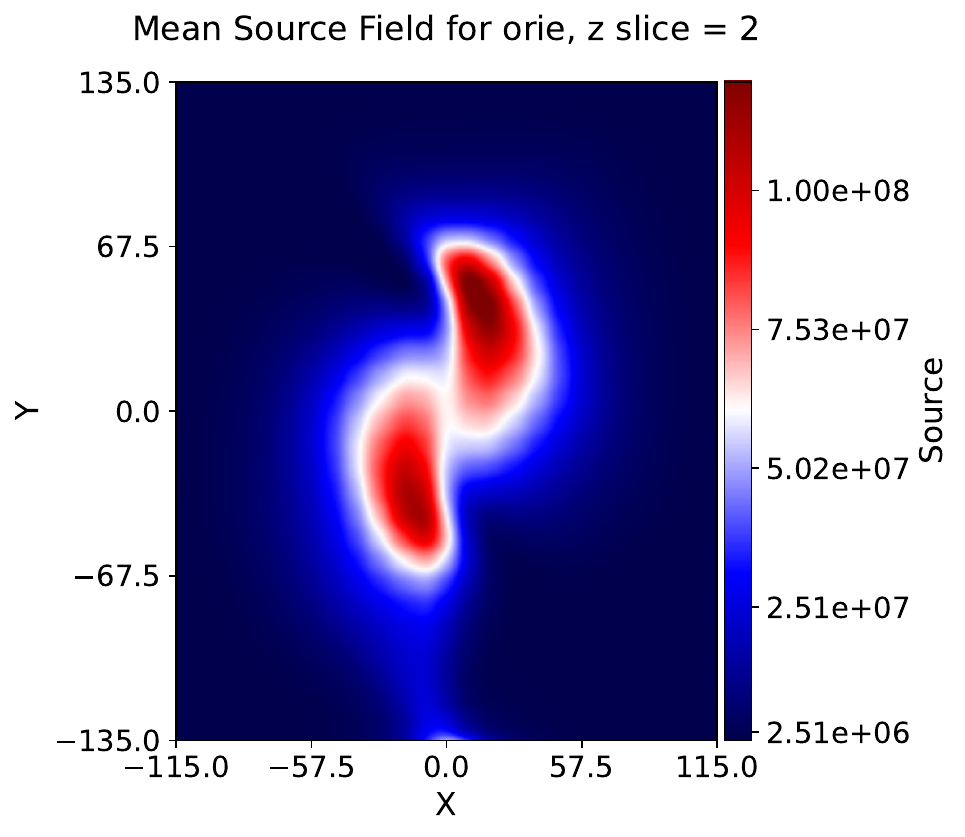}
        \caption{Mean of volumetric loss field for orientation uncertainty.}
        \label{fig:ref_qh_mean_ori}
    \end{subfigure} 
    \hfill    
    \begin{subfigure}[t]{0.32\textwidth}        
        \includegraphics[width=\textwidth, trim=0.0cm 0.0cm 0.0cm 0.7cm,clip]{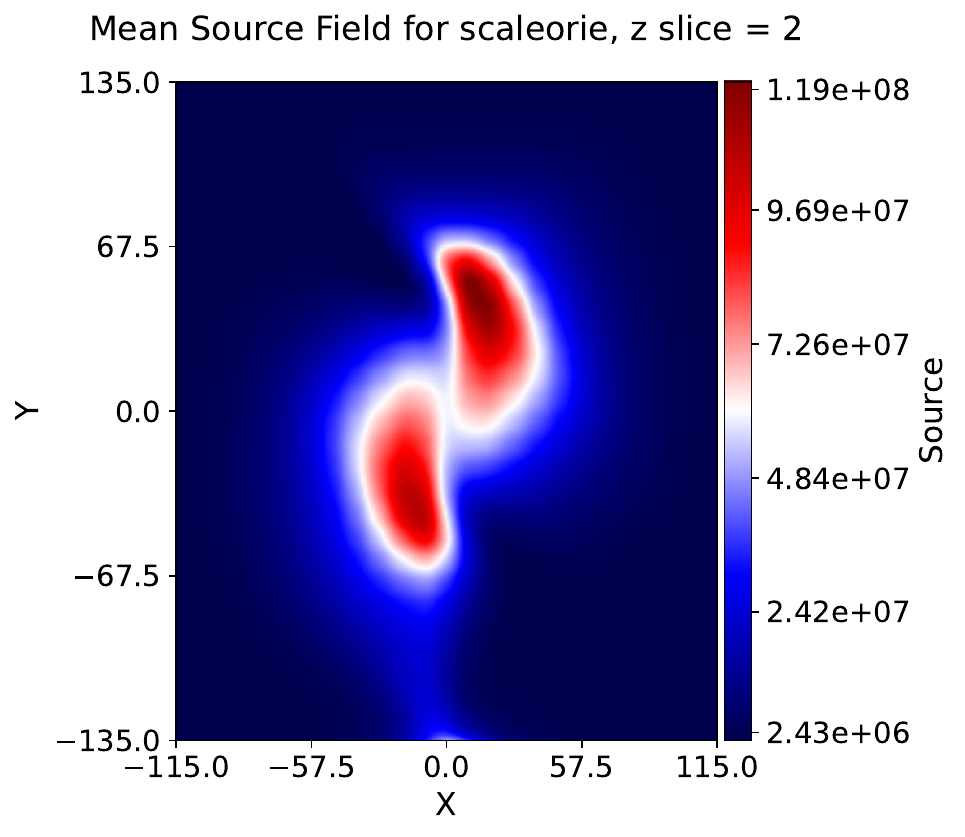}
        \caption{Mean of volumetric loss field for scaling-orientation uncertainty.}
        \label{fig:ref_qh_mean_sclori}
    \end{subfigure} 

    \begin{subfigure}[t]{0.32\textwidth}
        \includegraphics[width=\textwidth, trim=0.cm 0.0cm 0.cm 0.7cm,clip]{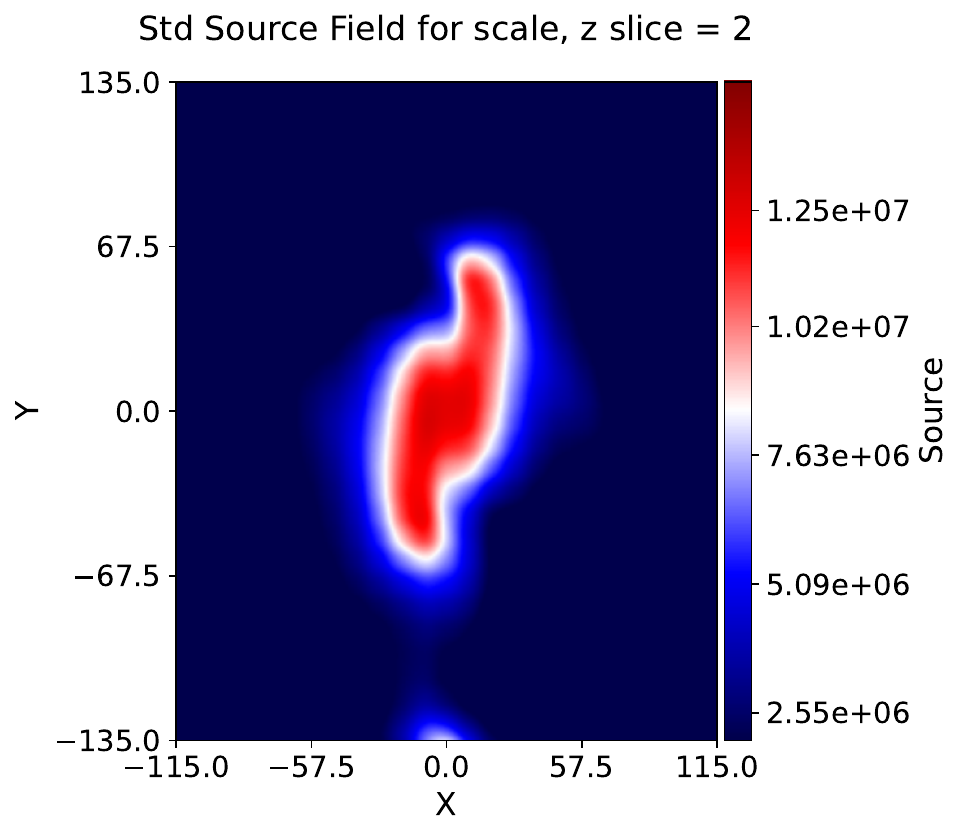}
        \caption{Std. in volumetric loss field for scaling uncertainty.}
        \label{fig:ref_qh_std_scl}
    \end{subfigure}    
    \hfill
    \begin{subfigure}[t]{0.32\textwidth}        
        \includegraphics[width=\textwidth, trim=0.cm 0.0cm 0.cm 0.7cm,clip]{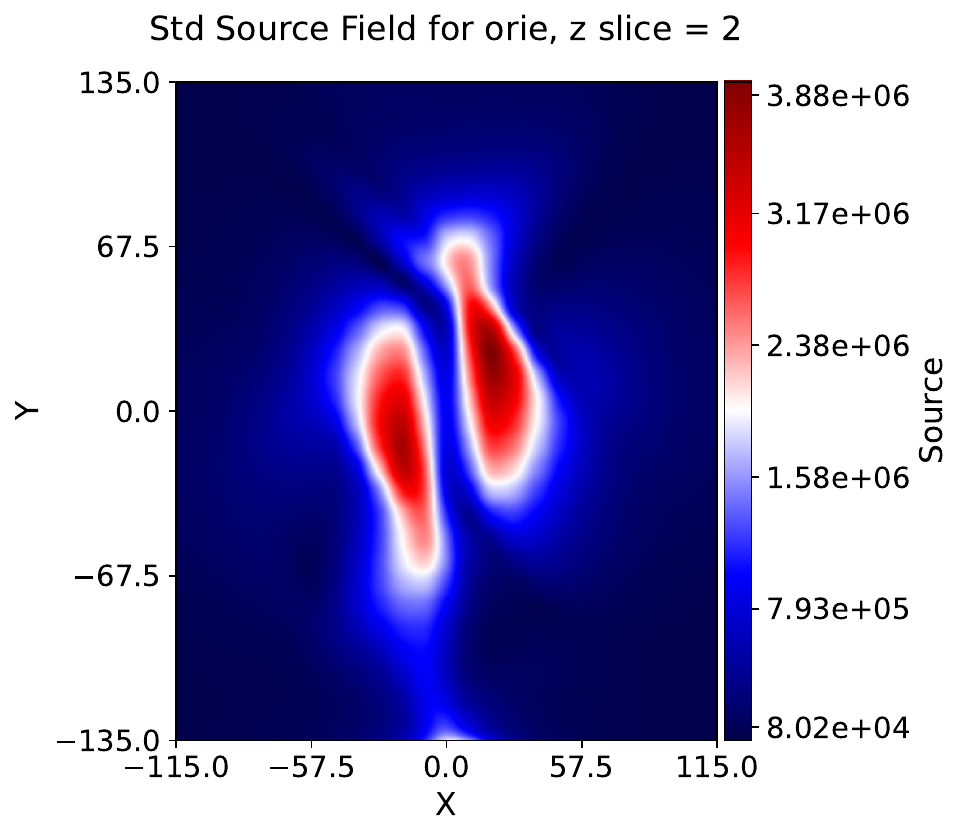}
        \caption{Std. in volumetric loss field for orientation uncertainty.}
        \label{fig:ref_qh_std_ori}
    \end{subfigure} 
    \hfill    
    \begin{subfigure}[t]{0.32\textwidth}        
        \includegraphics[width=\textwidth, trim=0.cm 0.0cm 0.cm 0.7cm,clip]{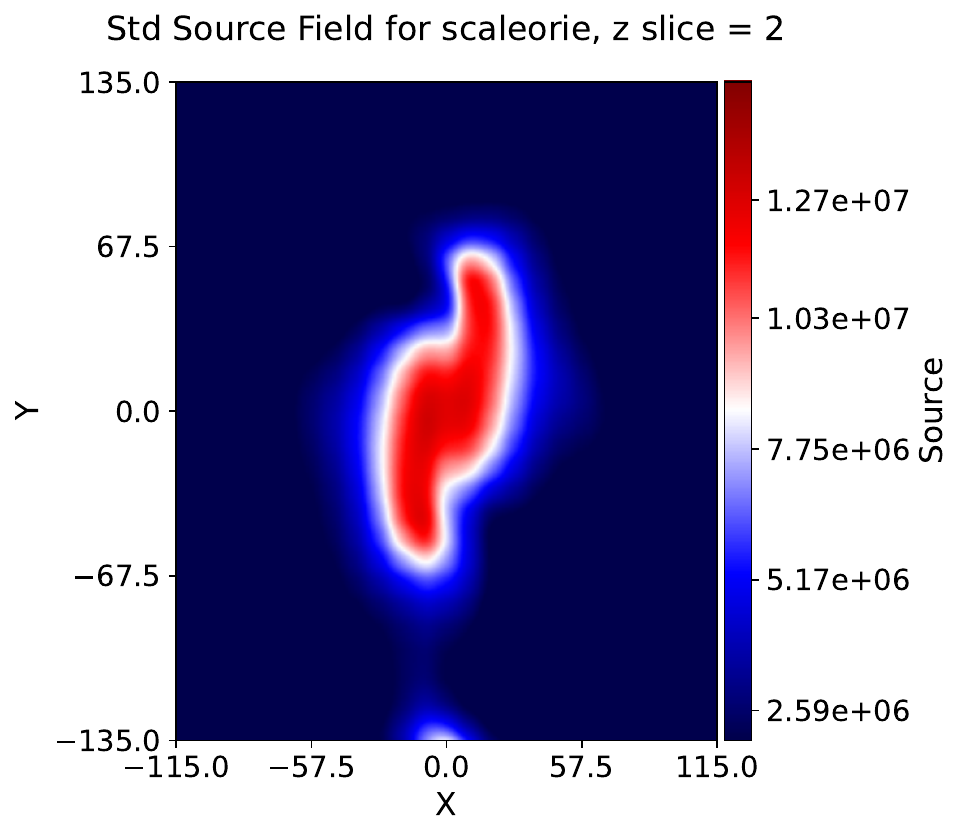}
        \caption{Std. in volumetric loss field for scaling-orientation uncertainty.}
        \label{fig:ref_qh_std_sclori}
    \end{subfigure} 
    
    \begin{subfigure}[t]{0.32\textwidth}
        \includegraphics[width=\textwidth, trim=0.cm 0.0cm 0.cm 0.7cm,clip]{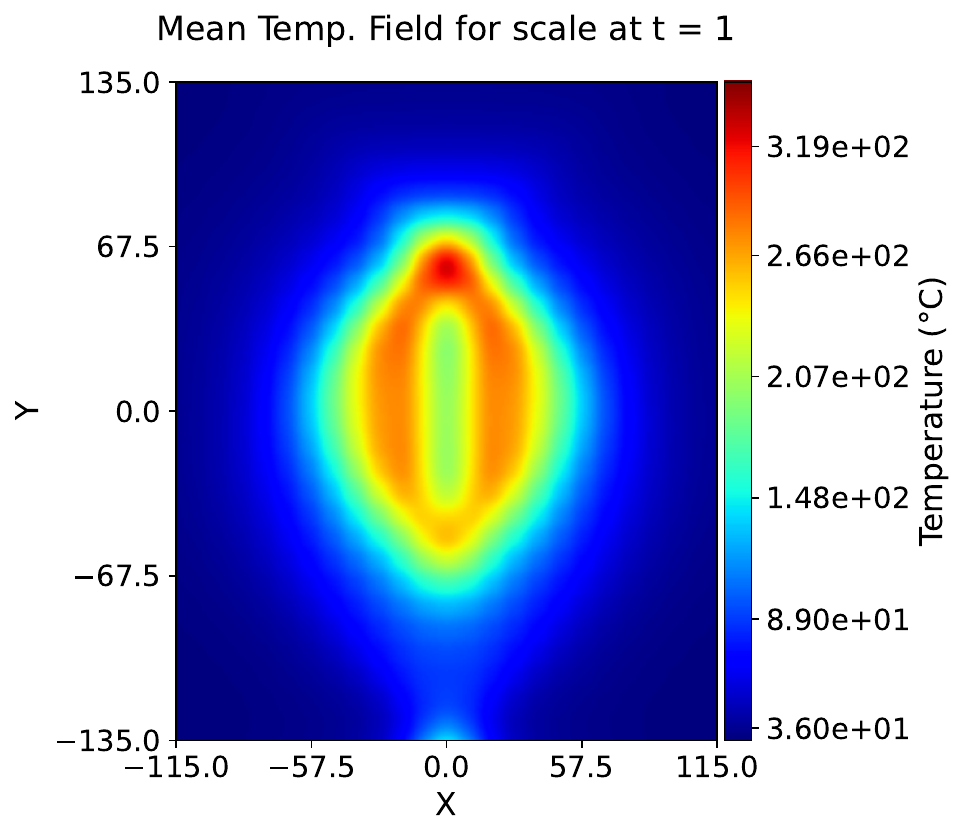}
        \caption{Mean of temperature field for scaling uncertainty.}
        \label{fig:ref_T_mean_scl}
    \end{subfigure}    
    \hfill
    \begin{subfigure}[t]{0.32\textwidth}        
        \includegraphics[width=\textwidth, trim=0.cm 0.0cm 0.cm 0.7cm,clip]{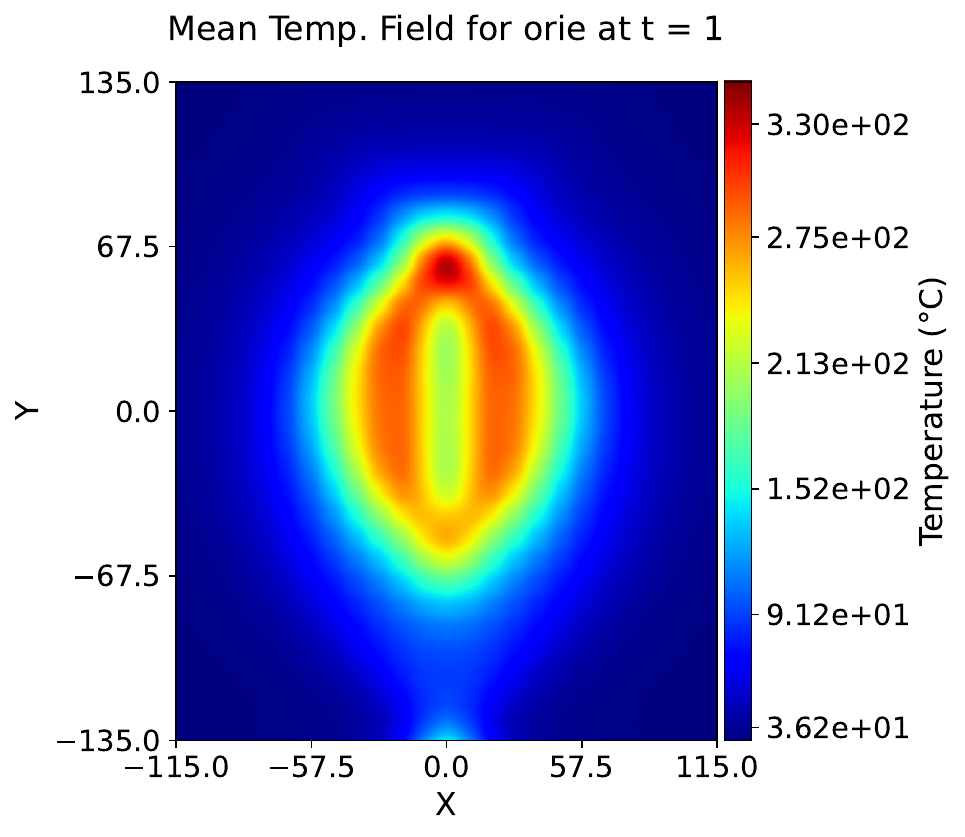}
        \caption{Mean of temperature field for orientation uncertainty.}
        \label{fig:ref_T_mean_ori}
    \end{subfigure} 
    \hfill    
    \begin{subfigure}[t]{0.32\textwidth}        
        \includegraphics[width=\textwidth, trim=0.cm 0.0cm 0.cm 0.7cm,clip]{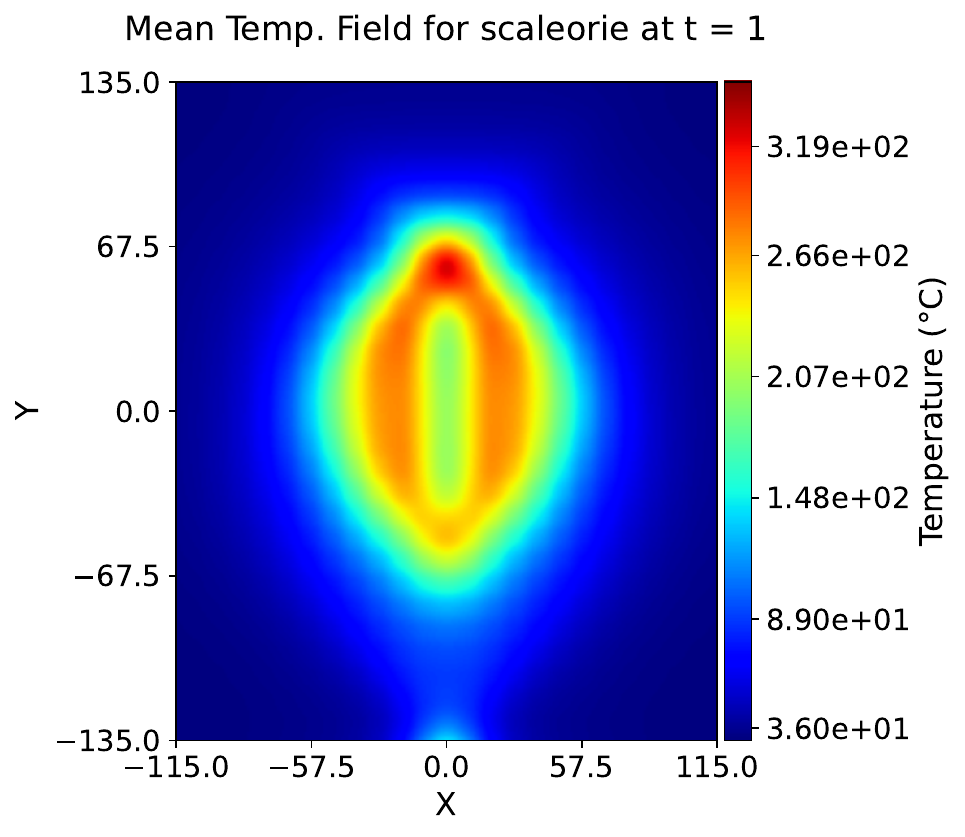}
        \caption{Mean of temperature field for scaling-orientation uncertainty.}
        \label{fig:ref_T_mean_sclori}
    \end{subfigure} 

    \begin{subfigure}[t]{0.32\textwidth}
        \includegraphics[width=\textwidth, trim=0.cm 0.0cm 0.cm 0.7cm,clip]{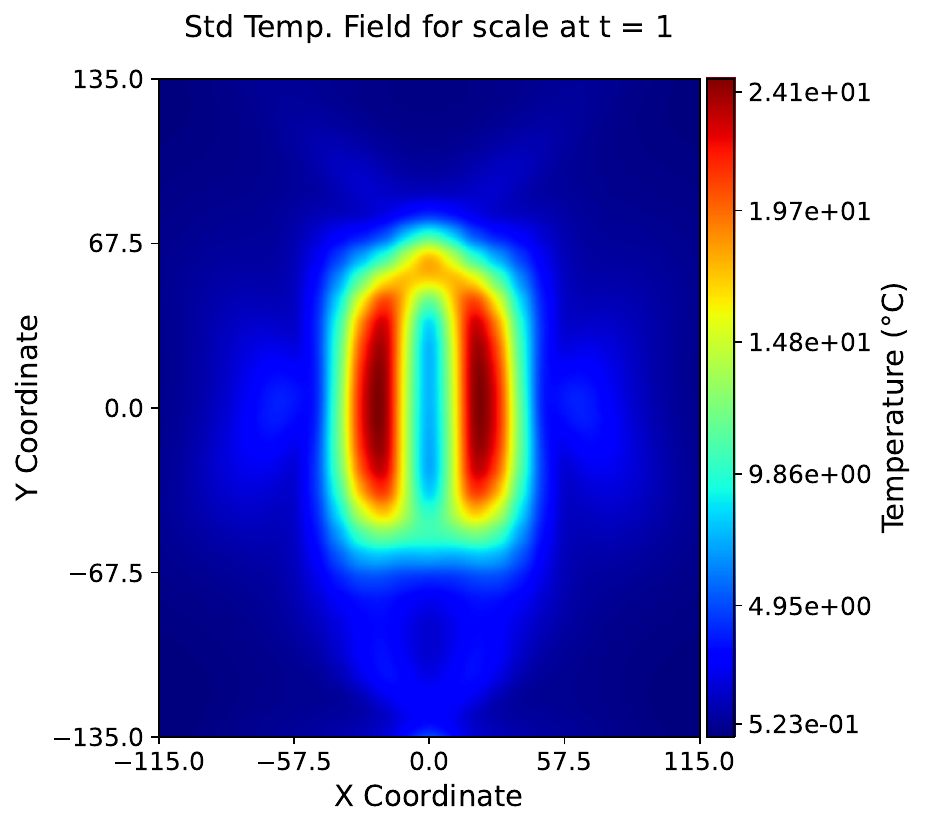}
        \caption{Std. in temperature field for scaling uncertainty.}
        \label{fig:ref_T_std_scl}
    \end{subfigure}    
    \hfill
    \begin{subfigure}[t]{0.32\textwidth}        
        \includegraphics[width=\textwidth, trim=0.cm 0.0cm 0.cm 0.7cm,clip]{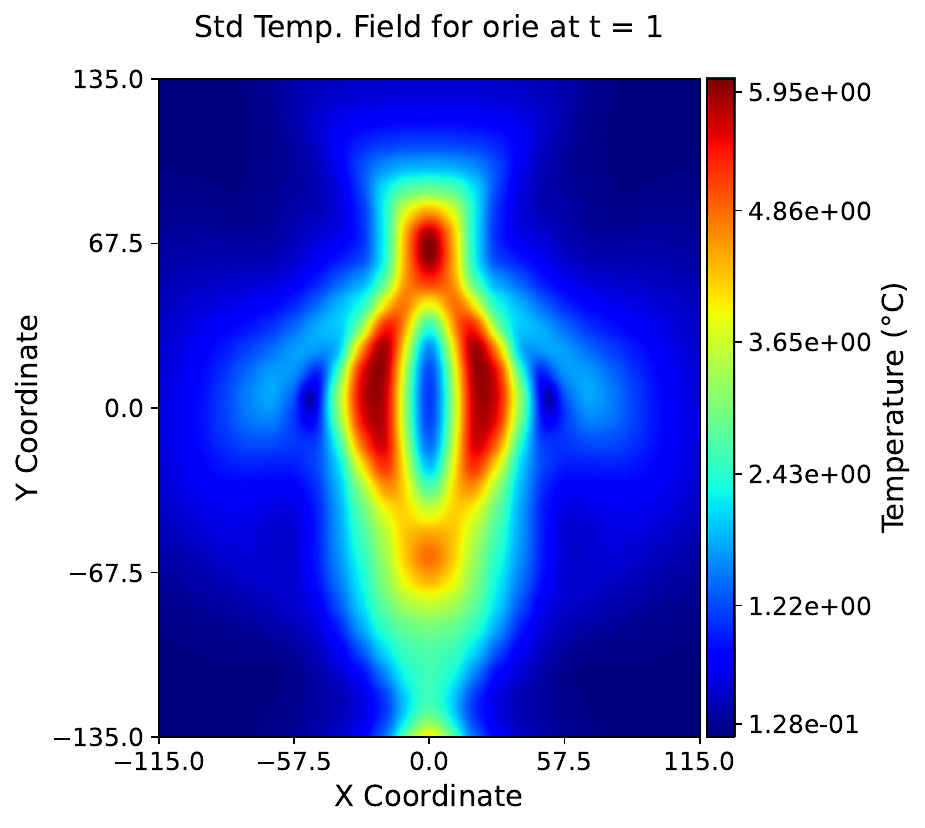}
       \caption{Std. in temperature field for orientation uncertainty.}
        \label{fig:ref_T_std_ori}
    \end{subfigure} 
    \hfill    
    \begin{subfigure}[t]{0.32\textwidth}        
        \includegraphics[width=\textwidth, trim=0.cm 0.0cm 0.cm 0.7cm,clip]{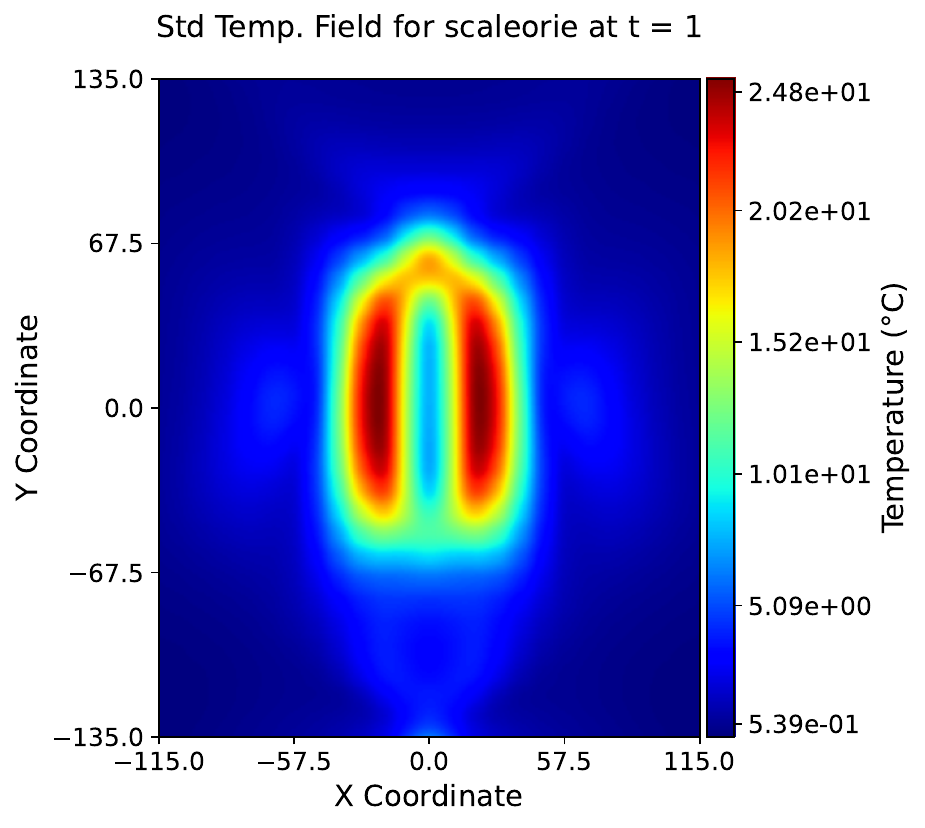}
        \caption{Std. in temperature field for scaling-orientation uncertainty.}
        \label{fig:ref_T_std_sclori}
    \end{subfigure} 
    
    \vspace{-0.25cm}
    \caption{Statistics for different types of uncertainties in the electrical conductivity during induction heating.}
    \label{fig:ref_mean_std_sclori}
\end{figure}

The mean and standard deviation of the volumetric loss $\varHsource$ and temperature fields $\varTemp$ for all three uncertainty cases are presented in \refFIG{\ref{fig:ref_mean_std_sclori}}. 
For the scaling uncertainty case, the mean fields in \refFIG{\ref{fig:ref_qh_mean_scl}} and \refFIG{\ref{fig:ref_T_mean_scl}} retain broadly the same contour structure as the deterministic solution, with only a slight modification of the spatial distribution. Because the three eigenvalues are sampled independently, each realisation has a slightly different anisotropy ratio rather than a common multiplicative factor, so averaging over the inputs does not exactly reproduce the deterministic field but nudges the mean while largely preserving its structure. The standard deviation fields shown in \refFIG{\ref{fig:ref_qh_std_scl}} and \refFIG{\ref{fig:ref_T_std_scl}} reveal elevated variability in the central region and along the parallel sections of the coil, respectively, likely due to the presence of higher volumetric loss thus high sensitivity to changes in electrical conductivity. 

For the orientation uncertainty case, the mean fields \refFIG{\ref{fig:ref_qh_mean_ori}} and \refFIG{\ref{fig:ref_T_mean_ori}} remain close to the deterministic solution, with slightly broader spatial distributions. The standard deviation of the volumetric loss \refFIG{\ref{fig:ref_qh_std_ori}} exhibits a modified contour along the parallel section of the coil, indicating a nonlinear interaction between fibre orientation variability and electromagnetic loss generation, though with considerably smaller standard deviation magnitude than in the scaling uncertainty case. More pronounced effects appear in the temperature standard deviation field \refFIG{\ref{fig:ref_T_std_ori}}, where larger deviations relatively to the scaling uncertainties are observed near the upper and lower sections of the coil.

 For the combined uncertainty case shown in \refFIG{\ref{fig:ref_qh_mean_sclori}, \ref{fig:ref_qh_std_sclori}, \ref{fig:ref_T_mean_sclori} and \ref{fig:ref_T_std_sclori}} the variance of the temperature $\varTemp$ is dominated by the scaling contribution, consistent with the comparatively large standard deviations of the scaling parameters. The spatial patterns reflect this dominance, with the orientation contribution introducing only a small increase to the field magnitudes. 

\begin{figure}[!]               
    \captionsetup{justification=centering,margin=0cm}
    \centering
    \begin{subfigure}[t]{0.45\textwidth}
        \includegraphics[width=\textwidth, trim=0.0cm 0.0cm 0.0cm 0.0cm,clip]{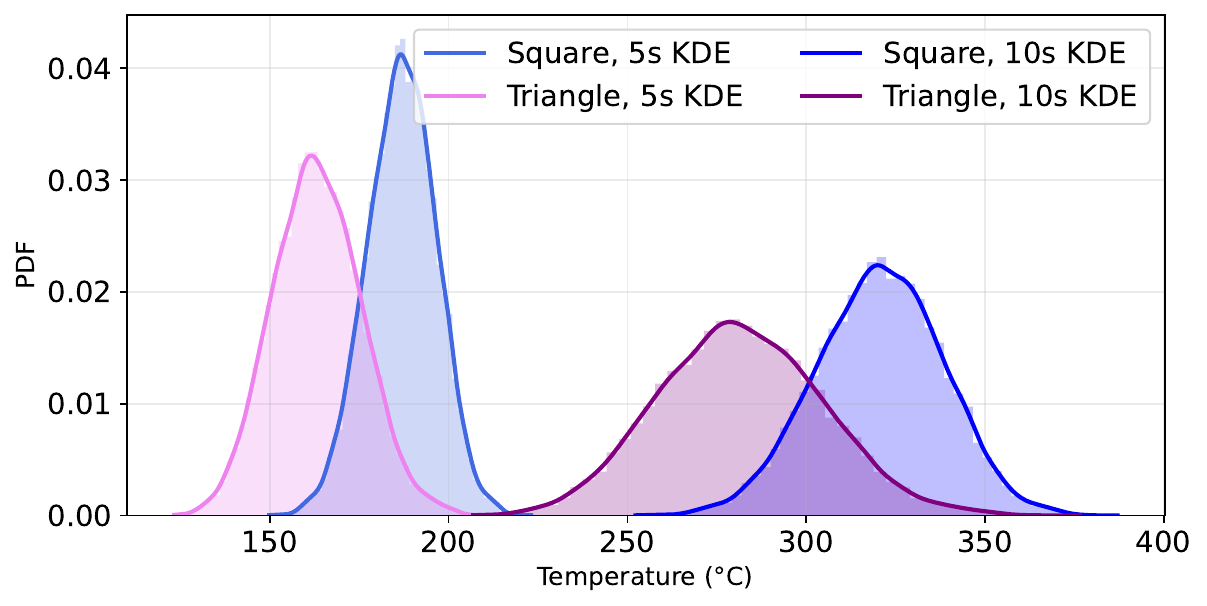}
        \caption{PDFs for scaling-only uncertainty.}
        \label{fig:pdf_marked_scl}
    \end{subfigure}    
    \hfill
    \begin{subfigure}[t]{0.45\textwidth}        
        \includegraphics[width=\textwidth, trim=0.0cm 0.0cm 0.0cm 0.0cm,clip]{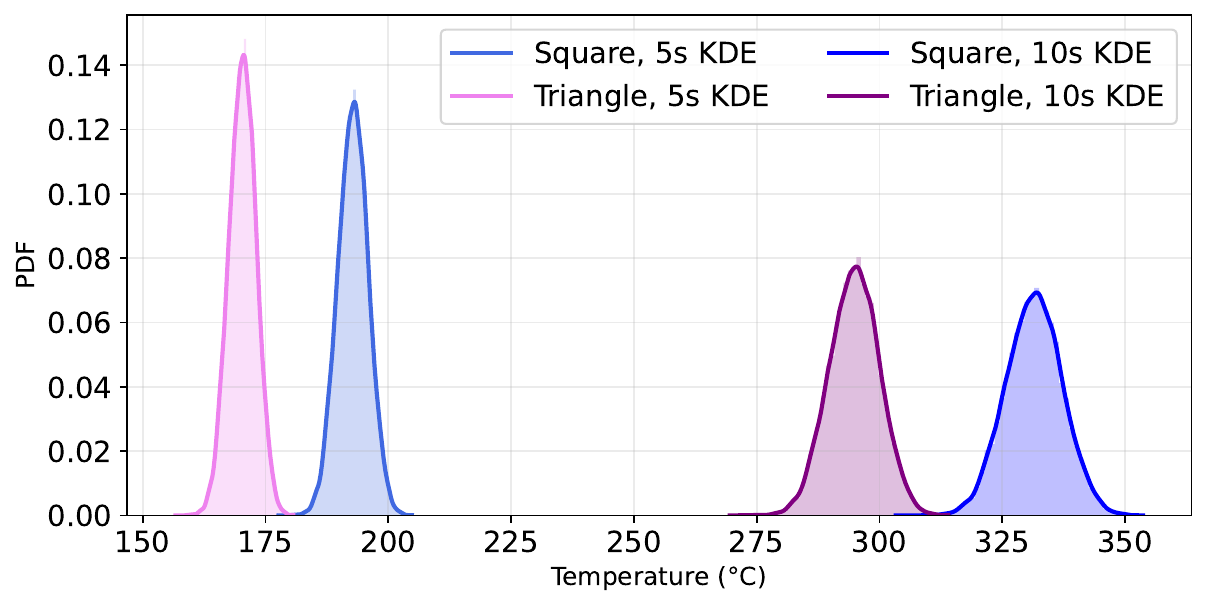}
        \caption{PDFs for orientation-only uncertainty.}
        \label{fig:pdf_marked_ori}
    \end{subfigure}  

    \begin{subfigure}[t]{0.45\textwidth}        
        \includegraphics[width=\textwidth, trim=0.cm 0.0cm 0.cm 0.0cm,clip]{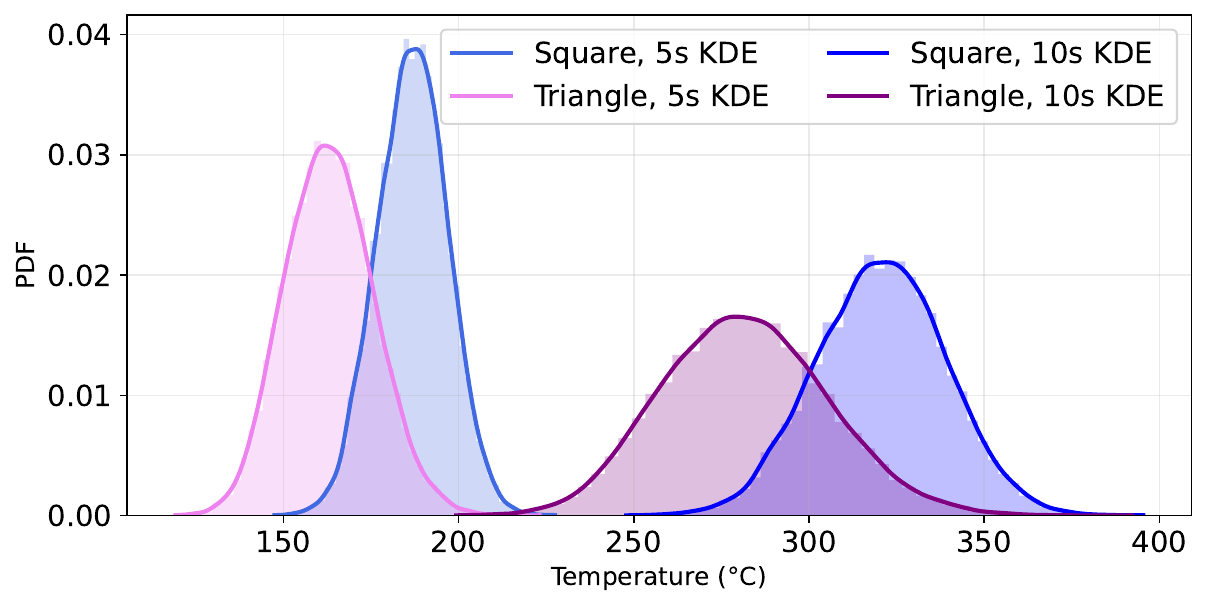}
        \caption{PDFs for scaling-orientation uncertainty}
        \label{fig:pdf_marked_sclori}
    \end{subfigure} 
    
    \vspace{-0.25cm}
    \caption{Probability density functions obtained by kernel density estimates at 5s and 10s for points marked in \refFIG{\ref{fig:det_temp}}.}
    \label{fig:pdf_marked}
\end{figure}

The stochastic temperature response is further analysed through probability density functions evaluated at $\vartime = 5\,\text{s}$ and $\vartime = 10\,\text{s}$ at the two locations indicated in \refFIG{\ref{fig:pdf_marked}}. An overall increase in variability over time is observed, consistent with the cumulative propagation of uncertainty through the transient heat-transfer process. The distributions associated with the scaling and orientation uncertainty cases exhibit noticeable right skewness, reflecting the asymmetric propagation of the lognormal scaling inputs through the thermal system. The combined uncertainty case exhibits a similar skewness character, dominated by the scaling contribution, though the aggregation of multiple independent sources introduces reduced skewness, consistent with the central limit effect. 

Overall, the results demonstrate a pronounced sensitivity of both the electromagnetic losses and the thermal response to the parameterisation of the electrical conductivity, with scaling uncertainty constituting the dominant source of variability in the predicted temperature field. 

%% file: 4_Result/Result_Surrogates.tex
\noindent
Part of the dataset generated for the Monte Carlo analysis is reused for training the surrogate model of Section~\ref{sec:method}. Only the combined scaling-and-orientation case is considered. The mean and standard deviation of the corresponding temperature fields used as the reference solution are shown in \refFIG{\ref{fig:ref_T_mean_sclori}} and \refFIG{\ref{fig:ref_T_std_sclori}}, and inputs and outputs are organised as in \refEQ{\ref{eq:EMHT_dataset}}.

The input vector $\boldsymbol{q}$ concatenates two components. The first is the current temperature field at the integration points, of dimension $24 \times 28 \times 4 = 2688$. The second is the StrAng embedding $f_\zeta(\varTensor)$ from \refEQ{\ref{eq:StrAng}}, of dimension $16$, comprising 3 eigenvalues and a single Z-axis rotation angle per ply across the four plies. The restriction of the rotation to the Z-axis compresses the $4 \times 6 = 24$ independent components of the unconstrained tensor field to this 16-dimensional manifold-aware representation. The output is the temperature field at the same quadrature points one time step ahead, of dimension $2688$.

To investigate the effect of the Neural ODE formulation on approximation performance, the Euler-based time integration approach shown in \refEQ{\ref{eq:euler_nn}} is compared with the explicit RK4 \refEQ{\ref{eq:rk4_network_discrete_time}} and implicit Adams--Moulton \refEQ{\ref{eq:adams_moulton2_nn}} Neural ODE integrators. These three networks are hereafter referred to as Euler, RK4, and AM, respectively.

\begin{figure}[!b]               
    \captionsetup{justification=centering,margin=0cm}
    \centering
    \vspace{0.5cm}
    \begin{subfigure}[t]{0.45\textwidth}
        \includegraphics[width=\textwidth, trim=0.0cm 0.0cm 0.0cm 0.75cm,clip]{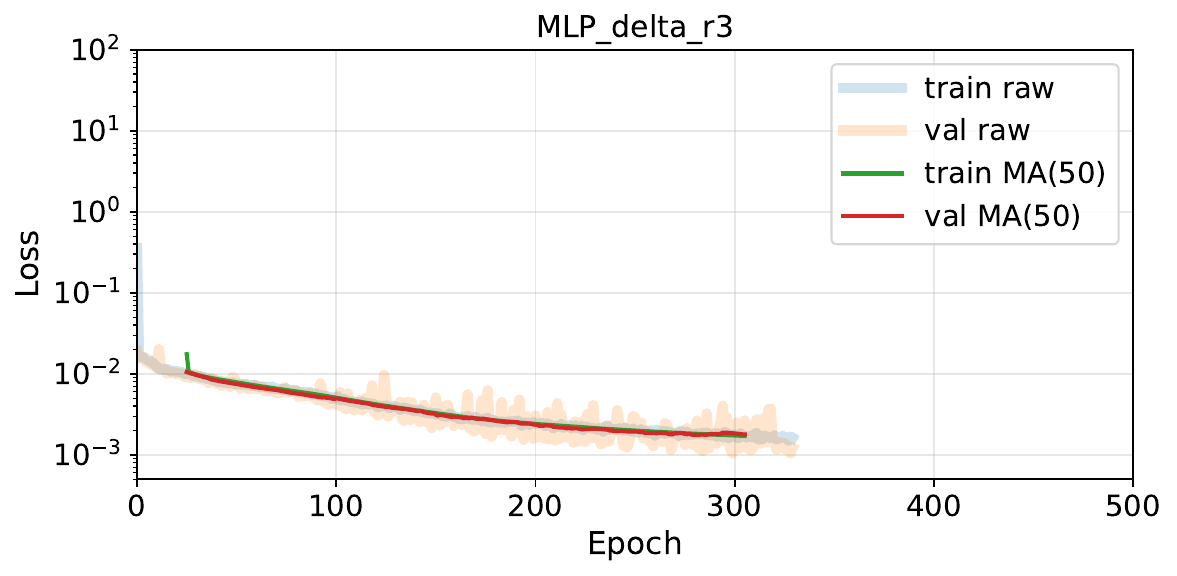}
        \caption{Training history for the Euler network.}
        \label{fig:MLP_train_loss}
    \end{subfigure}    
    \hfill
    \begin{subfigure}[t]{0.45\textwidth}        \includegraphics[width=\textwidth, trim=0.0cm 0.0cm 0.0cm 0.75cm,clip]{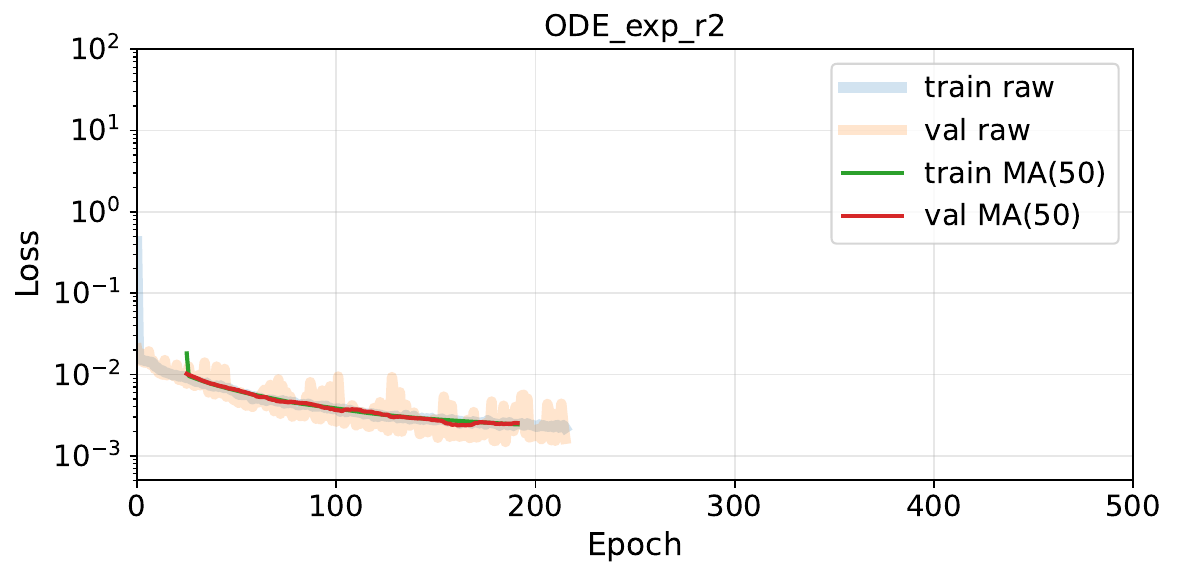}
        \caption{Training history for the explicit NODE network.}
        \label{fig:ODE_exp_train_loss}
    \end{subfigure} 
    
    \begin{subfigure}[t]{0.45\textwidth}        
        \includegraphics[width=\textwidth, trim=0.0cm 0.0cm 0.0cm 0.75cm,clip]{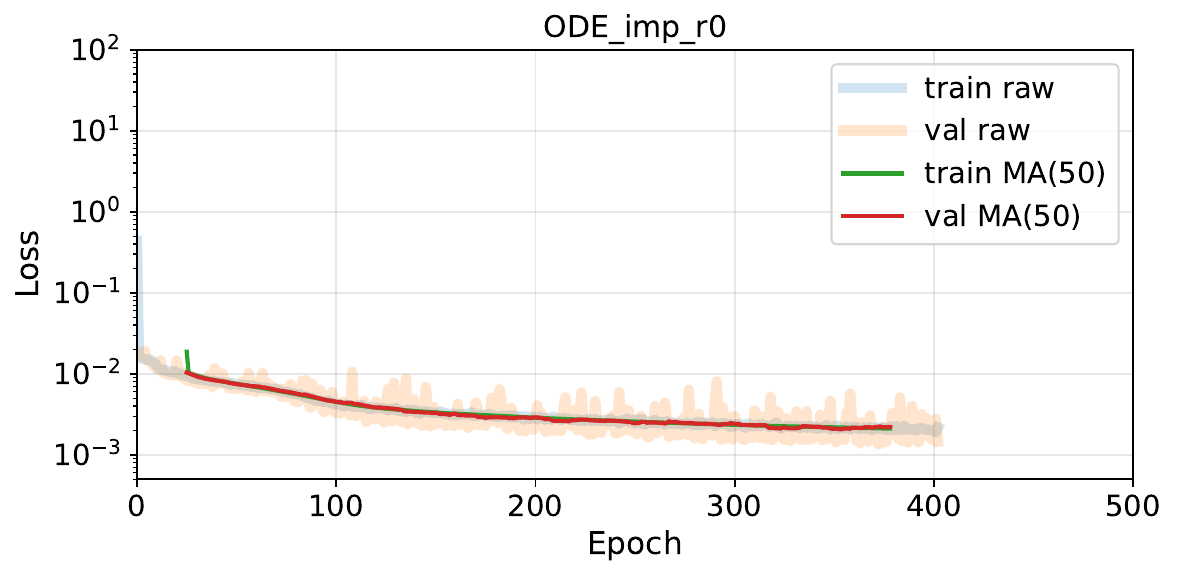}
        \caption{Training history for the implicit NODE network.}
        \label{fig:ODE_imp_train_loss}
    \end{subfigure}     
    \caption{Loss history for scaling and orientation uncertainty.}
    \label{fig:sclori_history}    
\end{figure}

The architectures of all networks are kept as similar as possible to ensure a fair comparison. Each model employs the StrAng layer and residual blocks within the hidden layers as presented in \refEQ{\ref{eq:StrAng}}. To determine a set of shared hyperparameters, an equal-weighted multi-objective Bayesian hyperparameter search was performed using a Tree-structured Parzen Estimator as implemented in \cite{akiba_optuna_2019}, with each network's validation loss contributing equally to the joint search objective. The tuneable parameters included the activation function, loss function, hidden layer width and depth, batch size, and learning rate. The resulting configuration consists of four hidden layers of width 12, the Sigmoid Linear Unit (SiLU) \cite{elfwing_sigmoid-weighted_2018} activation function, a batch size of 128, mean squared error as the training loss, and a learning rate of $7.5 \times 10^{-5}$. Fixed training parameters include an ODE solver step size of $0.1\,\text{s}$, corresponding to five sub-steps per $\Delta t = 0.5\,\text{s}$ integration interval. The training data is a subset of 5000 of the 20{,}000 Monte Carlo realisations, split 70--15--15\% into training, validation, and test sets, yielding $3500 \times 20 = 70{,}000$ training input--output pairs. All temperature values are normalised to $[0, 1]$ via min-max scaling fitted on the training set, with the same scaling applied to the validation and test data.

Surrogate accuracy is evaluated using the loss definition $\mathcal{E}$ introduced in \refEQ{\ref{eq:surrogate_target_MSE}}, consistent with the training objective. The loss is evaluated on the validation dataset over the reduced horizon $\vartime \in [2.5,\,7.5]\,\text{s}$ and on the test dataset over the full horizon $\vartime \in [0,\,10]\,\text{s}$. Comparisons of statistics and distributions are evaluated on the full set of 20{,}000 sample trajectories, generated with the same random seed as the Monte Carlo reference, so that the surrogate and reference solutions can be compared realisation-by-realisation. The training procedure follows a standard supervised setup. Network parameters $\varWeights$ are updated by minimising the loss \refEQ{\ref{eq:gen_loss_func}} on mini-batches drawn from the training subset using the Adam optimiser \cite{kingma_auto-encoding_2014} with gradient for the NODE-networks computed through an adjoint solve \cite{chen_neural_2018}. Mini-batches are reshuffled between epochs. Training terminates either at a fixed budget of $N_{\text{epoch}} = 1001$ epochs or when the validation loss fails to improve over $N_{\text{patience}} = 30$ consecutive epochs, whichever occurs first. To assess the robustness of the learning procedure, this process is repeated five times with different random network initialisations and dataset shuffling.

The training and validation histories of the best-performing networks in terms of test loss across the repeated experiments are shown in \refFIG{\ref{fig:sclori_history}}. The Euler network in \refFIG{\ref{fig:MLP_train_loss}}, together with the explicit and implicit NODE networks in \refFIG{\ref{fig:ODE_exp_train_loss}} and \refFIG{\ref{fig:ODE_imp_train_loss}}, exhibit comparable learning patterns, with the training losses decreasing steadily and settling into the low $10^{-2}$ range within the available number of epochs. The validation histories are decently noisy across all three networks, showing pronounced epoch-to-epoch jumps of similar magnitude while broadly tracking their respective training curves. A drawback of the NODE is a slower evaluation time. Under the current setting, inference for a single sample is on the order of $10^{-3}\,\mathrm{s}$ for the Euler network and $10^{-2}\,\mathrm{s}$ for the Neural ODE models, roughly an order of magnitude slower. This follows directly from the need to integrate the learnt vector field $f_{\varWeights}$ across each interval rather than apply a single explicit update. 


\begin{table}[t!]
  \caption{Performance comparison best performing networks for the Euler (EUL) , explicit (EXP) NODE and implicit (IMP) NODE networks.}
  \vspace{-0.25cm}
  \centering
  \setlength{\tabcolsep}{2pt}
  \hspace*{-0.25cm}
  \begin{tabular}{l|lll|lll|lll|lll}
    \toprule
    Error
    & \multicolumn{3}{c|}{MSE}
    & \multicolumn{3}{c|}{Mean L2}
    & \multicolumn{3}{c|}{Std L2}
    & \multicolumn{3}{c}{KL} \\
    \cmidrule(lr){2-4}
    \cmidrule(lr){5-7}
    \cmidrule(lr){8-10}
    \cmidrule(lr){11-13}
    & \multicolumn{1}{c}{EUL} & \multicolumn{1}{c}{RK4} & \multicolumn{1}{c}{AM}
    & \multicolumn{1}{c}{EUL} & \multicolumn{1}{c}{RK4}& \multicolumn{1}{c}{AM}
    & \multicolumn{1}{c}{EUL} & \multicolumn{1}{c}{RK4} & \multicolumn{1}{c}{AM}
    & \multicolumn{1}{c}{EUL} & \multicolumn{1}{c}{RK4} & \multicolumn{1}{c}{AM} \\
    \midrule
    Val set  
    & 0.10$e^{-2}$ & 0.15$e^{-2}$ & 0.13$e^{-2}$
    & - & - & -
    & - & - & -
    & - & - & - \\
    Test set 
    & 0.11$e^{-2}$ & 0.15$e^{-2}$ & 0.14$e^{-2}$
    & 0.36 & 0.44 & 0.33
    & 0.15 & 0.21 & 0.16
    & 0.22$e^{-1}$ & 0.42$e^{-1}$ & 0.83$e^{-1}$ \\
    t[5-15]  
    & 0.89$e^{-1}$ & 1.26$e^{-1}$ & 1.35$e^{-1}$
    & 3.01 & 7.65 & 4.27
    & 4.34 & 2.71 & 5.69
    & 0.11$e^{-1}$ & 0.28$e^{-1}$ & 0.54$e^{-1}$ \\
    t[0-20]  
    & 0.72 & 1.23 & 1.40
    & 6.89 & 35.76 & 19.08
    & 10.78 & 9.10 & 27.45
    & 0.05 & 0.17 & 0.22 \\
    \bottomrule
  \end{tabular}
  \label{tab:sclori_results}
\end{table}

\begin{table}[b!]
  \caption{Neural network performance over all repeated numerical experiments reported with mean $\pm$ std.}
  \vspace{-0.25cm}
  \centering
  \setlength{\tabcolsep}{3pt}
    \begin{minipage}{0.75\textwidth}
    \centering
    \setlength{\tabcolsep}{4pt}
    
    \begin{tabular}{l|ll|ll|ll}
    \toprule
    & \multicolumn{2}{c|}{EUL-$\delta$}
    & \multicolumn{2}{c|}{ODE-RK4}
    & \multicolumn{2}{c}{ODE-AM} \\
    
    & \multicolumn{1}{c}{mean} & \multicolumn{1}{c}{$\pm$}
    & \multicolumn{1}{c}{mean} & \multicolumn{1}{c}{$\pm$}
    & \multicolumn{1}{c}{mean} & \multicolumn{1}{c}{$\pm$} \\
    \midrule
    
    Val set
    & 1.59e$^{-3}$ & 5.2e$^{-4}$
    & 1.63e$^{-3}$ & 1.5e$^{-4}$
    & 1.52e$^{-3}$ & 1.3e$^{-4}$ \\
    
    Test set
    & 1.59e$^{-3}$ & 5.2e$^{-4}$
    & 1.64e$^{-3}$ & 1.6e$^{-4}$
    & 1.53e$^{-3}$ & 1.4e$^{-4}$ \\
    
    t[5-15]
    & 1.24e$^{-1}$ & 4.7e$^{-2}$
    & 1.38e$^{-1}$ & 1.8e$^{-2}$
    & 1.32e$^{-1}$ & 1.3e$^{-2}$ \\
    
    t[0-20]
    & 9.82e$^{-1}$ & 2.8e$^{-1}$
    & 1.07 & 0.11
    & 1.14 & 0.17 \\
    
    \bottomrule
    \end{tabular}
    \end{minipage}
    \vspace{2mm}
    \begin{minipage}{0.60\textwidth}
    \centering
    \setlength{\tabcolsep}{3pt}
    
    \begin{tabular}{l|ll|ll|ll}
    \toprule
    Metric
    & \multicolumn{2}{c|}{EUL-$\delta$}
    & \multicolumn{2}{c|}{ODE-RK4}
    & \multicolumn{2}{c}{ODE-AM} \\
    
    & \multicolumn{1}{c}{mean} & \multicolumn{1}{c}{$\pm$}
    & \multicolumn{1}{c}{mean} & \multicolumn{1}{c}{$\pm$}
    & \multicolumn{1}{c}{mean} & \multicolumn{1}{c}{$\pm$} \\
    \midrule
    
    Mean L2
    & 4.39e$^{-1}$ & 1.0e$^{-1}$
    & 4.75e$^{-1}$ & 7.8e$^{-2}$
    & 3.96e$^{-1}$ & 9.4e$^{-2}$ \\
    
    Std L2
    & 3.09e$^{-1}$ & 1.2e$^{-1}$
    & 2.09e$^{-1}$ & 4.9e$^{-2}$
    & 2.38e$^{-1}$ & 5.6e$^{-2}$ \\
    
    KL
    & 3.90e$^{-2}$ & 2.4e$^{-2}$
    & 6.35e$^{-2}$ & 2.4e$^{-2}$
    & 7.10e$^{-2}$ & 2.0e$^{-2}$ \\
    
    \bottomrule
    \end{tabular}
    \end{minipage}
  \label{tab:sclori_repeats}
\end{table}

The performance comparison is summarised in Table~\ref{tab:sclori_results}, while aggregated statistics over all repeated experiments are reported in Table~\ref{tab:sclori_repeats}. The three networks attain comparable accuracy, with the Euler network performing on par with, and frequently better than, the Neural ODE models. On the validation and test sets, all networks reduce the MSE to the order of $10^{-3}$, the Euler model attaining the lowest values ($1.0$--$1.1\times10^{-3}$) among the three. Trajectory-wise metrics on the test set remain closely clustered across all evaluated quantities. The mean $L_2$ error is $0.36\,^\circ\text{C}$ for the Euler network, against $0.44\,^\circ\text{C}$ and $0.33\,^\circ\text{C}$ for the explicit and implicit NODE models respectively, while the corresponding standard deviations lie between $0.15\,^\circ\text{C}$ and $0.21\,^\circ\text{C}$. The Kullback--Leibler (KL) divergence, used here as a measure of distributional agreement between the predicted and reference temperature distributions, is lowest for the Euler network, indicating that its predictions match the reference distribution at least as well as those of the Neural ODE models. The implicit scheme attains a marginally lower mean $L_2$ error than the Euler network, whereas the Euler network retains an advantage in KL divergence, indicating that the models trade off pointwise and distributional accuracy without a clear overall winner.

Over longer rollout horizons the networks remain closely matched, with the Euler network retaining a slight edge. Over the validation horizon $\vartime \in [2.5,\,7.5]\,\text{s}$ and the full test horizon $\vartime \in [0,\,10]\,\text{s}$, all three networks exhibit comparable error growth, and the Euler network attains the lowest rollout MSE and mean $L_2$ error over both horizons, indicating that its temporal stability is competitive with that of the Neural ODE models. 

The statistics of the repeated experiments shown in Table~\ref{tab:sclori_repeats} confirm these findings, but reveal a marked difference in run-to-run consistency. While all three networks attain comparable mean errors, the Euler network is substantially noisier across repeated training: its test-set MSE standard deviation is $5.2\times10^{-4}$, roughly a factor of four larger than the $1.4\times10^{-4}$ and $1.6\times10^{-4}$ obtained by the implicit and explicit NODE models. The same pattern holds for the trajectory-wise metrics, where the Euler network exhibits the largest spread in mean $L_2$ ($1.0\times10^{-1}\,^\circ\text{C}$, against $9.4\times10^{-2}\,^\circ\text{C}$ and $7.8\times10^{-2}\,^\circ\text{C}$ for the NODE models) and in standard deviation of the $L_2$ error ($1.2\times10^{-1}\,^\circ\text{C}$, versus $5.6\times10^{-2}\,^\circ\text{C}$ and $4.9\times10^{-2}\,^\circ\text{C}$). Thus, although the Euler network attains the lowest mean errors on several metrics, the Neural ODE models produce markedly more consistent results across repeated training, indicating greater robustness of the learned surrogate.

\begin{figure}[b!]      
    \captionsetup{justification=centering,margin=0cm}
    \centering
    \begin{subfigure}[t]{0.35\textwidth}
        \includegraphics[width=\textwidth, trim=0.0cm 0.0cm 0.0cm 0.8cm,clip]{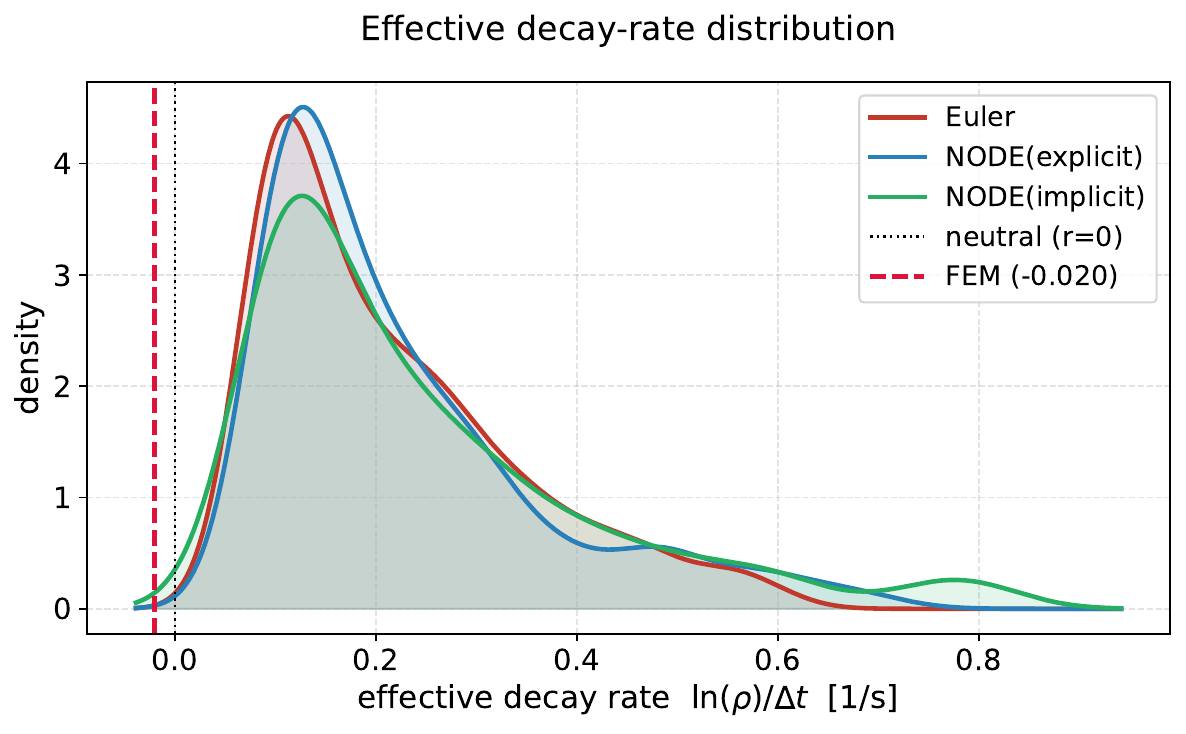}
        \caption{Distribution of the effective per-step amplification rate
        $r=\ln\rho/\Delta t$ across $500$ test states.}
        \label{fig:eigval_analysis}
    \end{subfigure}    
    \hfill
    \begin{subfigure}[t]{0.3\textwidth}
        \includegraphics[width=\textwidth, trim=0.0cm 0.0cm 0.0cm 0.9cm,clip]{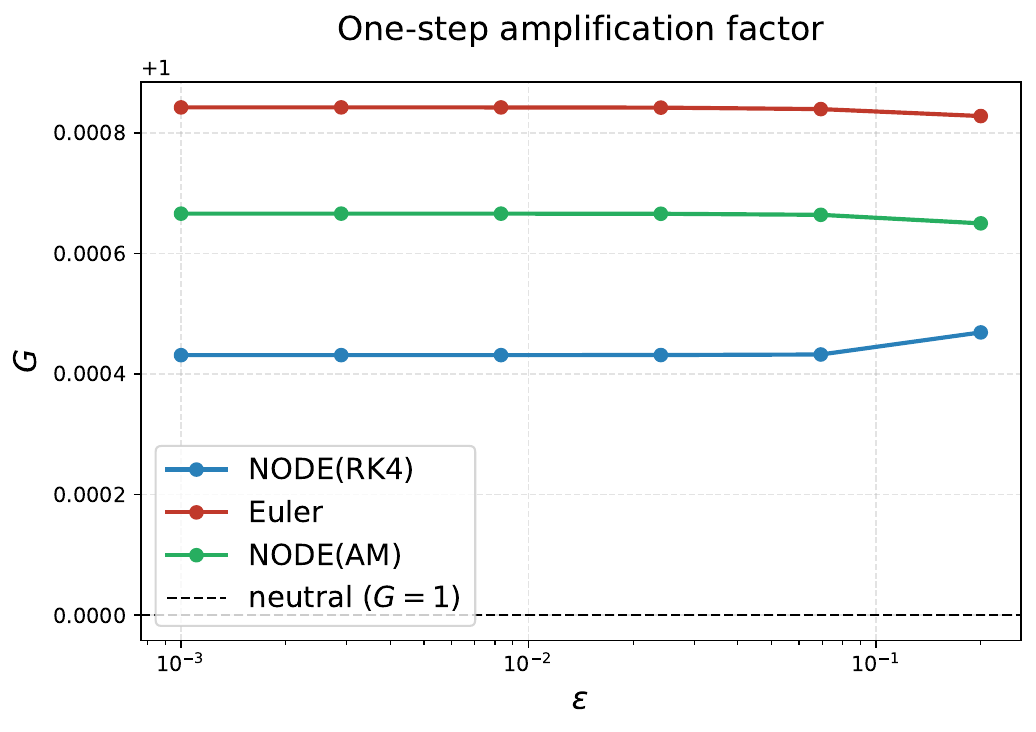}
        \caption{One-step amplification factor $G(\varepsilon)$ against input
        perturbation magnitude $\varepsilon$.}
        \label{fig:perturbation_amplification}
    \end{subfigure} 
    \hfill
    \begin{subfigure}[t]{0.3\textwidth}
        \includegraphics[width=\textwidth, trim=0.0cm 0.0cm 0.0cm 0.9cm,clip]{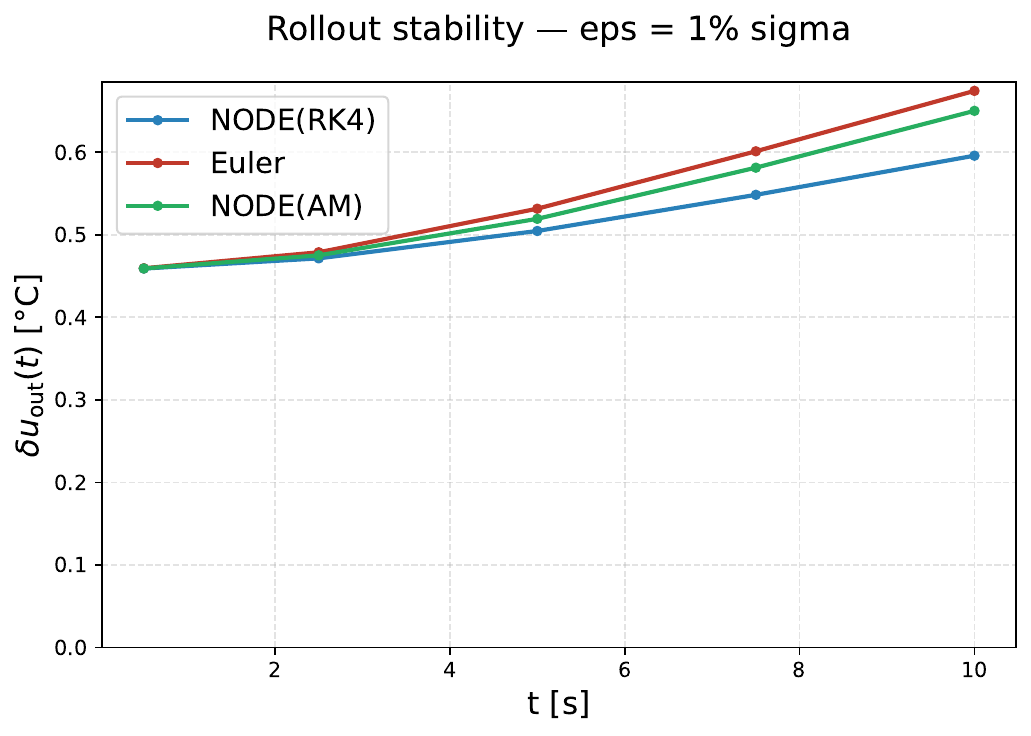}
        \caption{Growth of the initial-condition deviation
        $\lVert\boldsymbol{\delta}(t)\rVert$ over the rollout horizon.}
        \label{fig:perturbation_analysis}
    \end{subfigure}     
    \caption{Stability analysis of the learned surrogates.}
    \label{fig:stability_analysis}
\end{figure}

\begin{figure}[bh!]      
    \captionsetup{justification=centering,margin=0cm}
    \centering
    \begin{subfigure}[t]{0.32\textwidth}
        \includegraphics[width=\textwidth, trim=0.0cm 0.0cm 0.0cm 0.9cm,clip]{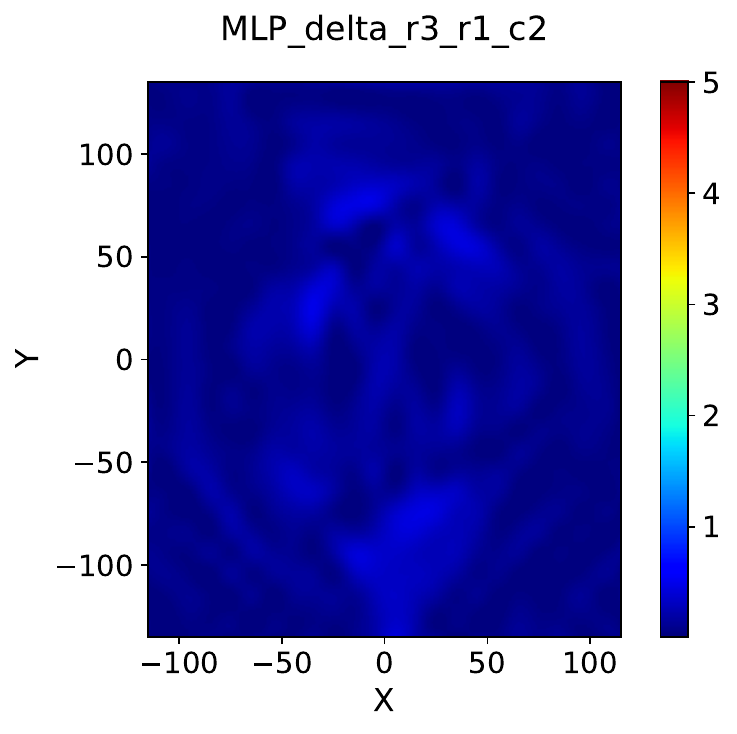}
        \caption{Euler NN absolute error in estimation of mean.}
        \label{fig:MLP_mean_abs}
    \end{subfigure}    
    \hfill
    \begin{subfigure}[t]{0.32\textwidth}
        \includegraphics[width=\textwidth, trim=0.0cm 0.0cm 0.0cm 0.9cm,clip]{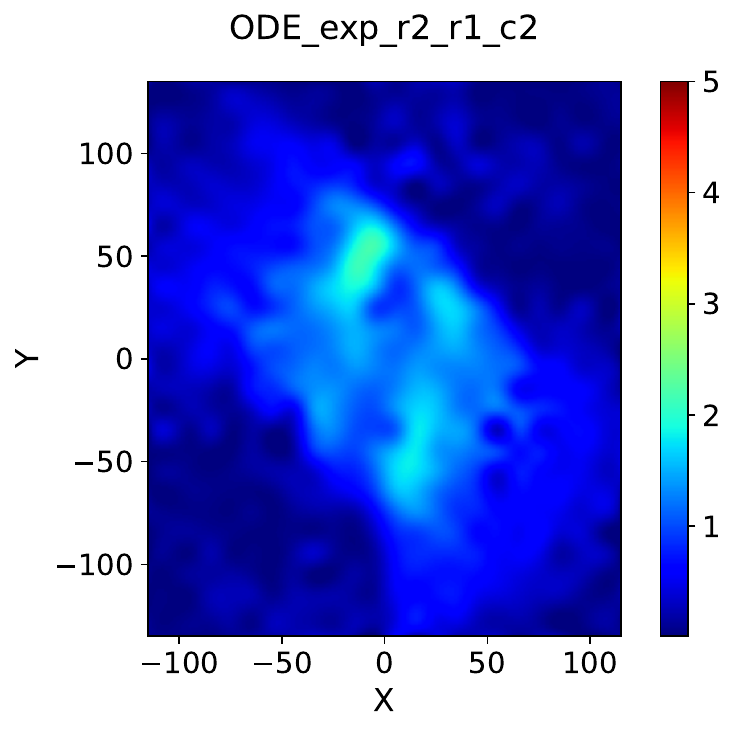}
        \caption{RK4 NN absolute error in estimation of mean.}
        \label{fig:EXP_mean_abs}
    \end{subfigure} 
    \hfill    
    \begin{subfigure}[t]{0.32\textwidth}        
        \includegraphics[width=\textwidth, trim=0.0cm 0.0cm 0.0cm 0.9cm,clip]{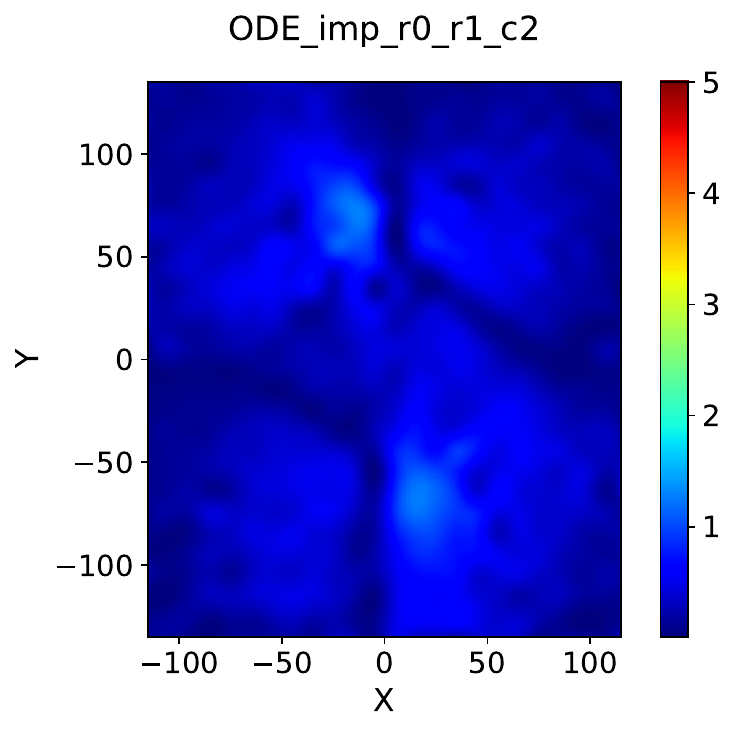}
        \caption{AM NN absolute error in estimation of mean.}
        \label{fig:IMP_mean_abs}
    \end{subfigure} 

    \begin{subfigure}[t]{0.32\textwidth}
        \includegraphics[width=\textwidth, trim=0.0cm 0.0cm 0.0cm 0.9cm,clip]{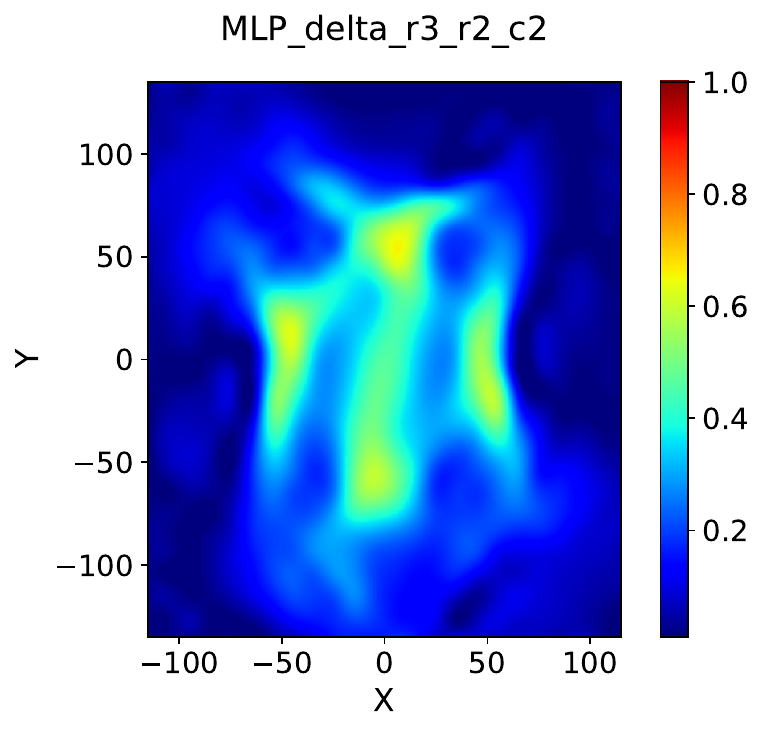}
        \caption{Euler NN absolute error in estimation of Std.}
        \label{fig:MLP_std_abs}
    \end{subfigure}    
    \hfill
    \begin{subfigure}[t]{0.32\textwidth}
        \includegraphics[width=\textwidth, trim=0.0cm 0.0cm 0.0cm 0.9cm,clip]{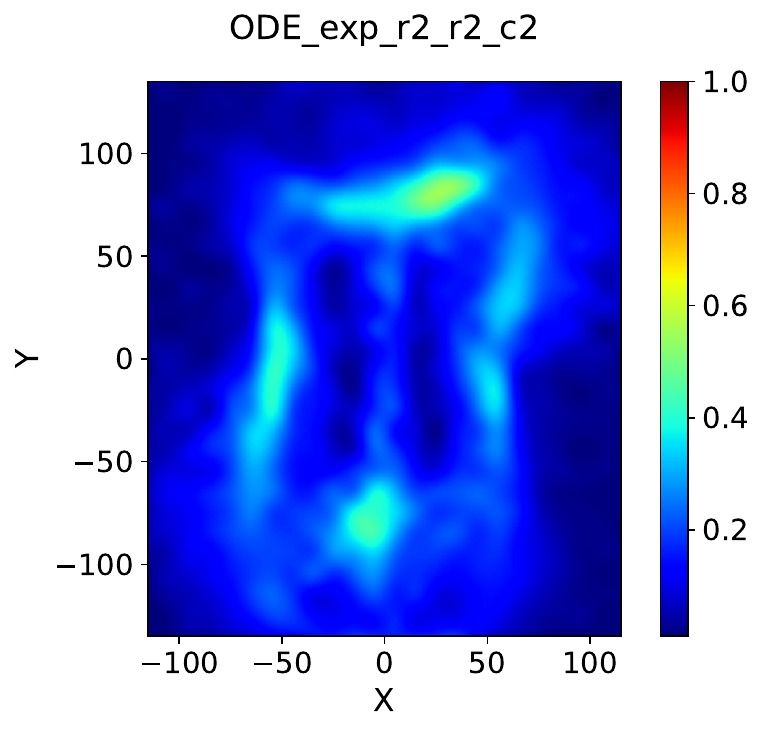}
        \caption{RK4 NN absolute error in estimation of Std.}
        \label{fig:EXP_std_abs}
    \end{subfigure} 
    \hfill    
    \begin{subfigure}[t]{0.32\textwidth}        
        \includegraphics[width=\textwidth, trim=0.0cm 0.0cm 0.0cm 0.9cm,clip]{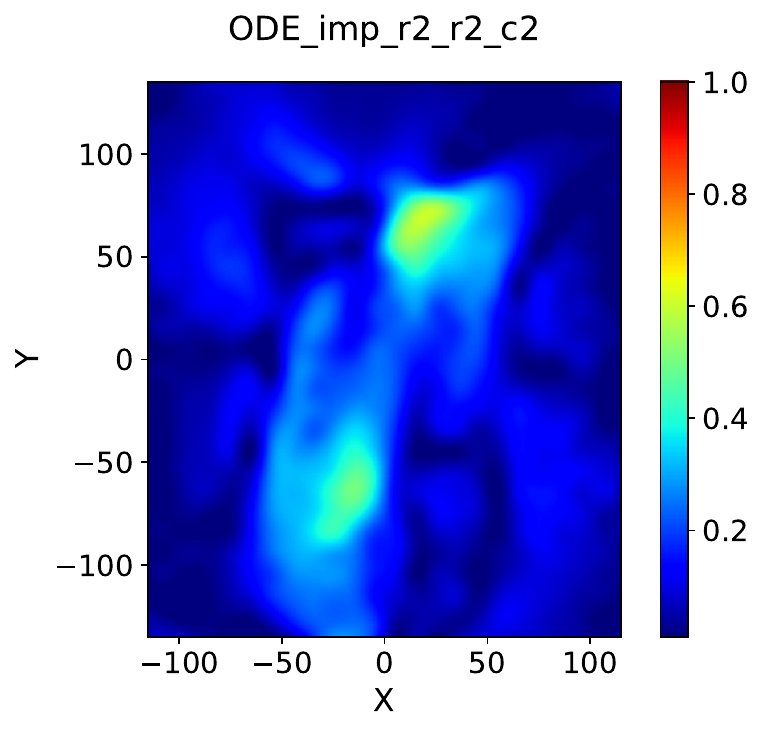}
        \caption{AM NN absolute error in estimation of Std.}
        \label{fig:IMP_std_abs}
    \end{subfigure} 

    \begin{subfigure}[t]{0.32\textwidth}
        \includegraphics[width=\textwidth, trim=0.0cm 0.0cm 0.0cm 0.9cm,clip]{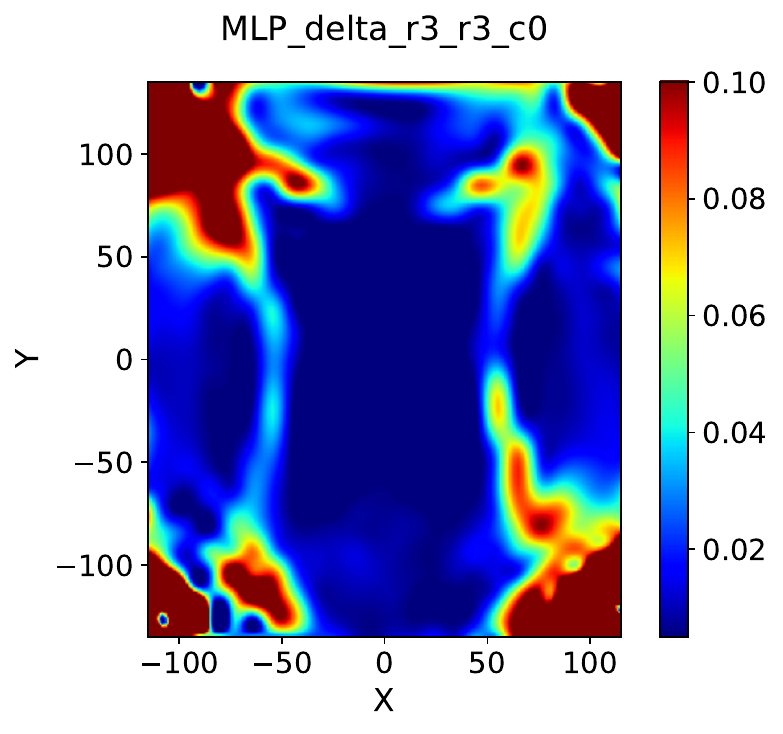}
        \caption{Euler NN absolute error in estimation of Std.}
        \label{fig:MLP_KLD_abs}
    \end{subfigure}    
    \hfill
    \begin{subfigure}[t]{0.32\textwidth}
        \includegraphics[width=\textwidth, trim=0.0cm 0.0cm 0.0cm 0.9cm,clip]{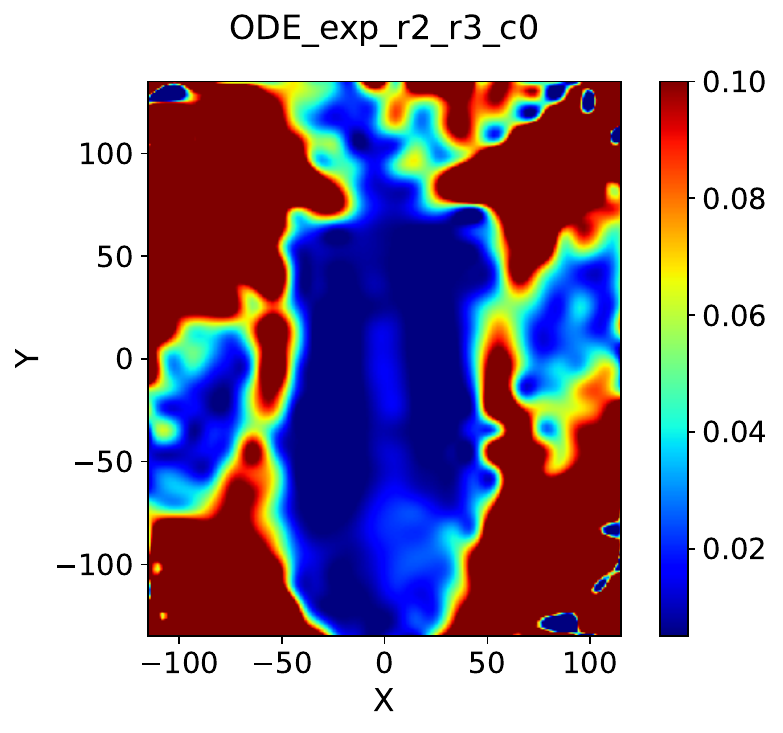}
        \caption{RK4 NN absolute error in estimation of Std.}
        \label{fig:EXP_KLD_abs}
    \end{subfigure} 
    \hfill    
    \begin{subfigure}[t]{0.32\textwidth}        
        \includegraphics[width=\textwidth, trim=0.0cm 0.0cm 0.0cm 0.9cm,clip]{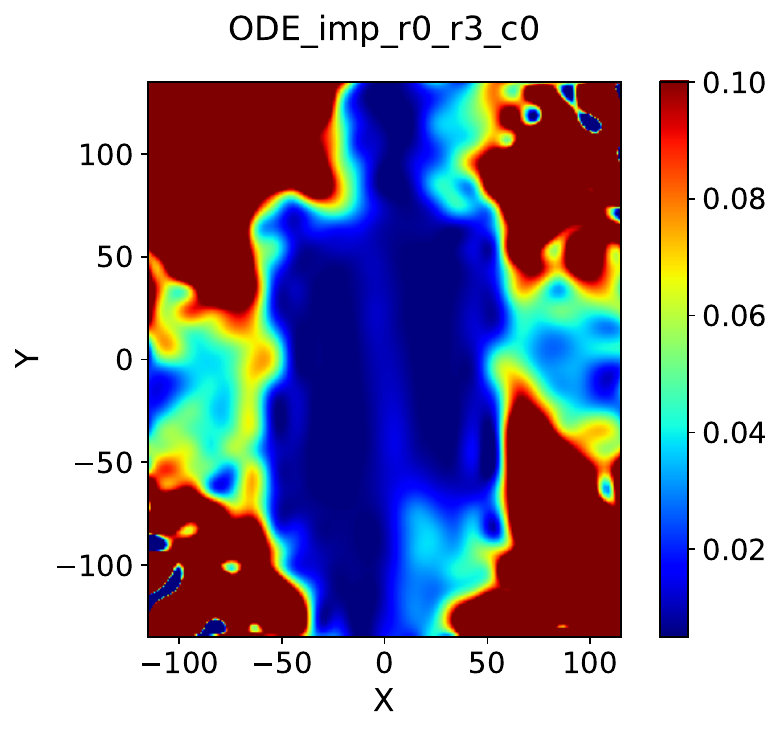}
        \caption{AM NN Kullbach leibner Divergence}
        \label{fig:IMP_KLD_abs}
    \end{subfigure} 

    \vspace{1.0cm}
    \caption{Errors in estimation of mean and std. for the different networks on rollouts from 0s to 10s.}
    \label{fig:mean_std_errors}
     \vspace{0.75cm}
\end{figure}

The local stability of the learned flow map is quantified in \refFIG{\ref{fig:eigval_analysis}}, which shows the distribution of the effective per-step amplification rate, $r = \ln\rho\!\left(\partial\Phi^{\xi}_{\Delta t}/\partial\boldsymbol{\varTemp}\right)/\Delta t$, over $500$ test states for each integrator. Here $r>0$ marks local amplification of perturbations and $r<0$ marks contraction. For all three networks the distribution lies almost entirely in the amplifying region, peaking near $r\approx0.12\,\mathrm{s}^{-1}$ with a tail reaching $r\approx0.8\,\mathrm{s}^{-1}$. The Euler network and the two Neural ODE variants are near-indistinguishable, so the amplification originates in the learned dynamics rather than in the time integrator. As the RK4, Adams--Moulton and Euler updates carry very different numerical damping yet give the same distribution.

This contrasts sharply with the reference finite-element operator $(-\boldsymbol{M}^{-1}\boldsymbol{K})$, whose slowest mode lies at $-2.0\times10^{-2}\,\mathrm{s}^{-1}$ (dashed line), i.e.\ weakly dissipative. The learned updates are therefore locally expansive across almost the entire state distribution. This local expansion carries through to the empirical probes as the one-step amplification factor exceeds unity across the full perturbation range (\refFIG{\ref{fig:perturbation_amplification}}), and the initial-condition deviation $\lVert\boldsymbol{\delta}(t)\rVert$ grows monotonically over the rollout (\refFIG{\ref{fig:perturbation_analysis}}) so the local instability compounds along the trajectory rather than averaging out.

This behaviour is a known failure mode of learned time-steppers, whose training objective rewards one-step accuracy but places no constraint on the spectrum of the update \cite{kim_stiff_2021}. A network can match the observed temperature increments closely while still acquiring a slightly amplifying Jacobian, since over the short training trajectories the resulting error growth stays small against the mean-temperature signal. Absent an explicit spectral or stability constraint, the fitted models are free to cross into the locally unstable regime \cite{manek_learning_2019}.

\subsection{Results Discussion}
\noindent
Several limitations of the present study warrant discussion. The surrogate models are trained and evaluated within the parameter ranges defined by the considered uncertainty scenarios, and their predictive capability outside these ranges remains untested. Extrapolation to longer time horizons or to larger deviations in electrical conductivity has not been assessed, and no claim is made regarding generalisation beyond the training distribution.

The spatial representation used during training introduces a potential source of error. The surrogate is trained on temperature values at a single integration point per element, whereas the FEM solver enforces accuracy at a $2\times2$ integration point configuration per element. This reduces memory and training cost but removes fine-scale information from the reference fields, and may obscure local temperature effects, potentially degrading accuracy of the reconstructed field relative to the full-order solution.

Error accumulation is stronger over $\vartime \in [0,\,10]\,\text{s}$ than over $\vartime \in [2.5,\,7.5]\,\text{s}$, reflecting the rapid gradient formation during the initial $0.5\,\text{s}$ heating stage, in which the temperature field transitions from a spatially uniform state to one with pronounced gradients. Errors introduced during this qualitatively distinct change propagate through the subsequent rollout.

Finally, the electrical conductivity is modelled as a spatially homogeneous random variable, which likely overestimates effective variability relative to realistic material behaviour in which spatial random field would be present. Furthermore, no experimental validation has been performed for the parameter ranges considered. The results should therefore be interpreted as a quantification of relative surrogate performance under the adopted modelling assumptions, rather than as absolute predictions of physical variability in induction heating processes.

%% file: 5_Conclusion/Conclusion.tex
\noindent
A manifold-aware and temporally stable surrogate framework was developed in this work for uncertainty propagation in induction heating of carbon-fibre-reinforced thermoplastics. It enables efficient stochastic analysis under manufacturing-induced variability in electrical conductivity at a reduced cost compared to full Monte Carlo simulation. The stochastic material model was introduced in the electromagnetic--heat-transfer model to represent this variability, respecting the SPD nature of the conductivity tensor and explicitly separating scaling and orientation uncertainties arising from local manufacturing imperfections in fibre placement and fibre-fibre contact quality.

The resulting stochastic conductivity model was first used to quantify uncertainty in the induction heating process through Monte Carlo simulations, providing detailed insight into induced currents and the resulting temperature field over time. The simulation data were subsequently used to train mainfold-aware surrogate models formulated as Constitutive Manifold Neural Networks (CMNNs), which explicitly respect the SPD manifold structure of the
conductivity tensor through the StrAng input layer, integrated with a Neural Ordinary Differential Equation (NODE) framework to capture temporal dynamics. The results reveal substantial temperature variability driven by stochastic conductivity, with uncertainty increasing continuously over time, consistent with the cumulative propagation of uncertainty through the transient heat-transfer process.

The numerical results demonstrate that the CMNN-Euler, CMNN-RK4, and CMNN-AM surrogates achieve comparable accuracy on unseen test cases and long-horizon rollouts, with mean $L_2$ errors of $0.36\,^\circ\text{C}$, $0.44\,^\circ\text{C}$, and $0.33\,^\circ\text{C}$ respectively. The Euler baseline attains the lowest mean errors on several metrics but exhibits markedly larger run-to-run variability, whereas the Neural ODE integrators produce more consistent results across repeated training, indicating greater robustness of the learned surrogate. The principal advantage of the Neural ODE formulation is the flexibility it affords in the choice of time-step and integration scheme at inference, decoupling the temporal discretisation from that seen during training and in principal enabling adaptive or variable step sizes without retraining. Between the two Neural ODE integrators, RK4 achieves lower mean error while AM better preserves the distributional character of the temperature field, indicating a trade-off between pointwise and distributional accuracy rather than a clear overall winner.

While demonstrated here for induction heating of CFRTPs, aspects of the surrogate pipeline are broadly applicable to any stochastic PDE system governed by SPD-valued material parameters, including diffusion, elasticity, and coupled multiphysics problems where manufacturing-induced anisotropy introduces parameter uncertainty. Future work should prioritise extending the conductivity parameterisation to spatially varying random tensor fields, which would allow the surrogate to capture local manufacturing variations and microstructural heterogeneity in CFRTPs that are not represented in the current homogeneous formulation. The rapid evaluation afforded by the surrogate opens further opportunities in induction heating process engineering, including real-time temperature control during welding and optimisation of coil geometry and process parameters.

%% file: 6_Appendix/mat_param.tex
\section{Unique Spectral Decomposition} \label{app:unique_decomp}

A unique spectral decomposition is required for StrAng, as eigenvectors can flip sign and eigenvalues can permute without changing the tensor. Such ambiguities may break physical consistency across samples. To enforce a consistent orientation, each decomposition is aligned with a reference tensor, for instance the mean, $\varTensorDet = \varEigVecDet\,\varEigValDet\,\varEigVecDet^{\!\top}$, by solving
\begin{equation} \label{eq:decomp_min}
(P, S) = \underset{P \in \mathcal{P},\, S \in \mathcal{S}}{\operatorname{argmin}}\, d_R(P,S),
\end{equation}
where $P$ represents one of six eigenvalue permutations, $S$ one of eight sign configurations for the eigenvectors, and $d_R(P,S)$ measures the misalignment with $\overline{\varTensor}$. The selected $(\hat P,\hat S)$ provides consistent alignment and is stable for rotational deviations up to approximately $45^\circ$, exceeding typical variability observed in advanced materials.